\documentclass[times, twocolumn]{aastex63}

\usepackage{amsmath}
\usepackage{graphicx}
\usepackage{natbib}
\usepackage{morefloats}
\usepackage{color}
\defcitealias{milli17}{M17}
\newcommand{\uatnum}[1]{\href{http://vocabs.ands.org.au/repository/api/lda/aas/the-unified-astronomy-thesaurus/current/resource.html?uri=http://astrothesaurus.org/uat/#1}{#1}}

\usepackage{ amssymb }

\begin{document}

\newlength{\figwidth}

\title{Multi-Band GPI Imaging of the HR 4796A Debris Disk}
ArXiv pre-print version (accepted in ApJ)

\author[0000-0002-8382-0447]{Christine Chen} \email{cchen@stsci.edu}
\affiliation{Space Telescope Science Institute (STScI), 3700 San Martin Drive, Baltimore, MD 21218, USA}
\affiliation{Department of Physics and Astronomy, The Johns Hopkins University, 3701 San Martin Drive, Baltimore, MD 21218, USA}

\author[0000-0002-9133-3091]{Johan Mazoyer}
\altaffiliation{NFHP Sagan Fellow}
\affiliation{NASA Jet Propulsion Laboratory, California Institute of Technology, Pasadena, CA 91109, USA}
\affiliation{LESIA, Observatoire de Paris, Université PSL, CNRS, Sorbonne Université, Université de Paris, 5 place Jules Janssen, 92195 Meudon, France}

\author[0000-0003-4845-7483]{Charles A. Poteet} 
\affiliation{Space Telescope Science Institute (STScI), 3700 San Martin Drive, Baltimore, MD 21218, USA}

\author[0000-0003-1698-9696]{Bin Ren}
\affiliation{Department of Astronomy, California Institute of Technology, 1200 East California Boulevard, Pasadena, CA 91125, USA}
\affiliation{Department of Physics and Astronomy, The Johns Hopkins University, 3701 San Martin Drive, Baltimore, MD 21218, USA}
\affiliation{Department of Applied Mathematics and Statistics, The Johns Hopkins University, 3400 North Charles Street, Baltimore, MD 21218, USA}

\author[0000-0002-5092-6464]{Gaspard Duch\^ene}
\affiliation{Astronomy Department, University of California, Berkeley, CA 94720, USA}
\affiliation{Universit\'e Grenoble-Alpes, CNRS Institut de Plan\'etologie et d'Astrophysique (IPAG), F-38000 Grenoble, France}

\author[0000-0001-9994-2142]{Justin Hom}
\affiliation{School of Earth and Space Exploration, Arizona State University, PO Box 871404, Tempe, AZ 85287, USA}

\author[0000-0001-6364-2834]{Pauline Arriaga}
\affiliation{Department of Physics \& Astronomy, 430 Portola Plaza, University of California, Los Angeles, CA 90095, USA}

\author[0000-0001-6205-9233]{Maxwell A. Millar-Blanchaer}
\affiliation{Department of Astronomy, California Institute of Technology, 1200 East California Boulevard, Pasadena, CA 91125, USA}
\affiliation{NASA Jet Propulsion Laboratory, California Institute of Technology, 4800 Oak Grove Drive, Pasadena, CA 91109, USA}

\author[0000-0001-7824-5372]{Jessica Arnold}
\affiliation{Department of Terrestrial Magnetism, Carnegie Institution for Science, 5421 Broad Branch Road, Washington, DC 20015, USA}

\author[0000-0002-5407-2806]{Vanessa P. Bailey}
\affiliation{NASA Jet Propulsion Laboratory, California Institute of Technology, 4800 Oak Grove Drive, Pasadena, CA 91109, USA}

\author[0000-0002-2731-0397]{Juan Sebasti{\'a}n Bruzzone}
\affiliation{Department of Physics and Astronomy, The University of Western Ontario, London, ON, N6A 3K7, Canada}

\author[0000-0001-6305-7272]{Jeffrey Chilcote}
\affiliation{Department of Physics, University of Notre Dame, 225 Nieuwland Science Hall, Notre Dame, IN 46556, USA}

\author[0000-0002-9173-0740]{\'Elodie Choquet} 
\affiliation{Aix Marseille Univ., CNRS CNES, LAM, Pôle de l'Etoile Site de Château-Gombert 38, rue Frédéric Joliot-Curie 13388 Marseille Cedex 13 France}

\author[0000-0002-4918-0247]{Robert J. De Rosa}
\affiliation{European Southern Observatory, Alonso de C\'{o}rdova 3107, Vitacura,
Santiago, Chile}

\author[0000-0002-1834-3496]{Zachary H. Draper}
\affiliation{University of Victoria, 3800 Finnerty Rd, Victoria, BC, V8P 5C2, Canada}

\author[0000-0002-0792-3719]{Thomas M. Esposito}
\affiliation{Astronomy Department, University of California, Berkeley, CA 94720, USA}

\author[0000-0002-0176-8973]{Michael P. Fitzgerald}
\affiliation{Department of Physics \& Astronomy, 430 Portola Plaza, University of California, Los Angeles, CA 90095, USA}

\author[0000-0002-7821-0695]{Katherine B. Follette}
\affiliation{Physics and Astronomy Department, Amherst College, 21 Merrill Science Drive, Amherst, MA 01002, USA}

\author[0000-0003-3726-5494]{Pascale Hibon}
\affiliation{European Southern Observatory, Alonso de C\'{o}rdova 3107, Vitacura,
Santiago, Chile}

\author[0000-0003-4653-6161]{Dean C. Hines}
\affiliation{Space Telescope Science Institute (STScI), 3700 San Martin Drive, Baltimore, MD 21218, USA}

\author[0000-0002-6221-5360]{Paul Kalas}
\affiliation{Astronomy Department, University of California, Berkeley, CA 94720, USA}
\affiliation{SETI Institute, Carl Sagan Center, 189 Bernardo Ave.,  Mountain View, CA 94043, USA}
\affiliation{Institute of Astrophysics, FORTH, GR-71110 Heraklion, Greece}

\author[0000-0001-7016-7277]{Franck Marchis}
\affiliation{SETI Institute, Carl Sagan Center, 189 Bernardo Avenue,  Mountain View, CA 94043, USA}

\author{Brenda Matthews}
\affiliation{SETI Institute, Carl Sagan Center, 189 Bernardo Avenue,  Mountain View, CA 94043, USA}

\author{Julien Milli}
\affiliation{European Southern Observatory (ESO), Alonso de Córdova 3107, Vitacura, Casilla, 19001, Santiago, Chile}

\author{Jennifer Patience}
\affiliation{School of Earth and Space Exploration, Arizona State University, PO Box 871404, Tempe, AZ 85287, USA}

\author[0000-0002-3191-8151]{Marshall D. Perrin} 
\affiliation{Space Telescope Science Institute (STScI), 3700 San Martin Drive, Baltimore, MD 21218, USA}

\author{Laurent Pueyo}
\affiliation{Space Telescope Science Institute (STScI), 3700 San Martin Drive, Baltimore, MD 21218, USA}

\author[0000-0002-9246-5467]{Abhijith Rajan}
\affiliation{Space Telescope Science Institute (STScI), 3700 San Martin Drive, Baltimore, MD 21218, USA}

\author[0000-0002-9667-2244]{Fredrik T. Rantakyr\"o}
\affiliation{Gemini Observatory, Casilla 603, La Serena, Chile}

\author[0000-0002-7535-2997]{Timothy J. Rodigas}
\affiliation{Department of Terrestrial Magnetism, Carnegie Institution for Science, 5421 Broad Branch Road, Washington, DC 20015, USA}

\author[0000-0002-7402-7797]{Gael M. Roudier}
\affiliation{NASA Jet Propulsion Laboratory, California Institute of Technology, 4800 Oak Grove Drive, Pasadena, CA 91109, USA}

\author[0000-0002-4511-5966]{Glenn Schneider}
\affiliation{Steward Observatory, The University of Arizona, 933 North Cherry Avenue, Tucson, AZ 85721, USA}

\author[0000-0003-2753-2819]{R\'emi Soummer} 
\affiliation{Space Telescope Science Institute (STScI), 3700 San Martin Drive, Baltimore, MD 21218, USA}

\author{Christopher Stark}
\affiliation{Space Telescope Science Institute (STScI), 3700 San Martin Drive, Baltimore, MD 21218, USA}

\author[0000-0003-0774-6502]{Jason J. Wang}
\altaffiliation{51 Pegasi b Fellow}
\affiliation{Department of Astronomy, California Institute of Technology, 1200 East California Boulevard, Pasadena, CA 91125, USA}

\author[0000-0002-4479-8291]{Kimberly Ward-Duong}
\affiliation{School of Earth and Space Exploration, Arizona State University, PO Box 871404, Tempe, AZ 85287, USA}
\affiliation{Physics and Astronomy Department, Amherst College, 21 Merrill Science Drive, Amherst, MA 01002, USA}

\author[0000-0001-6654-7859]{Alycia J. Weinberger}
\affiliation{Department of Terrestrial Magnetism, Carnegie Institution for Science, 5421 Broad Branch Road, Washington, DC 20015, USA}

\author[0000-0003-1526-7587]{David J. Wilner}
\affiliation{Harvard-Smithsonian Center for Astrophysics, 60 Garden Street, Cambridge, MA 02138, USA}

\author[0000-0002-9977-8255]{Schuyler Wolff}
\affiliation{Leiden Observatory, Leiden University, P.O. Box 9513, 2300 RA Leiden, The Netherlands}

\begin{abstract}
We have obtained Gemini Planet Imager (GPI) J-, H-, K1-, and K2-Spec observations of the iconic debris ring around the young, main sequence star HR 4796A. We applied several PSF subtraction techniques to the observations (Mask-and-Interpolate, RDI-NMF, RDI-KLIP, and ADI-KLIP) to measure the geometric parameters and the scattering phase function for the disk. To understand the systematic errors associated with PSF subtraction, we also forward modeled the observations using an MCMC frame work and a simple model for the disk. We found that measurements of the disk geometric parameters were robust with all of our analyses yielding consistent results; however, measurements of the scattering phase function were challenging to reconstruct from PSF subtracted images, despite extensive testing. As a result, we estimated the scattering phase function using disk modeling. We searched for a dependence of the scattering phase function with respect to the GPI filters but found none. We compared the H-band scattering phase function with that measured by \textit{HST} STIS at visual wavelengths and discover a blue color at small scattering angles and a red color at large scattering angles, consistent with predictions and laboratory measurements of large grains. Finally, we successfully modeled the SPHERE H2 HR 4796A scattered phase function using a Distribution of Hollow Spheres composed of silicates, carbon, and metallic iron.
\end{abstract}


\section{Introduction}
During the past twenty years, visual, infrared, and millimeter observations have provided incontrovertible evidence that most stars are surrounded at birth by circumstellar accretion disks, and at least some disks have built the $\sim$3000 extrasolar planets that have been discovered thus far \citep{williams11}. In general, high angular resolution visual to near-infrared images show starlight that is scattered off of predominantly sub-micron sized dust grains while those at mid-infrared to millimeter wavelengths show thermal emission from increasingly larger and larger dust grains \citep{hughes18}. At visual to near-infrared wavelengths, observations suffer from the complication that host stars are bright compared to circumstellar disks; thus, these observations benefit from the application of high contrast imaging techniques to reveal faint scattered light \citep{schneider99}. High contrast imaging techniques typically use Point Spread Function (PSF) subtraction to remove light in conjunction with coronagraphs that suppress light from the bright central star \citep{soummer12,lafreniere09}. Indeed, high contrast imaging studies have discovered planetary mass companions in a handful of protoplanetary and debris disks thus far \citep{meshkat15,marois10,lagrange10}. 

In disk imaging, disk geometry and surface brightness measurements can shed light on the dynamics and composition of the circumstellar dust and therefore the planets in the underlying planetary systems. High resolution images of debris disks have revealed that the dust in many systems appears sculpted into rings. Some rings are elliptical. In these systems, undetected planets on elliptical orbits may force the eccentricies of the parent bodies and dust into elliptical orbits \citep{wyatt99}. High resolution images have also revealed local brightness enhancements, or clumps, and warps in some disks; the properties of which have been used to effectively predict the presence of an exoplanet \citep{mouillet97}. However, these clumps may also be generated by giant collisions. In our Solar System, giant collisions between forming planets and Mars-sized impactors are believed to have ejected Mercury's mantle, formed the Moon, and created the large crater on Mars' northern hemisphere. Such events are expected to produce copious quantities of dust initially in very localized areas. Finally, high resolution imaging enables measurements of the scattering phase function. Measurements of the scattering phase function have been used to constrain grain size \citep{milli17} [hereafter \citetalias{milli17}]. Unfortunately, PSF subtraction techniques impact the field-of-view differently depending on the angular distance from the central star and the position angle. For companions, the impact on point source photometry and astrometry is relatively minor. However, for disks that are spatially extended (and can fill the field-of-view), the impact can be significant. For example, Angular Differential Imaging \citep[ADI,][]{Marois06} produces regions of self-subtraction around a disk which strongly impact the disk shape and local photometry \citep{milli12}.

In 2014, Gemini Observatory commissioned the Gemini Planet Imager (GPI), a second-generation high contrast imaging instrument that provides near-infrared integral field spectroscopy and polarimetry \citep{macintosh14, perrin15} on the Gemini South Telescope. Although GPI was primarily designed to search for and characterize Jovian mass planets, it has spatially resolved a handful of debris disks for the first time, particularly in the nearby ScoCen OB Association \citep{kalas15, draper16}. To date, GPI studies have focused on characterizing the gross morphology of disks because understanding the impact of PSF subtraction on high contrast imaging data is challenging. We are carrying out a multi-filter, integral field spectroscopy and polarimetry study of approximately one dozen bright debris disks that have been spatially resolved using \textit{Hubble Space Telescope} (\textit{HST}) as part of the 2015B Gemini Large and Long Program ``Characterizing Dusty Debris in Exoplanetary Systems'' (PI Chen). Our goal is to not only extract measurements of the disk geometry and surface brightness but also to combine multi-filter observations to provide the best constraints on the dust grain properties (e.g. size, porosity, shape, and composition). In 2014, the European Southern Observatory commissioned the Spectro-Polarimetic High contrast imager for Exoplanets REsearch \citep[SPHERE,][]{beuzit19} instrument, a second-generation high contrast imaging instrument for the Very Large Telescope. SPHERE addresses many of the same science goals as GPI with slightly different instrument capabilities.

The A0V star HR 4796A at a distance $d$ $\sim$ 72.8 pc possesses a spectacular narrow, inclined ring ($i$ = 76$\arcdeg$) imaged in scattered light \citep{schneider09, rodigas15, perrin15, schneider18} and thermal emission \citep{telesco00, kennedy18} with a semi-major axis $a$ $\sim$ 72 AU. An age of $\sim$8$\pm$2 Myr has been estimated for the central star based on the Lithium abundance and isochrone fitting of its M-type companion \citep{stauffer98} that is located 7.7$\arcsec$ from the primary star. Since the collisional and Poynting-Robertson drag lifetimes of the circumstellar dust are substantially shorter than the age of the star, the dust in this system is believed to be replenished by collisions among parent bodies \citep{jura93}. The lack of circumstellar dust close to the star has led to speculation that there are planetary mass companions sculpting the disk \citep{jura95}. However, no companions have yet been detected to a completeness limit of $\sim$4 $M_{Jup}$ \citep{milli17}; for reference, a Neptune-mass planet is sufficient to maintain the sharp inner edge of this disk. Recent, deep \textit{HST} STIS imaging has revealed that the HR 4796A ring is located within a substantially more spatially extended population of circumstellar dust that may be interacting with the local interstellar medium \citep{schneider18}. Recent SPHERE IRDIS and ZIMPOL observations \citep[\citetalias{milli17},][]{milli19} have provided exquisite measurements of the disk geometry, infrared scattering phase function, and visual polarized intensity phase function. Most interestingly, \cite{milli19} show that the North-South asymmetry in the HR 4796 disk can not be explained exclusively by a geometrical effect such as pericenter glow \citep[introduced by][]{wyatt99}, but is probably due to a dust density enhancement at pericenter \cite{olofsson19}. Finally, no circumstellar gas has yet been detected \citep{chen04, kennedy18} around HR 4796 A.

We report the results from a GPI integral field spectroscopy study of the iconic HR 4796A debris ring. We PSF subtract our observations using multiple algorithms leveraging both Angular Differential Imaging (ADI) and Reference Differential Imaging (RDI) techniques to explore the benefits and weaknesses of each algorithm. For comparison, we also fit our total intensity images using a simple disk model and a Markov Chain Monte Carlo (MCMC) approach. We recover measurements of the disk geometry and scattering phase function that are consistent with results from other groups using other high contrast imaging instruments and PSF subtraction techniques. We find that the HR 4796A disk may be slightly less forward scattering than previously reported and evidence that the near-infrared-to-visual color of the HR 4796A scattering phase function varies as a function of phase angle with a slightly blue color at relatively small scattering angles ($<$60$\arcdeg$) and a slightly red color at relatively large scattering angles ($>$120$\arcdeg$), consistent with measurements of large particles in the laboratory. We demonstrate that a population of large, irregular particles can reproduce the total intensity scattering phase function and the measured near-infrared-to-visual color of the scattering phase function as a function of phase angle.

\section{Observations and Data Reduction}
\subsection{Observations}
We obtained GPI Integral Field Spectrograph (Spec) observations of the debris disk around HR 4796A using the J-Spec, H-Spec, K1-Spec, and K2-Spec observing modes. The J-Spec data was obtained as part of the ``Debris Characterization in Exoplanetary Systems'' Gemini Large and Long Program (PI C. Chen; GS-2016A-LP-6); the H-Spec data was obtained as part of the ``GPI Exoplanet Survey'' (PI B. Macintosh; GS-2016A-Q-500); and the K1- and K2-Spec data were obtained as part of a Principal Investigator Program ``Does the HR 4796 Debris Disk Contain Icy Grains?'' (PI C. Chen; GS-2015A-Q-27). GPI provides 14 mas sampling for all coronagraphic observing modes. The IFS provides observations at 1.12-1.35 $\mu$m, 1.50-1.80 $\mu$m, 1.9-2.19 $\mu$m, and 2.13-2.4 $\mu$m, respectively, with spectral resolutions, R (=$\lambda/ \Delta \lambda$) $\sim$ 35-39, 44-49, 62-70, and 75-83, respectively.

\begin{figure*}[!ht]
\includegraphics[scale=0.30]{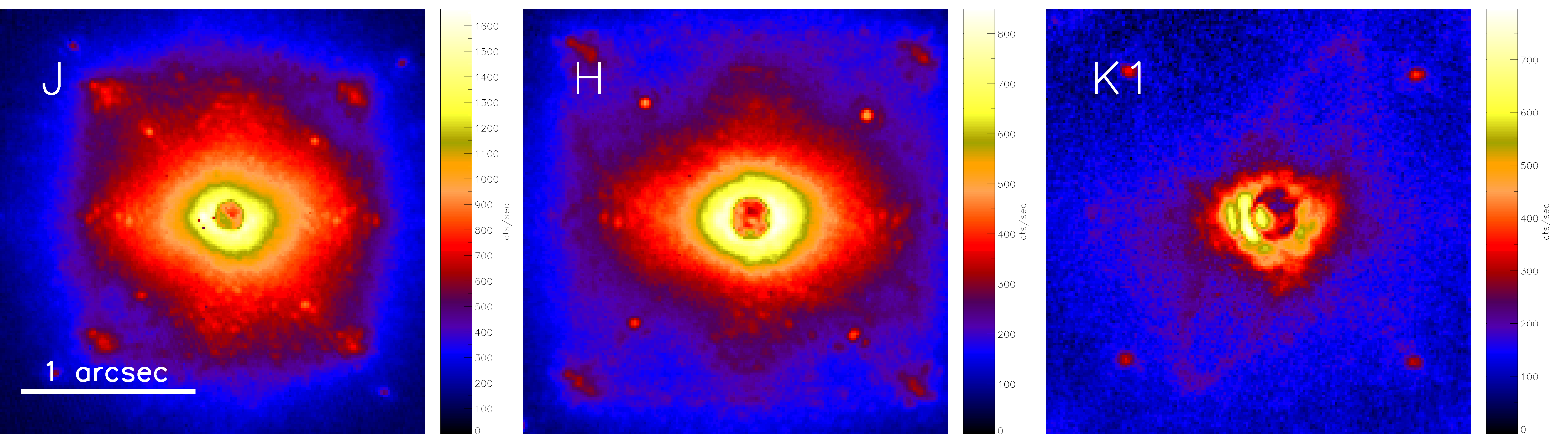}
\figcaption{Slices of calibrated GPI exposures of the HR 4796A disk obtained using (a) J-Spec (left), (b) H-Spec (middle), (c) K1-Spec (right). The contrast between the disk and the bright speckles is poorer in J- and H-bands where the speckles are brighter. The faint ring visible in the calibrated K1-Spec exposure is the bright HR 4796A ring. The J-band diffraction pattern is complicated with the first-order satellite, waffle, and second-order satellite spots visible as three spots near the corners of the detector. At longer wavelengths, the angular size of the diffraction pattern increases. Only the first order and waffle spots are visible in H. Only the first-order spots are visible in K1 and K2. 
\label{fig:jhk1exp}}
\end{figure*}

For each observation, the bright, primary star was centered behind the focal plane mask. On-axis light from the star was not only reduced but also diffracted by a grid imprinted on the pupil plane mask, generating a diffraction pattern including astrometric reference or satellite images of the primary star that are $\sim$10,000 times fainter than the star itself. These spots are used not only to determine the exact location of the occulted star but also to flux calibrate the astronomical scene \citep{Sivaramakrishnan06,wang14}. The distance of the satellite spots from the primary star scales with wavelength such that the spots appear closer to the star at shorter wavelengths. For J-band observations, the coronagraphic images are complex, revealing not only lower Strehl-ratios but also additional PSF structure such as AO ``waffle mode'' spots, second-order satellite spots, and diffractive spots from the Deformable Mirror (DM) actuator print-through (see Figure~\ref{fig:jhk1exp}). To facilitate ADI analysis of the data, the instrument field of view was allowed to rotate during an observation, producing diversity in the orientation of the science target with respect to the instrument reference frame. In some of our individual exposures, the satellite and waffle spots overlapped with the disk, making PSF subtraction more challenging. 

Each of our observations was composed of several tens of individual exposures with integration times between $\sim$30 and $\sim$90 sec. The frame times were selected to be long enough such that the detector readout noise did not dominate the disk signal but short enough to avoid saturation of residual speckles and angular smearing of the disk. K1-Spec and K2-Spec observations typically tolerate longer integration times because the star and therefore the speckles are fainter and the PSF is more oversampled. In general, most of our GPI observations were obtained in contiguous time blocks near transit to maximize the field rotation. However, our 25 March 2014 and 2 April 2015 K1 datasets were not. In 2014, the target was observed for 10 minutes; the observations were paused 15 minutes before transit and resumed again 5 minutes after transit; after transit, the target was observed for an additional 10 minutes. In 2015, the target was observed for an hour; the observations were paused two hours before transit and resumed two hours after transit; after transit, the target was observed for an additional hour. Since the night sky is bright at K-band, we obtained dedicated sky observations (20$\arcsec$ offset from our target) for thermal/sky background subtraction in K1- and K2-bands. We typically took five sky exposures after every hour of K1- or K2-band observing.

\input{table1.tex}

A GPI observing mode usually corresponds to a choice of filter, apodizer, and focal plane and Lyot plane masks. However, GPI lost the ability to reliably change apodizers from July 2015 until March 2016; therefore, the majority of the observations obtained during this time used the H-band optimized APOD$\_$H$\_$G6205 independent of the wavelength of the observations made. For spectroscopic observations, using the APOD$\_$H$\_$G6205 apodizer for any mode other than H-Spec is considered ``Nonstandard''. Nonstandard observations possess slightly different throughputs and inner working angles (IWAs). While our H, K1, and K2 observations were obtained in Standard configurations, our J observations were not. Since we estimate the PSF from the satellite spots in our forward modeling analysis, the nonstandard J-band observation does not affect our analysis. We list the integration times, the number of science exposures, the average airmass of the target, the seeing, and the amount of field rotation in Table~\ref{tab:observations}. 

We note that HR 4796 was observed using the K1 IFS on three different nights spanning a period of more than one year. Thus, our observations provide some empirical insight into the repeatability of GPI K1 IFS disk observations.

\subsection{Data Reduction}
\begin{figure*}
\plotone{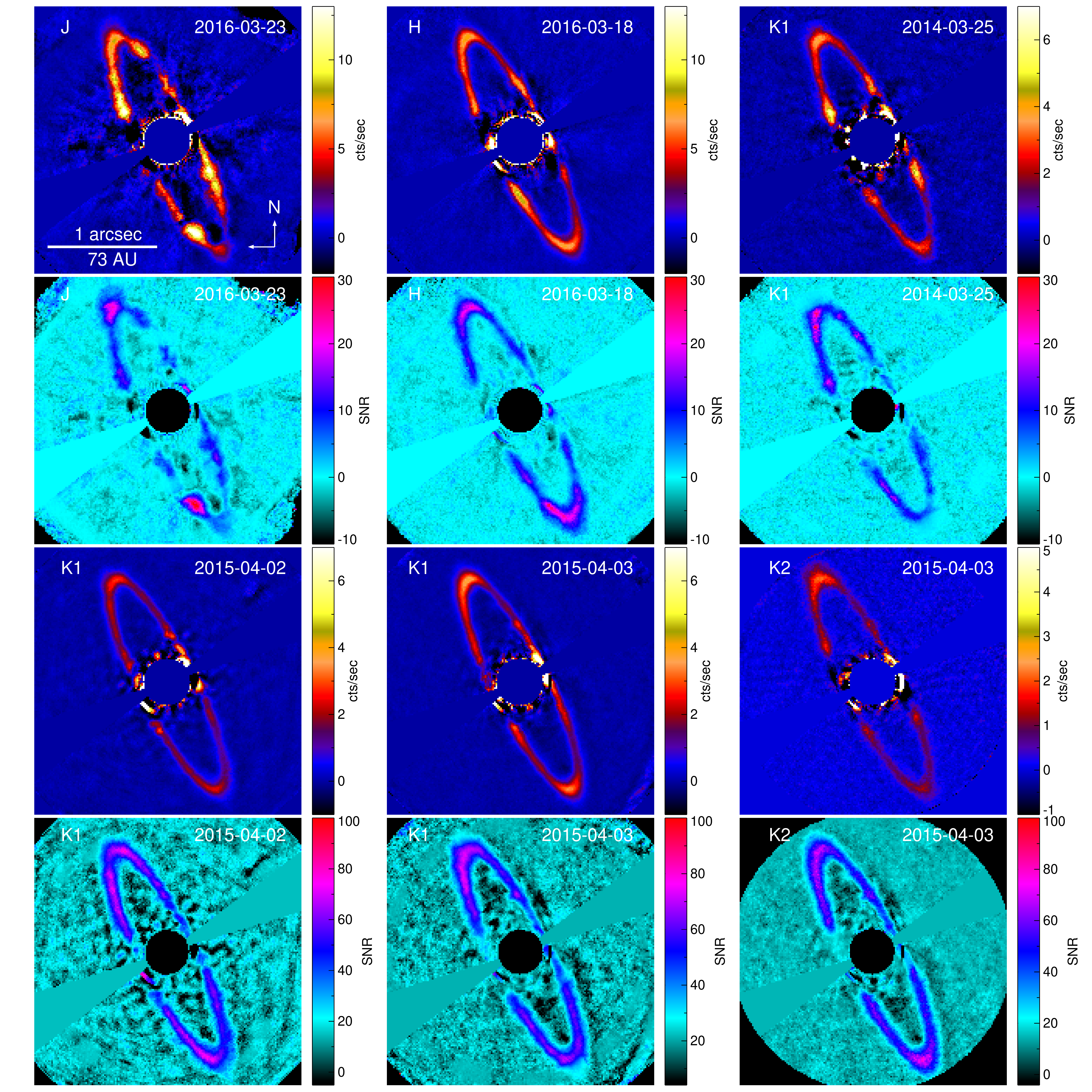}
\figcaption{Mask-and-interpolate total intensity images and SNR maps of the HR 4796A debris disk, oriented with north-up and east-left. \label{fig:maskandinterpolate}}
\end{figure*}

Reduction and interpretation of ground-based coronagraphic observations are challenging because observing conditions change on rapid timescales. The disk around HR 4796A is so bright that it can be seen in the individual exposures before PSF subtraction (Fig.~\ref{fig:jhk1exp}). We used the GPI Data Reduction Pipeline\footnote{\url{http://www.stsci.edu/~mperrin/software/gpidata/}} (DRP,  \citealp{perrin14}) to reduce, wavelength calibrate, and assemble spectral datacubes (which contain 37 2D images) at the individual exposure level. Specifically, for the J and H data, we used the DRP to subtract the dark background, update spot shifts for flexure, interpolate bad pixels in the 2D frame, assemble the IFS spectral datacube, interpolate bad pixels in the cube, correct distortion, and measure satellite spot locations. For the K1 and K2 data, we also subtracted dedicated thermal/sky background images and destriped the science images after dark subtraction but before interpolating bad pixels in the 2D frame. For K1- and K2-bands, with high thermal noise, we removed some spectral slices from the datacubes if the SNR of the satellite spots was lower than 2. We determined the location of the occulted star in each image cube using a least squares fit to all of the satellite spots’ positions. We then used the star's position to align all the images to a common frame before PSF subtraction. Our method allowed us to estimate the position of the star in each frame to a precision 0.05 pixels or 0.7 mas \citep{wang14}. Next, we collapsed our IFS cubes in wavelength space to produce a single broad band image for each exposure and removed exposures in which the disk overlaps with one or two satellite spots. Next, we generated a single high SNR empirical PSF for each observation, using all of the remaining satellite spots. Finally, we rotated the images so that North was up and East was to the left.

\subsection{PSF subtraction}

Several methods have been developped to remove stellar speckles from coronagraphic images in post-processing. These techniques may severely impact circumstellar objects, especially extended objects (debris and protoplanetary disks), for which the photometry and the shape are impacted \citep{milli12}. In this work, we used four different methods to estimate the PSF for PSF subtraction. Our analysis allowed us to quantify the relative strengths and weaknesses of each algorithm for measuring the geometry and phase function of a bright, narrow, inclined ring: (1) Mask-and-Interpolate \citep{perrin15}, (2) Reference Differential Imaging (RDI) - Non-negative Maxtrix Factorization (NMF, \citealp{ren18}), (3) RDI - Karhunen-Lo\'{e}ve Image Projection (KLIP, \citealp{soummer12}), and (4) Angular Differential Imaging (ADI, \citealp{Marois06}) - KLIP.

\begin{figure*}
\plotone{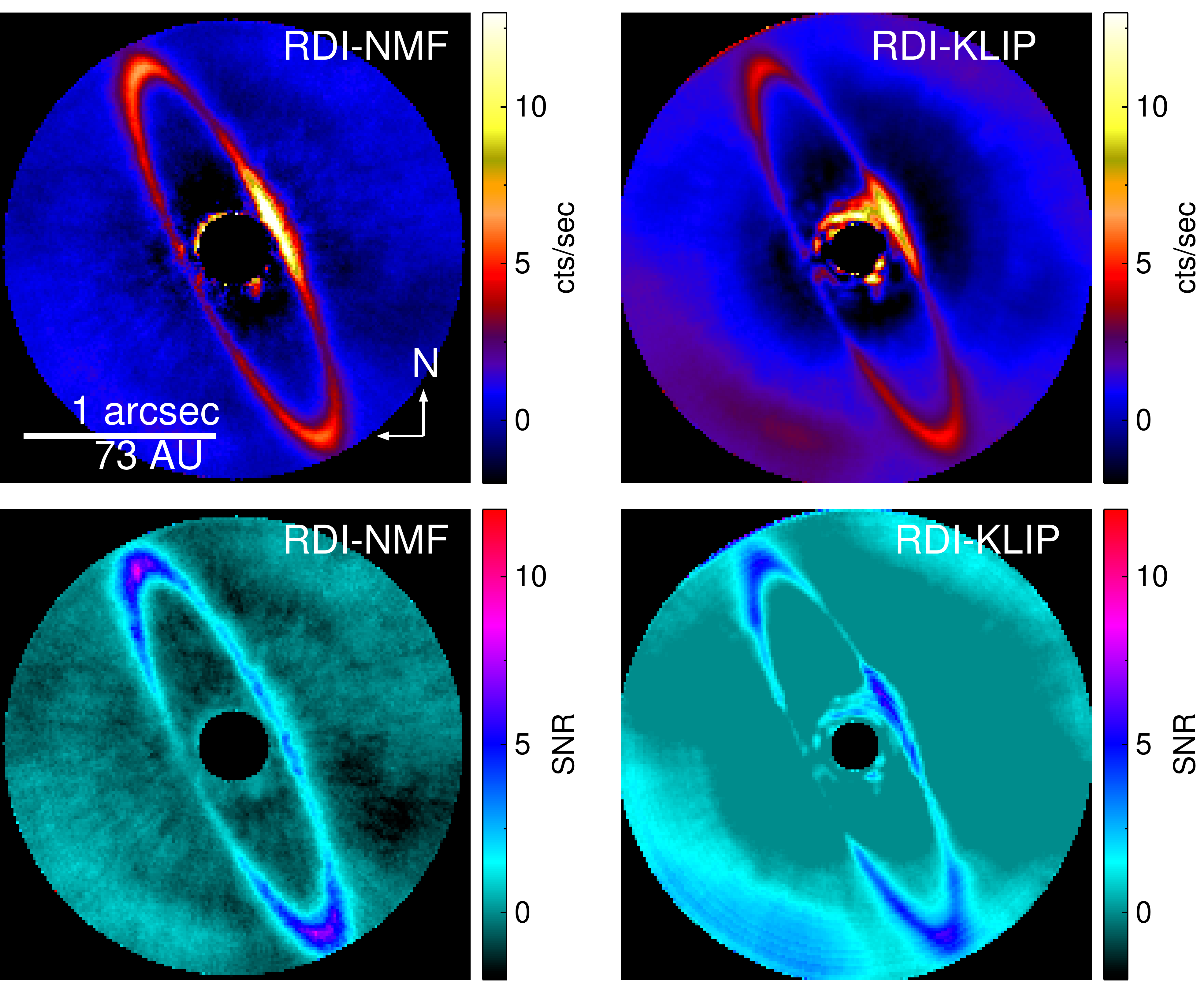}
\figcaption{(Top-Left) RDI-NMF and (Top-Right) RDI-KLIP H-band total intensity images and SNR maps of the HR 4796A debris disk, oriented with north-up and east-left. The RDI-NMF reduction includes 15 NMF components. The RDI-KLIP reduction includes 5 KL mode basis vectors.
\label{fig:rdi}}
\end{figure*}

The Mask-and-Interpolate method takes advantage of the high signal-to-noise and rather smooth nature of the residual PSF halo. For each exposure, (1) The disk is masked, assuming that it can be circumscribed by 2 nested ellipses with a maximum radial width of 15 pixels (210 mas) at the ansae. In addition, the first, second, and waffle-mode satellite spots are also masked. (2) The PSF is estimated under the mask using a fourth-order polynomial interpolation in both the x- and y-directions. The masked disk is replaced by the interpolated PSF. (3) The resulting PSF is smoothed using a median filter with a smoothing length of 7 pixels and subtracted from the original image. (4) The PSF subtracted images are median-combined.
(5) The remaining background in each PSF subtracted image is fit in concentric annuli within a wedge $\pm$85$\arcdeg$ from the major axis or excluding $\pm$5$\arcdeg$ from the minor axis using a 4th order azimuthal polynomial (see Figure~\ref{fig:maskandinterpolate}). This background model is then subtracted from the smoothed PSF subtracted image to produce the final image. 
We present the total intensity images for each of our observations and their corresponding Signal-to-Noise Ratio (SNR) maps (formed by dividing our total intensity image by our uncertainty image) in Figure~\ref{fig:maskandinterpolate}.

In RDI, independent observations of bright stars (ideally without companions or disks) are used to create a PSF library from which a representative PSF is estimated. The GPI Exoplanet Survey (GPIES) team is carrying out a search for planetary mass companions around $\sim$600 young stars using the GPI H-Spec mode. To date, the team has published observations of the first $\sim$300 stars \citep{nielsen19}. The majority of the GPIES observations contain isolated stars, and all the observations for which no companion, disk or background object have been detected can be used as a PSF reference library. We cross-correlate our HR 4796A Spec H-Spec observations with those in the library to select the 100 exposures that are most correlated with our target exposures. Then, we perform both an NMF and a KLIP principal component analysis on the empty reference observations to reconstruct the PSF. For RDI-NMF, the main parameter is the number of components used. We use 15 NMF components. For RDI-KLIP, the main parameter is the number of Eigen modes used in the Karhunen-Lo\'{e}ve (KL) basis. We use 5 KL modes basis vectors. For both RDI-NMF and RDI-KLIP, we perform the analysis on the whole image rather than concentric annuli to enable the best measurements of relative surface brightness within the disk. We present the H-band total intensity images for our RDI-NMF and RDI-KLIP reductions in Figure~\ref{fig:rdi}. 

\begin{figure*}
\plotone{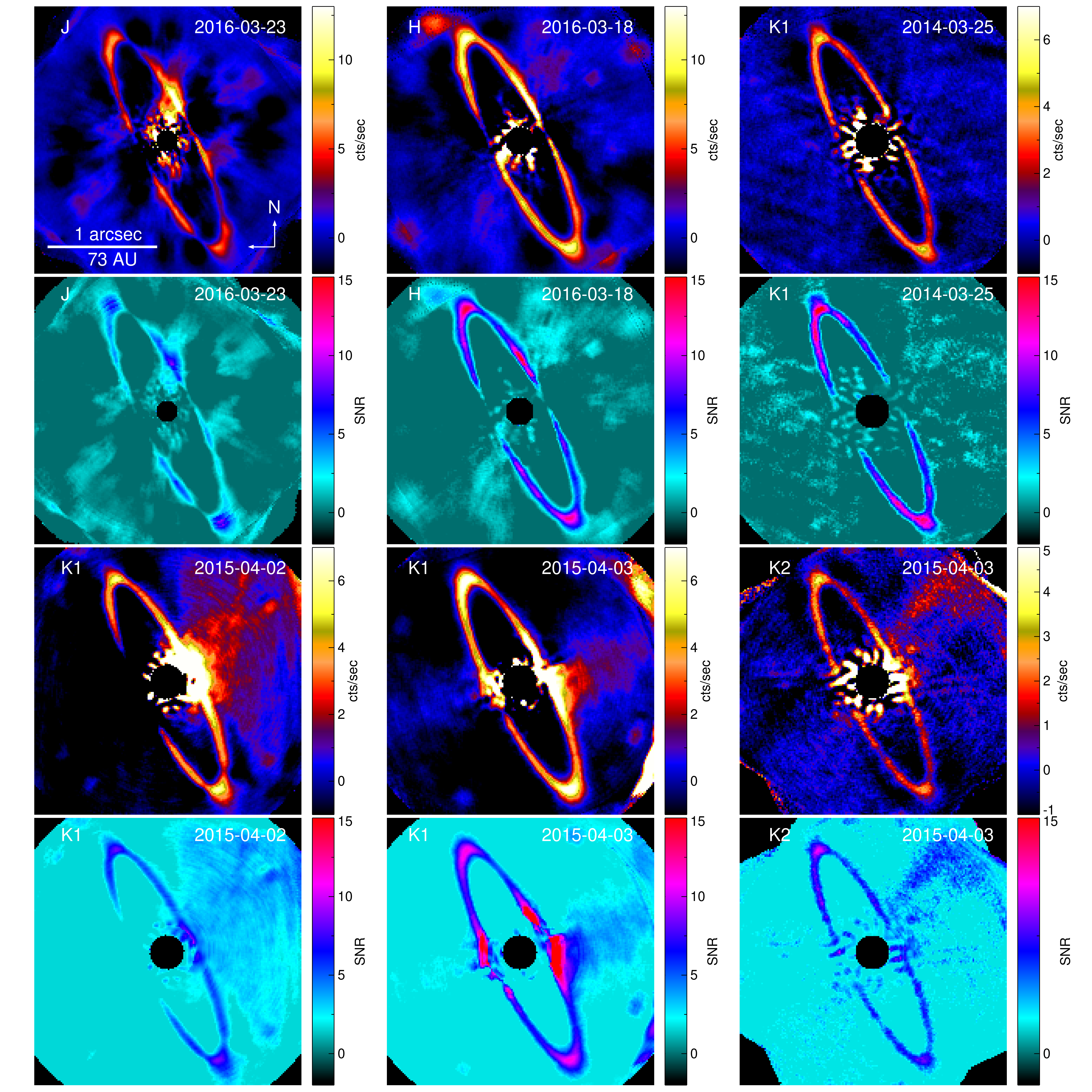}
\figcaption{ADI-KLIP total intensity images and SNR maps of the HR 4796A debris disk, oriented with north-up and east-left. These reductions were made using one angular zone and 3 KL mode basis vectors.
\label{fig:adi}}
\end{figure*}

In ADI, the astrophysical scene is allowed to rotate with respect to the diffraction pattern of the occulted star. In this case, a representative PSF can be estimated directly from the observing sequence including the astrophysical target. To minimize self-subtraction, we added an additional parameter compared with RDI-KLIP: the exclusion criterion (equivalent to $N_\delta$ in \citealp{lafreniere07}), the minimum number of pixels a hypothetical point-like astrophysical source must move azimuthally to avoid overlap in two images. For all bands, we used 3 KL modes and an exclusion angle of $N_\delta$ =6$\arcdeg$, which is a good compromise between SNR of detection and disk impact. To estimate the PSF using ADI-KLIP, we used the {\tt pyKLIP}\footnote{\url{https://bitbucket.org/pyKLIP/pyklip}} package \citep{wang15}, a Python implementation of the KLIP algorithm. We applied {\tt pyKLIP} globally (with the same parameters for the whole image) to avoid discontinuities in the reduction along the object. 

We assumed that the uncertainty in our observations was dominated by speckle noise. Therefore, we estimated the uncertainty by masking out the disk and calculating the standard deviation of the remaining pixels in concentric annuli, assuming an annular width of 3 pixels, approximately the PSF FWHM for all observations and reduction methods. We present the total intensity images for each of our observations (collapsed along the wavelength axis) and their corresponding SNR maps in Figure~\ref{fig:adi}. 

We generally detect the disk with SNR $\sim$ 5 - 10. The SNR with which the disk is detected is higher in H and K1 bands than in J and K2 bands. The J data suffer from the largest speckle noise and the most overlap between the disk and PSF structures (e.g. satellite spots). The K2 data suffer from a combination of lower throughput in the instrument (particularly at wavelengths 2.26-2.4 $\mu$m) and higher sky backgrounds. In addition, our K2 observations have horizontal stripes, created by a misalignment of the Focal Plane Mask. Consistent with previous ground- \citep[][\citetalias{milli17}]{wahhaj14, rodigas15} and space-based observations \citep{schneider09}, we find that (1) the disk appears bright at the ansae, with the NE ansa $\sim$10\% brighter than the SW ansa and (2) the host star is offset from the center of the disk.  Our J and H observations indicate that the disk surface brightness near the forward scattering peak appears asymmetric with the Northern side of the disk brighter than the Southern side. 

The exact properties of the disk recovered are dependent on the PSF subtraction method used. For example, the disk recovered using ``Mask-and-Interpolate'' is (1) More spatially extended. ADI analyses produce disk self-subtraction for spatially extended disks with a finite spatial extent \citep{esposito14}. The detection of disk self-subtraction in our ADI-KLIP reduction is consistent with other studies. (2) More affected by speckles. Interpolation is not expected to reproduce the complex GPI PSF structure with very high fidelity, particularly near the satellite and waffle spots. The lack of PSF fidelity is particularly problematic at J-band where the first- and second-order satellite spots, waffle spots, and DM spots are within the instrument field-of-view and overlap with the disk. To minimize the impact of satellite and waffle spots on the disk, we masked out areas of the disk where the satellite and waffle spots overlap when calculating our average PSF subtracted image. However, we note that despite taking this pre-caution, some satellite and waffle spot artifacts may still remain. For example, the waffle spots are immediately adjacent to the disk ansae in approximately half of the J-band exposures, leading to the inteprolation being based on more distant pixels.
Thus, the bright spots near the ansae in the J-band Mask-and-Interpolate image, that do not appear in the KLIP image, are suspect.

\section{Analysis Tools}
\label{sec:tools}
Historically, scattered light images have provided some of the highest angular resolution images of debris disks, enabling detailed measurements of disk geometry and scattering phase function (SPF). The latter measures the change in total intensity of the scattered light as a function of scattering angle, $\theta$, the angle of deviation from forward scattering (\citetalias{milli17}). 
However, accurately extracting the SPF for a disk is challenging because PSF subtraction techniques alter the disk surface brightness differently, depending on the apparent angular separation from the star and the position angle. For example, ADI produces regions of over-subtraction around inclined disks, resulting in flux losses \citep{milli12}. Therefore, we used two techniques to estimate the disk geometry and empirical SPF: (1) Reconstruction - we measure the disk geometry and the SPF from our PSF subtracted images. For the SPF, we correct for over-subtraction. (2) Modeling - we model the disk using a simple prescription for the geometry and SPF and used an MCMC method to explore the disk parameters until the residuals between the disk model and observation are minimized.

\subsection{Reconstruction}

\subsubsection{Geometric Parameters}
We used the {\tt Debris Ring Analyzer} ({\tt DRA}) to measure the projected disk parameters using the techniques described in \cite{stark14}. Briefly, the {\tt DRA} divides the disk into pie-shaped wedges centered on the star, then iteratively measures the peak of the radial surface brightness distribution in each wedge using a polynomial fit. The peak coordinates are transformed to cartesian coordinates and then fit with an ellipse using the mpfitellipse code (IDL). The peak coordinates and ellipse fit are shown in Figure~\ref{fig:ellipsefit} for the H band for all the PSF subctraction methods. The {\tt DRA} requires a mask to select  the areas of the image to be included in the geometric fit. We created one mask for each PSF-subtracted image of HR 4796A approximately following the area of the ring peak surface brightness, excluding wedges near forward and backward scattering where the speckle noise is the highest. We overlaid the measurements of peak surface \sout{density} brightness on our total intensity images and excluded points that did not lie on the disk. Our Mask-and-Interpolate images required the largest exclusion angles around forward and backward scattering while our RDI-KLIP images required the smallest. For example, for our H-Spec observations, Mask-and-Interpolate required 60$\arcdeg$ exclusion regions around forward and backward scattering; RDI-NMF required 0$\arcdeg$ and 15$\arcdeg$ exclusion regions around forward and backward scattering, respectively; RDI-KLIP required no exclusion regions; and ADI-KLIP required 10$\arcdeg$ and 30$\arcdeg$ exclusion regions (see Figure~\ref{fig:ellipsefit}). We divided the remaining disk area into 2$\fdg$7 wedges in the sky plane, centered on the star. For each wedge, we iteratively fit a third order polynomial to the disk surface brightness as a function of distance to determine the radial location of the peak surface brightness. The initial fit used the full radial extent of the wedge and progressively selected smaller and smaller radial sections of the wedge around the peak surface brightness until its profile was well fit. We estimated the uncertainty in the peak position by randomly varying the disk surface brightness within a wedge according to the observed 1$\sigma$ uncertainties assuming a Gaussian distribution, refitting the radial peak, and calculating the standard deviation of 50 unique instances. Finally, we fitted an ellipse to the collection of peak surface brightness coordinates and their uncertainties by minimizing the perpendicular distance from the ellipse to the coordinates. 

\subsubsection{Scattering Phase Function (SPF)}

We hypothesize that we can reconstruct the intrinsic SPF by correcting the SPF extracted from our PSF subtracted images for the effects of ADI-subtraction at the 1-dimensional level. We note that the disk total intensity image appears bright near the ansae as a result of limb brightening. This limb brightening is generated by the large column density of scatterers near the ansae in this optically thin disk, compared with that along the minor axis, corresponding to the front and back sides of the disk, as well as by the finite resolution of the telescope. When extracting the SPF, we deproject the disk image and divide it by the assumed model SPF to correct for limb brightening and over-subtraction and therefore reveal the intrinsic SPF. This approach has been used in previous studies of the HR 4796A SPF (\citetalias{milli17}). 

In our study, we use {\tt MCFOST} \citep{pinte06, pinte09} to simulate a disk with the same geometric parameters as HR 4796A and a known SPF. We match the disk  surface brightness to our observations of the HR 4796A ansa. We convolve our idealized model disk with an empirical GPI PSF to mimic the observed broadening of the disk ansae. Specifically, we use unblocked Spec observations of the bright ($K_{s}$=7.6 mag), nearby A-type star HD 118335 obtained using the J, H, K1, and K2 filters and appropriate matching focal plane masks on 2014 March 26 to approximate the PSF. We inject our convolved model disk into an empty GPI sequence (without a companion or a disk) with similar noise properties, fixing the parallactic angles in the empty data set to be the same as in our target observations. We PSF subtract the injected disk sequence using the same PSF subtraction algorithm and parameters that we used for the HR 4796A sequence.  We extract the SPF from the PSF subtracted, simulated disk image by performing aperture photometry on the deprojected, 1/r$^2$ corrected disk image. Finally, we multiply the SPF extracted from the PSF subtracted HR 4796A image by the SPF assumed in our simulated disk and divide by the SPF extracted from our PSF subtracted, simulated disk image. For simplicity, we simulate a disk with an isotropic SPF or a Henyey-Greenstein (HG) scattering parameter, $g$=0. In this case, the SPF is constant as a function of scattering angle. 

\begin{figure}
\plotone{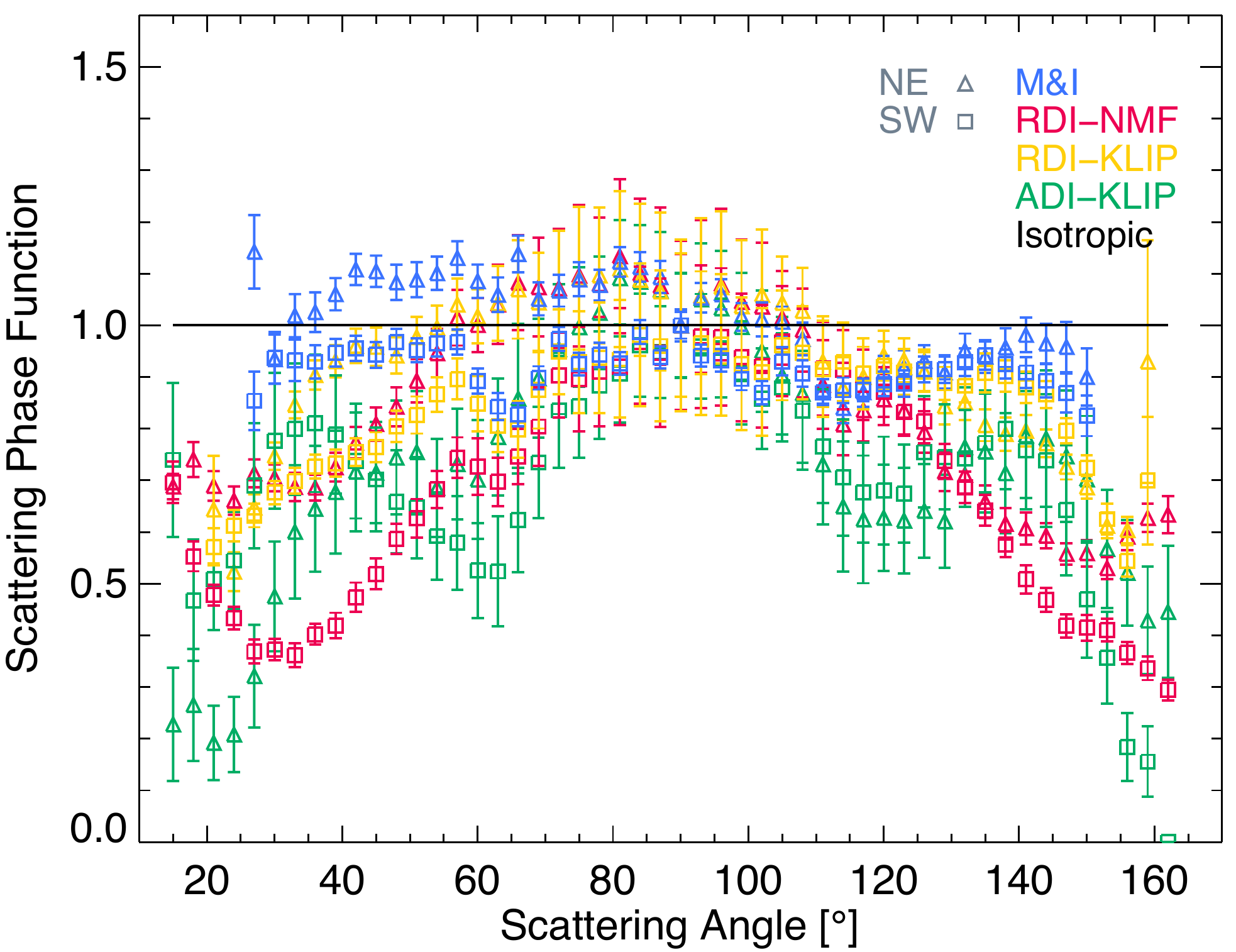}
\figcaption{Scattering phase functions extracted from a simulated observing sequence with an injected isotropic scattering disk ($g$ = 0) and a HR 4796A-like geometry, corrected for limb-brightening, using several different PSF subtraction techniques. The disk model is constructed so that the disk surface brightness SNR is the same as the HR 4796A ansa. An isotropic disk is expected to have a constant scattering phase function; thus, regions with values $<$1 are affected by over-subtraction.  
\label{fig:isotropicspf}}
\end{figure}

We show how the various PSF subtraction algorithms impact the SPF extracted from a disk with a HR 4796A-like geometry by plotting the SPFs extracted from PSF subtracted, isotropic disk sequences and corrected for limb brightening effects in Figure~\ref{fig:isotropicspf}. We find that Mask-and-Interpolate alters the SPF the least with relatively little over-subtraction across a broad range of scattering angles (40$\arcdeg$-145$\arcdeg$). Mask-and-Interpolate produces the least over subtraction because it uses an interpolation to estimate the PSF within the disk mask. RDI-NMF and RDI-KLIP produce more over-subtraction at a larger range of angles near forward and backward scattering. ADI-KLIP produces the most over-subtraction at a larger range of angles near forward and backward scattering and generates additional wavy structure beyond the over-subtraction at forward and backward scattering. ADI-KLIP relies on the observing sequence to estimate the PSF. In this case, large-scale, diffuse, disk scattered light could be mis-identified as part of the PSF. We tested our method on a simulated dataset in Section \ref{appendix-injection-with-exctraction} and demonstrated that SPF reconstruction should be feasible at scattering angles of 40$\arcdeg$ - 140$\arcdeg$ for H- and K1-bands.

\subsection{Modeling} 
\label{subsec:image_fitting}

An alternate method for estimating the scattering phase function is fitting the total intensity image using an MCMC approach combined with a forward modelling technique described in \cite{pueyo16} and assuming a simple prescription for the 3D dust density distribution. Similar analyses have been used to estimate the geometric properties of the several debris disks from GPI observations \citep{maxmb16}. We fit our ADI-KLIP total intensity images to estimate the the SPFs from our GPI total intensity images.

\subsubsection{Model Description} 
\label{sec:desc_modelling}
We assume that the dust density can be well described using a flared disk with a power-law surface density distribution 
\begin{equation}
\eta(r,z) \propto \left( \frac{r}{R_1} \right) ^{-\beta} \exp \left[ -\frac{1}{2} \left( \frac{z}{h_0 r}\right)^2 \right]
\end{equation}
\citep{maxmb15} where $r$ is the radial distance of the dust from the star, $z$ is the height of the dust above the disk mid-plane, and $\beta$ is the power-law index of the dust density distribution. Since we study the relative brightness of the disk as a function of the disk and not the total disk mass, we do not solve for the porportionality constant in our analysis. Since scattered light imaging of the HR 4796A debris disk indicates that it is a flat, narrow ring, we assume that the dust surface density, $\eta(r,z)$ = 0 interior to the ring ($r$ $<$ $R_{1}$) and exterior to the ring ($r$ $>$ $R_{2}$) and $h_0$ = 0.01. We note that $R_{1}$ (inner radius) and $R_{2}$ (outer radius) are distinct from the ellipse radius $a$ measured in Section \ref{sec:disk_geometry} that is extrapolated from points of maximum surface brightness calculated in pie-shaped wedges from ADI-KLIP, PSF-subtracted images. We initially left the scale height a free parameter and our best-fit models favored a very thin ring $h_0$ $<$ 0.001 because the vertical extent of the disk was challenging to resolve. Unfortunately, the locations in the disk that provide the most leverage on the scale height measurements were along the disk minor axis where our data were most impacted by ADI self-subtraction. Since the disk scale height would be degenerate with the disk radial width parameters ($R_1$ and $R_2$) if the scale height were resolved, we decided to fix $h_{0}$ to a small value that was consistent with measurements on other disks. We note that small variations in $h_0$ would produce slightly different values for the disk radial width ($R_1$ and $R_2)$.

Following \citetalias{milli17}, we combine the dust density distribution with a two-component HG scattering phase function to generate a 2D projection of the disk from the observer's point of view. The two-component HG function
\begin{equation}
\label{eq:2g_spf}
p_2(g_1,\alpha,g_2,\theta) = \alpha HG(g_1, \theta) + (1-\alpha) HG(g_2, \theta)
\end{equation}
where $\theta$ is the scattering angle, $HG(g,\theta)$ is the single component HG function, $g_1$ and $g_2$ are the asymmetric scattering parameters for the two separate components, and $\alpha$ is the relative weight for the two components (0 $\leq$ $\alpha$ $\leq$ 1). We note that the HG SPF function does not carry any physical meaning in our fits. Instead, we use HG SPFs to reproduce the essential behavior of complex SPFs using a small number of free parameters. For example, a single component HG has one free parameter: the asymmetric scattering parameter $g$. A two component HG function has three free parameters: the two asymmetric scattering parameters, $g_1$ and $g_2$, and the relative weight between them, $\alpha$. The HG SPF was first used to empirically model the SPF observed for interstellar dust \cite{henyey41}. The majority of debris disk SPF studies in the literature use a single component HG SPF because most SPFs are not well measured over a large range of scattering angles. A two component HG SPF has been successfully used to reproduce the SPFs of zodiacal dust \cite{hong85} and Saturn's rings \cite{hedman15}) over a large range of scattering angles. We explore SPFs including additional complexity in \ref{appendix-3g}.

The intensity for a given pixel $(x',y')$ integrated along the line of sight is \citep{maxmb15}:
\begin{equation}
I(x',y') = I_0 + \cos i \int^{R_2}_{z'=-R_2}\frac{N_0}{r^2}\eta(r,z)p_2(g_1,\alpha,g_2,\theta)dz'
\end{equation} 
where $\eta(r,z)$ is the dust density distribution of the disk, inclined, $i$, from face-on and rotated so that the semi-major axis is $PA$ degrees West of North. In this model, the disk offset is defined in the plane of the disk with $dx$ (in AU) in the disk minor axis direction towards the North-West and $dy$ (in AU) in the disk major axis direction towards the South-West. Using, the disk-plane offset, we can measure the eccentricity ($e = \sqrt{dx^2 + dy^2}/R_1$) and argument of pericenter ($\omega = \arctan(dy/dx)$). Consistent with the definition, the scattering phase function is implemented at the grain level, in the disk midplane before the image is inclined and offset. The observed azimuthal surface brightness distribution is dependent not only on the SPF but also on the inclination (creating limb brightning) and the stellar offset (more light shines on the pericenter than the apocenter, making the reflected light from the pericenter brighter than the apocenter). We assume $N_0$ is the flux normalization and $I_0$ is a constant offset. If the ADI-KLIP algorithm is applied correctly, then the average sky background is expected to be zero \citep{soummer12}. As a result, we assume that $I_0$ = 0. In summary, our model has 11 free parameters, including 7 geometric parameters ($R_{1}$, $R_{2}$, $\beta$, $i$, $PA$, $dx$ and $dy$), 3 SPF parameters ($g_1$, $g_2$, $\alpha$) and the normalization $N_0$.

\subsubsection{Bayesian Estimation of the Parameters} 

We used Bayesian parameter estimation to derive the best fit values and the Posterior Distribution Functions (PDFs) for our model of the HR 4796A total intensity images. Our model images (Fig. \ref{fig:gpihmodel}, Top-Left) are forward modeled to include the instrument and PSF-subtraction artifacts. This process, illustrated in Fig. \ref{fig:gpihmodel} for our GPI H-Spec total intentsity image, is as follows:

\begin{figure*}
\centering \includegraphics[width=0.8\textwidth, trim=0 80 0 0, clip]{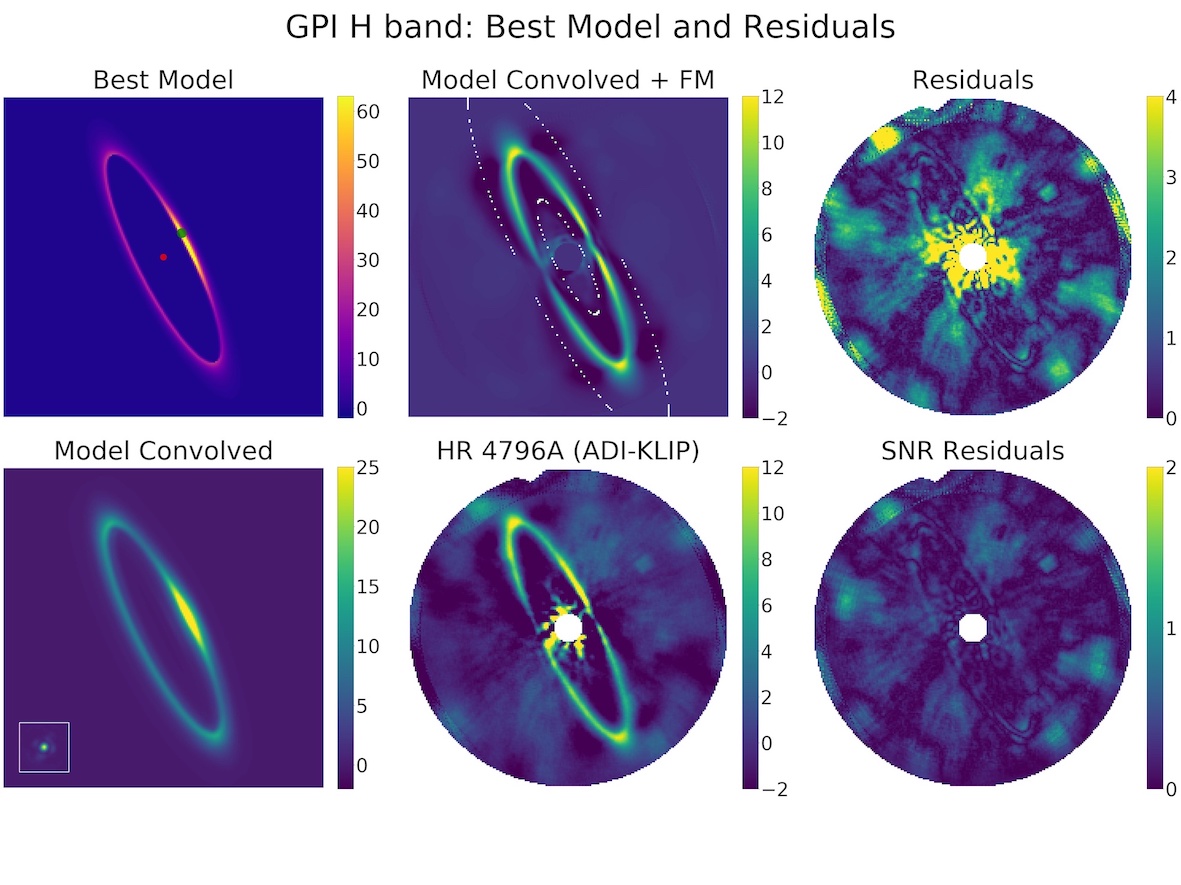}
\figcaption{
 \label{fig:gpihmodel}  Best-fit model resulting from the MCMC: GPI H-band (Top-Left) Best-fit model image; the red and green spots mark the position of the star and the disk pericenter, respectively. (Bottom-Left) Best-fit model image after convolution with an observed GPI PSF (shown in a small vignette at the Bottom-Left). (Top-Middle) Best-fit model image after convolution and Forward-Modelling to reproduce KLIP-ADI effect. The white dashed ellipses indicate the inner and outer edges of the zone over which the likelihood for the MCMC is estimated. (Bottom-Middle) Image showing the KLIP-ADI reduced dataset. (Top-Left) Residuals from the MCMC. 
 (Bottom-Right) SNR of the residuals of the MCMC.}
\end{figure*} 

\begin{itemize}
\item First, we generated a model image of the disk (described in Section~\ref{sec:desc_modelling}) with 11 free parameters. 
\item Second, we estimated the PSF for the observing sequence being analyzed. Each GPI Spec observation included the four satellite spots. We approximated the PSF during each observation by averaging the images of the four satellite spots to increase the SNR. In some exposures, the disk rotated onto the satellite spots. In these cases, we omitted the affected satellite spot from the PSF estimate and masked the portion of the disk that overlapped with the satellite spot. The resulting PSF is shown in a small vignette on the Bottom-Left of Fig. \ref{fig:gpihmodel}, Bottom-Left. We convolved this PSF with the model.
\item Third, we forward modeled the observing sequence \citep{pueyo16} using the DiskFM pipeline\footnote{\url{https://pyklip.readthedocs.io/en/latest/diskfm_gpi.html}} of the {\tt pyKLIP} package \citep{wang15} to simulate the impact of the ADI-KLIP on the model. The forward model was constructed by simulating the convolved disk at each of the parallactic angles within our observing sequence and combining these images using the same pyKLIP parameters as the observations (Fig. \ref{fig:gpihmodel}, Top-Middle). 
\item Fourth, we estimated the uncertainty map using a technique described in \cite{gerard16}. To estimate the noise map, we reduced the data using ADI-KLIP reduction with the same parameters but back-rotated to produce a final reduced image in which the disk is averaged out. We calculated the standard deviation of the pixel values within concentric annuli, centered on the star, with a constant width (3 pixels) in the combined image with the disk averaged out. We used this azimuthally symmetric image showing the standard deviation of the pixel values as our noise map.
\item Finally, we compared the forward model to the reduced image (Fig. \ref{fig:gpihmodel}, Bottom-Middle) by measuring:
\begin{equation}
\chi^2 =\sum_{S} \frac{(Data - ForwardModel)^2}{Uncertainty^2}
\end{equation} 
where $S$ is the zone, corresponding to the disk. $S$ is defined as an ellipsoidal annulus with the same inclination and position angle as the HR 4796A disk and with inner and outer radii of 40 au and 130 au, respectively (shown in a white dashed line in Figure \ref{fig:gpihmodel}, Top-Middle). We applied an empirical scalar correction to the noise map to retrieve realistic error bars. 
\end{itemize}

We performed these steps hundred of thousands of times within an MCMC wrapper that maximizes $e^{-\chi^2/2}$ until the chains had converged, using the {\tt emcee} package \cite{foreman-mackey13}. We used 192 parallel walkers and removed some iterations during the burn in phase. We noted the total number of iterations for each band at the top of each corner plot. After the MCMC converged, we plotted the residals ($|Data - BestForwardModel|$, Fig. \ref{fig:gpihmodel}, Top-Right) and the SNR of the residuals ($|Data - BestForwardModel|/Uncertainty$, Fig. \ref{fig:gpihmodel}, Bottom-Right) as well as the Posterior Distribution Functions. All corner plots are shown in \ref{appendix-figsmcmc}. We derived uncertainties based on the 16th ($- 1\sigma$), 50th (median value), and 84th ($+1\sigma$) percentiles of the samples in the distributions (plotted as vertical lines in the corner plots). We demonstrated our method on a simulated dataset in Section \ref{appendix-injection-with-mcmc}. We found that the model priors could be recovered with Gaussian PDFs to withn 1$\sigma$ (with the exception of $R_2$) if the noise map was artificially multiplied by a scalar.

We found that the best value for $R_2$ often depends on the KLIP parameters used; higher KL modes, consistent with more aggressive KLIP reduction, leads to smaller $R_2$. We hypothesized that ADI-KLIP has difficulty recovering $R_2$ because the diffuse extended structures were severely impacted by ADI self-subtraction. Indeed, the STIS coronagraphic image of this disk revealed diffuse, low surface brightness scattered light, extending hundreds of au from HR 4796A, extending well outside of both the GPI and SPHERE fields-of-view  \citep{schneider18}. Therefore, we concluded that the values estimated for $R_2$ with this method are not physically relevant, a fact confirmed by our test on a simulated dataset in \ref{appendix-injection-with-mcmc}. As a result, we assumed a prior for $\log R_2$ that was uniform from 82au to 100au with an inverse exponential function $1/(1 + \exp(k(R-R_c))$ and $k = 40$ beyond 100au to avoid wasting computational time on an irrelevant parameter. We assumed that all of the other priors were uniform between our minimum and maximum values. For example, we assumed uniform priors for $\log R_1$ between 60au and 80au, for $\beta$ between 1 and 30, for $g_1$ between 5\% and 99.99\%, for $g_2$ between -5\% and -99.99 \%, for $\alpha$ between 1\% and 99.99\%, for $i$ between $70\arcdeg$ to $80\arcdeg$, for $\theta$ between $20\arcdeg$ to $30\arcdeg$, for $dx$ and $dy$ between -10au and 10au and finally for $\log N$ between 0.5 and 50000. 

\section{Results: Disk Geometry}
\label{sec:disk_geometry}

\begin{figure*}
\plottwo{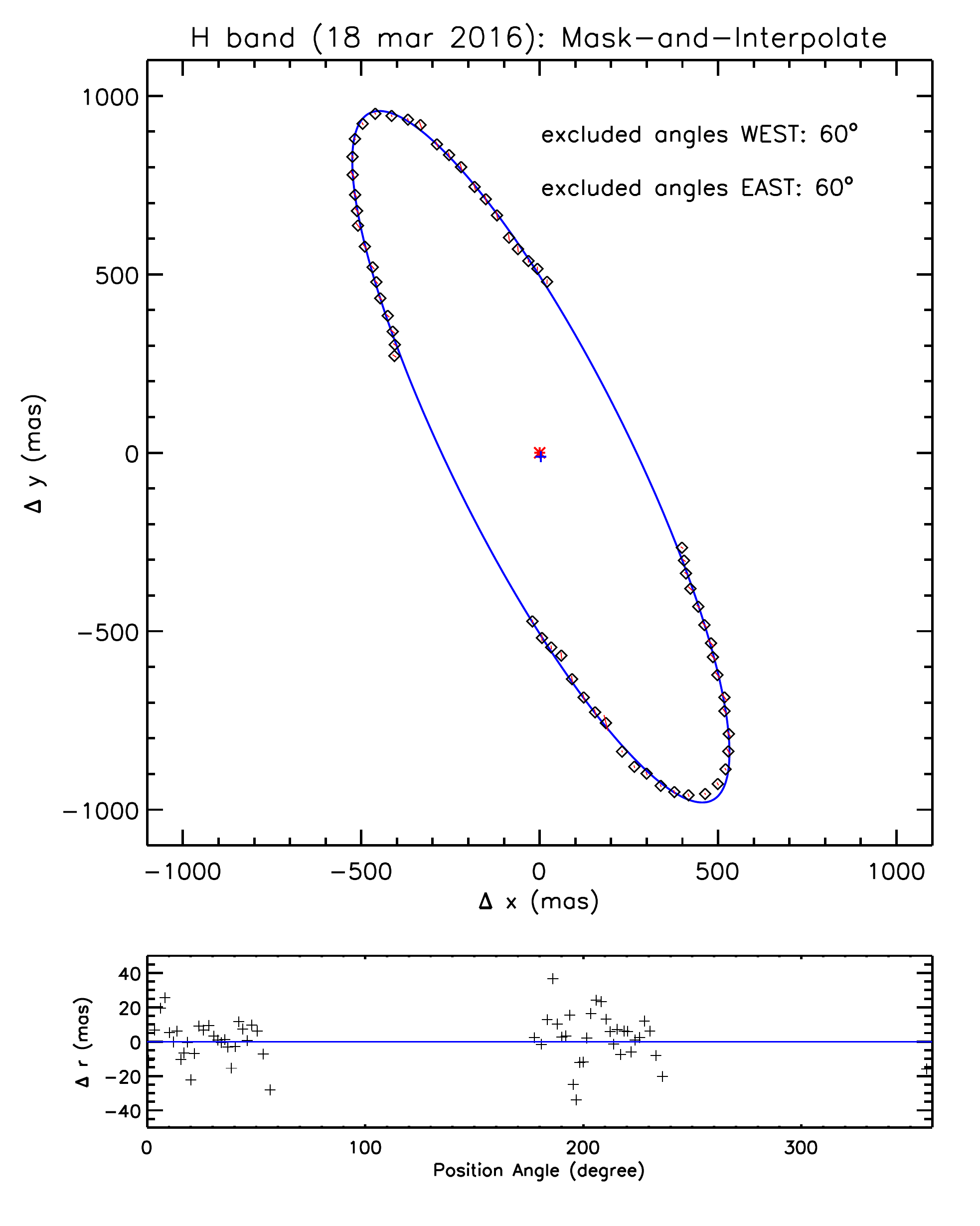}{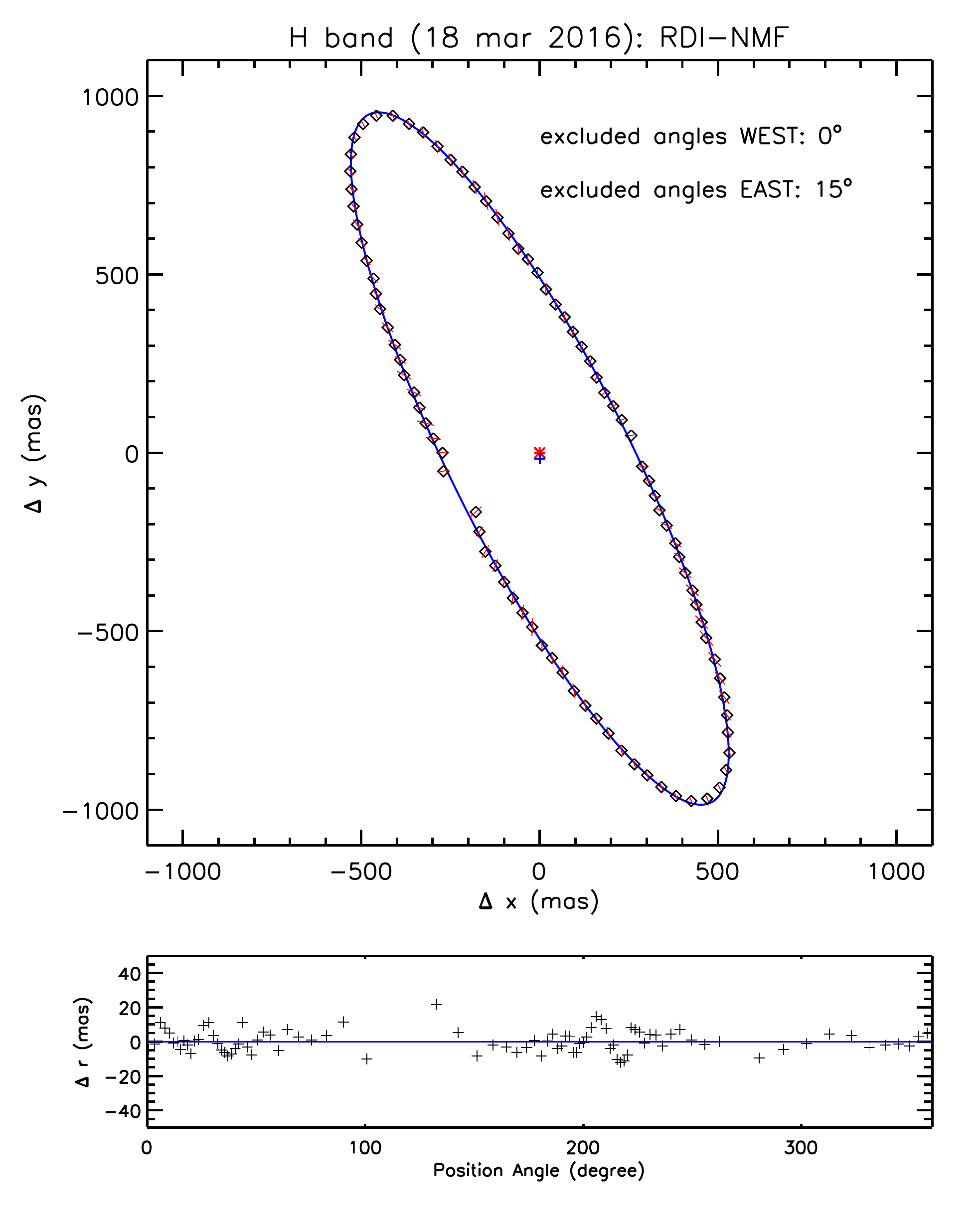}
\plottwo{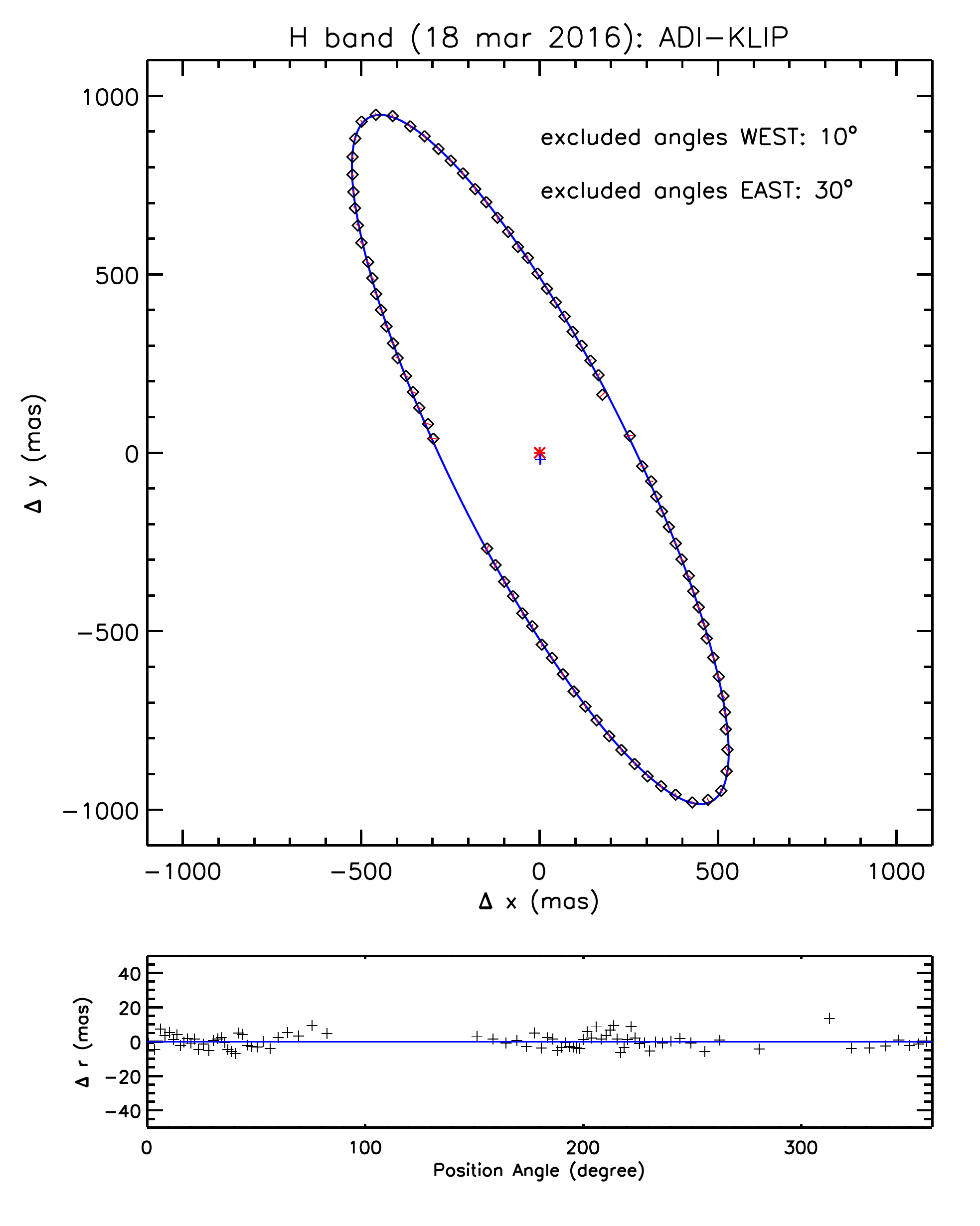}{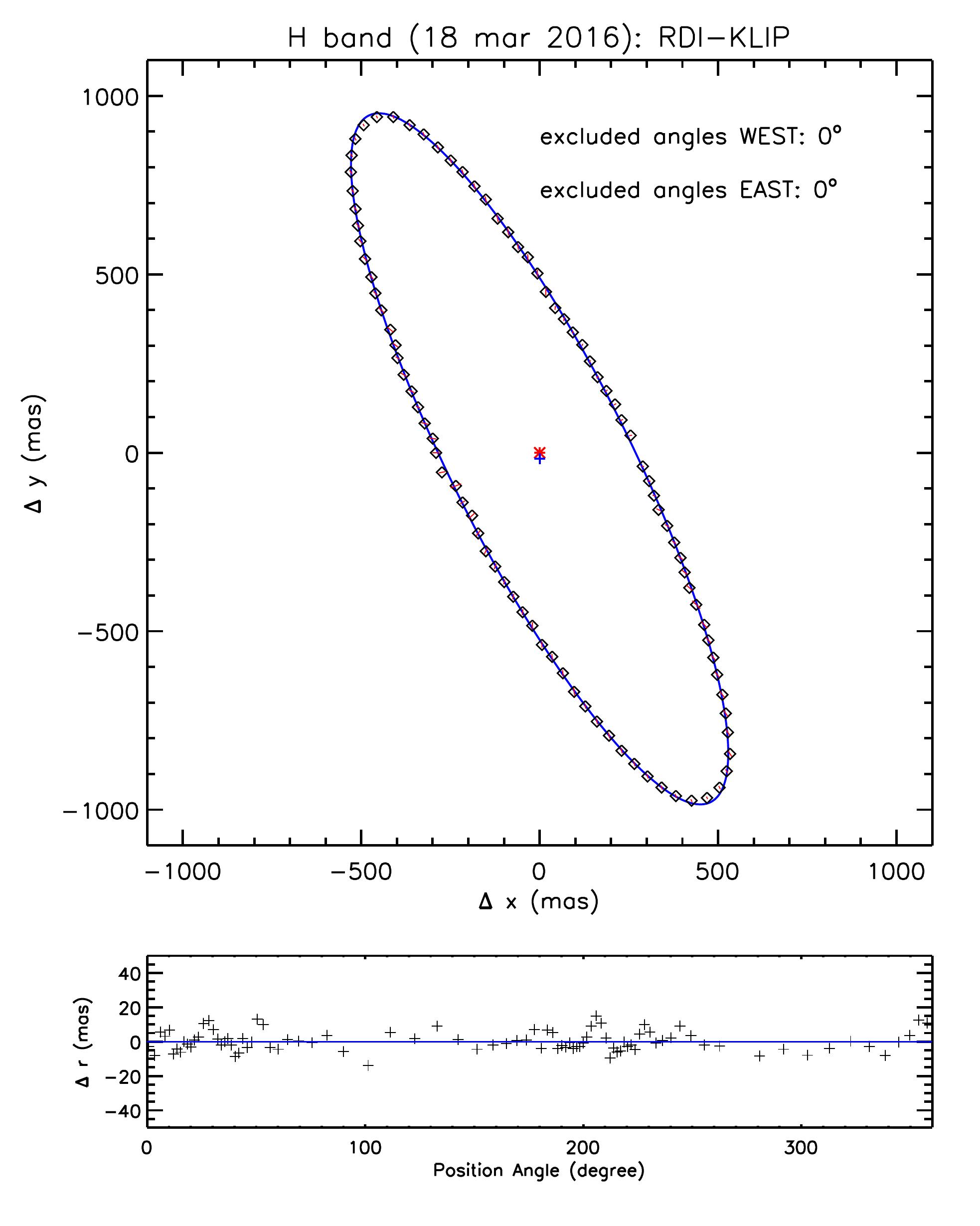}
\figcaption{Demonstration of the projected ellipse fitting procedure for our H-band observation for each of the PSF subtraction techniques: (top-left) Mask-and-Interpolate, (top-right) RDI-NMF, (bottom-left) ADI- KLIP and (bottom-right) RDI-KLIP. Diamonds show the positions of peak disk surface brightness calculated in 3$\arcdeg$ wedges centered on the star. The red star marks the position of the star and blue cross marks the center of the ellipse. We excluded wedges near forward and backward scattering because the residual speckle noise in this region was large. The size of the excluded wedges depends on the PSF subtraction technique with Mask-and-Interpolate requiring the largest exclusion wedges and RDI-KLIP the smallest. The excluded angles WEST and EAST give the angular extent of the excluded region near forward and backward scattering, respectively.
\label{fig:ellipsefit}}
\end{figure*}

The dust in the HR 4796 disk is produced by collisions among parent bodies on Keplerian orbits. Thus, we idealize the narrow dusty ring using a Keplerian orbit, specified by five orbital elements: (1) the semi-major axis, $a$, (2) the eccentricity, $e$, (3) the inclination from face-on viewing, $i$, (4) the $P.A.$ of the ascending node, $\Omega$, and (5) the argument of pericenter, $\omega$. We infer the orbital elements from our measurements of the projected semi-major axis, $a'$, the projected eccentricity, $e'$, the position angle (measured East of North), $PA'$, and the projected offset of the host star from the center of the ellipse, $\Delta$R.A.$'$ and $\Delta$Dec.$'$ (see Table~\ref{tab:geoparam}). We adopted the same convention as \citetalias{milli17}: a positive $\Delta$R.A.$'$ means that the ellipse center (blue cross in Figure~\ref{fig:ellipsefit}) is located West of the star (red star in Figure~\ref{fig:ellipsefit}), a positive $\Delta$Dec.$'$ means the ellipse center is located North of the star. The uncertainty in the measurement of True North is 0.03$\arcdeg$ for GPI \citep{konopacky14}.

\subsection{Debris Ring Analyzer Analysis}
\label{sec:disk_geometry_dra}
We used the {\tt DRA} code to estimate the projected parameters and then to apply the Kowalsky deprojection routine to transform the projected disk parameters to deprojected orbital parameters \citep{smart30, stark14}. We measured the projected disk properties and inferred the orbital parameters for each of our observations using our Mask-and-Interpolate, RDI-NMF, RDI-KLIP, and ADI-KLIP PSF subtracted images to better understand the uncertainties in our derived quantities.

\input{table2.tex}

We measured projected disk parameters (see Table~\ref{tab:geoparam}) that are generally consistent from night-to-night and filter-to-filter using all of our PSF subtraction techniques; however, we observed some minor systematic differences. In general, the standard deviation of our KLIP measurements is smaller than that of our Mask-and-Interpolate measurements. For example, the average and standard deviation of the KLIP and Mask-and-Interpolate are $\Delta$RA$'$=0$\pm$2\,mas and $\Delta$Dec$'$=-20$\pm$3\,mas, and $\Delta$RA$'$=2$\pm$5\,mas and $\Delta$Dec$'$=-15$\pm$5\,mas, respectively. Thus, we conclude that the residual speckle noise within the Mask-and-Interpolate reductions makes it more difficult to measure the geometric parameters reliably. Despite this caveat, our measurement for the positional offset of the star $\Delta$RA$'$ and $\Delta$Dec$'$ are broadly consistent with those obtained from other observations \citep[][\citetalias{milli17}]{schneider09, thalmann11, wahhaj14, rodigas15, perrin15}.

Consistent with previous studies, we estimated a small eccentricity, suggesting that the ring is very circular. Specifically, we estimated an orbital eccentricity ($e\sim0.03$) that is comparable to that reported from GPI commissioning observations ($e\sim0.02$, \citealp{perrin15}), Magellan AO ($e \sim0.06$, \citealp{rodigas15}) and SPHERE ($e\sim0.06$, \citetalias{milli17}). The almost circular nature of the ring makes it challenging to constrain the argument of pericenter, espcially with this geometrical fit method. Comparing our measured value for $\omega$ to others published in the literature is further complicated by inconsistencies in the definition of this parameter. We adopted the same definition as \citetalias{milli17} and \cite{milli19}, who found $\omega$=-74$\fdg$3$\pm$6$\fdg$2 and $\omega$=-74$\fdg$2$\pm$11$\fdg$9. These results are in good agreement with \cite{olofsson19}, who adopted a different convention ($\omega$= -254$\fdg$3$\pm$1$\fdg$8 = -180$\arcdeg$ - 74$\fdg$3$\pm$1$\fdg$8) and with \cite{rodigas15}, who adopted yet another convention ($\omega$=110.6$\fdg$6$\pm$13$\fdg$7 = 180$\arcdeg$ - 70.6$\fdg$6$\pm$13$\fdg$7). Finally, \cite{perrin15} found  $\omega$=-16$\fdg$9$\pm$1$\fdg$9. Since they did not discuss this value in detail, we could not confirm the convention they used; however, they found a relative position of the ring center (small offset to the South-West relative to the star) that would place the position of the pericenter closer to the North ansae, in slight disagreement with more recent studies. Using the Debris Ring Analyzer, we estimated $\omega$ between -14$\arcdeg$ and -80$\arcdeg$ with a typical uncertainty of a few tens of degrees. If we exclude the low SNR mask and interpolate dataset values, we are finding values between -59$\arcdeg$ and -69$\arcdeg$, in very good agreement with the most recent analysis. The Debris Ring Analyzer is not precise enough to provide an accurate estimation of the ring center compare with the stellar position, especially for low SNR observations, such as our J and K2 obervations. For this reason, we used the modeling analysis to derive more precise geometric parameters that are described in the next section.
 
\subsection{Disk Modeling Analysis} 

\input{table3.tex}

We fitted our GPI J-, H-, K1-, and K2-Spec total intensity images using the procedure described in Section \ref{subsec:image_fitting}. We plotted the Posterior Distribution Functions for the parameters estimated in Figure \ref{fig:gpihposteriors} for H-Spec (in the Appendix) and in \ref{appendix-figsmcmc} for the other wavelengths. The best fit model for the H-Spec observation and its forward model are shown in Figure \ref{fig:gpihmodel} and the ones for the other observations are shown in \ref{appendix-figsmcmc}. We marked the position of the star with a red point and the position of pericenter with a green point. The SNR residuals for the individual pixels were typically less than 2; however, visual inspection of the residual SNR image revealed some disk structure indicating that the disk is not perfectly subtracted. Such residuals are not surprising because HR 4796A is observed with high SNR and the model for the dust geometry is simple with relatively few free parameters compared with the number of pixels over which the disk is resolved.

We extracted the best fit disk geometric parameters (disk inner radius $R_1$, eccentricity $e$, inclination $i$, argument of the pericenter $\omega$ and principal angle $\Omega$) from the MCMC fit. We listed the best fit parameters in Table~\ref{tab:mcmc_geometry_param}. Since the time needed to optimize the fit for a single band was long, we decided to optimize only the fit for the highest SNR K1-Spec dataset (2015/04/03). The disk geometric parameter uncertainties were smaller but consistent with those estimated from DRA fitting (in Section \ref{sec:disk_geometry_dra}). The disk semi-major axis, $a$, estimated from DRA fitting should not be directly compared with the disk inner radius, $R_1$, estimated from model fitting because the two quantities are defined differently. The {\tt DRA} semi-major axis $a$ is the position of the maximum of the disk in the convolved image and model $R_1$ is the position of the inner radius before convolution. 

We found that most of the GPI best fit parameters are approximately consistent with one another. However, we noted that the stellar offset (eccentricity and argument of the pericenter) was slightly inconsistent because the uncertainties estimated in model fitting did not include the errors in the registration of the star and the direction of True North. Our geometric parameters are generally consistent with those previously measured by \cite{olofsson19} and \citep{milli17,milli19}. Specifically, we also found that the disk pericenter was located on the front side of the disk (north west of the star). However, we measured an eccentricity and pericenter that were slightly different than previously reported in \cite{milli19} ($e = 0.072\pm0.037$, $\omega = -74\fdg2\pm11.9$). We hypothesized that some of these difference may be the result of uncertainties in the measurement of stellar positions due to differences in the instrument and even the observing conditions. However, using the same SPHERE IRDIS H data, we also found differences between our measurements ($e = 0.047\pm0.002$, $\omega = -65\arcdeg\pm1$) and those of \cite{milli17} ($e = 0.070\pm0.011$, $\omega = -72\fdg4\pm5.1$). As a result, we concluded that these small differences in the measured position of the disk center are the result of different measurement methods. Finally, in our disk modeling, we assumed that the disk was circular while \cite{olofsson19} assumed that the disk was elliptical. Since the HR 4796A disk has such a small eccentricity, we do not expect that these differences in assumed eccentricity should impact the comparison of our analyses.

Multiwavelength observations indicate that the NE side of the disk is $\sim$5\%-20\% brighter than the SW side of the disk \citep{telesco00, wahhaj14,olofsson19}. We generated a variety of disk models, varying the disk geometry and scattering phase function, to explore the origin of this effect. Since the scattering phase function is always symmetric with respect to the minor axis (and is studied in more detail in the next section), we did not anticipate that changes in the scattering phase function could reproduce the observed brightness asymmetry. Instead, we expected that geometric effects would make one side of the disk appear brighter than the other if one side was closer to the star than the other. We created a hypothetical disk model with a stellar offset that produced a pronounced North-South asymmetry at the forward-scattering part of the disk in Figure \ref{fig:testmodels} (Bottom-Middle). Once this asymmetry was subtracted from the observations, we detected 2$\sigma$ residuals at pericenter in the fits to the GPI-H and GPI-K1 observations. These remaining residual suggested that the disk pericenter may be bright not only because it is closer to the star but also because it has a higher dust density, consistent with the hypothesis of \cite{olofsson19}.

Recent simulations of disks composed of collisionless particles show that the apocenter is expected to be 1+$e$ brighter than the pericenter at far-infrared and submillimeter wavelengths because the larger number of particles at apocenter overcomes the brightness difference between the apocenter and pericenter at these wavelengths \citep{pan16}. However,
existing ALMA observations have too low a signal-to-noise to detect the predicted apocenter glow for HR 4796 \citep{kennedy18}.

\section{Results: Scattering Phase Function}
\subsection{Reconstruction Analysis} 
\label{sec:spf_extracted_from_image}
\begin{figure*}
\plottwo{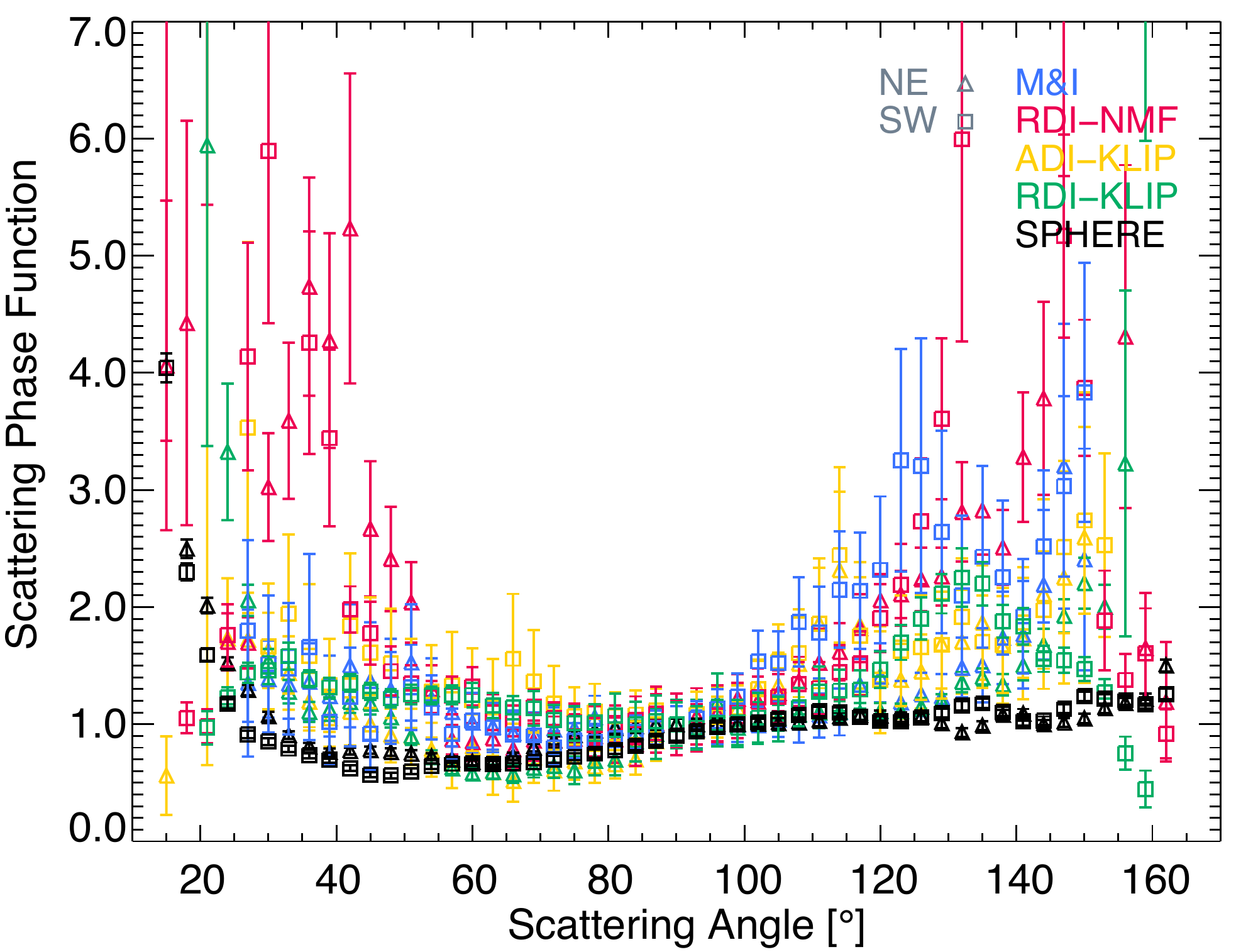}{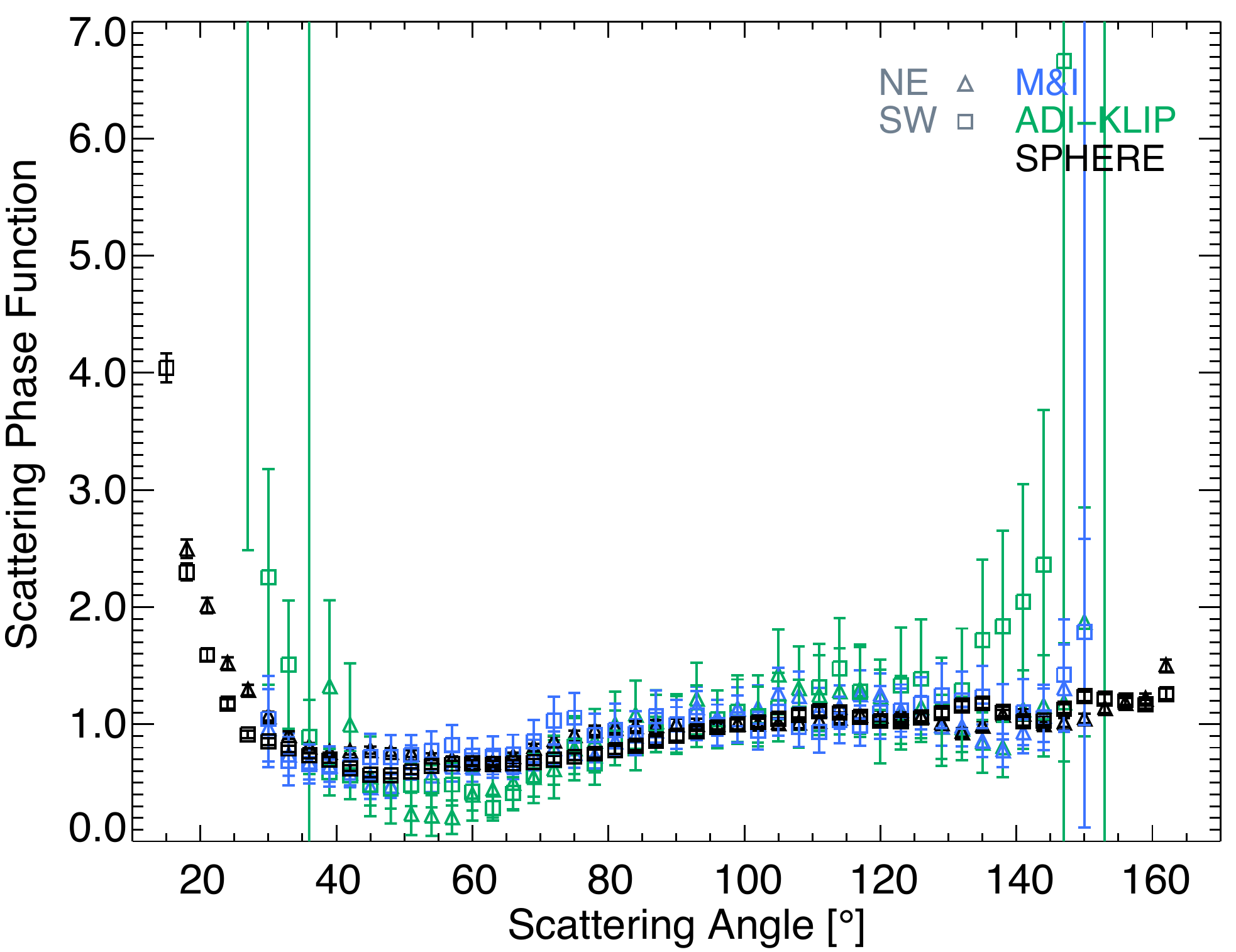}
\figcaption{Scattering phase functions for the HR 4796A disk obtained at (left) H-band and (right) K1-band. The phase functions extracted using Mask-and-Interpolate (blue), RDI-NMF (pink), ADI-KLIP (yellow), and RDI-KLIP (green) are overplotted with SPHERE H2 scattering phase function (black, \citetalias{milli17}) for comparison. Unfortunately, the library of reference PSFs is substantially smaller in J and K1 than in H; therefore, reference PSF subtraction for J- and K1 bands is currently not feasible. For K1, we plot the scattering phase function extracted from the 2015 April 3 observation because it had a large number of individual frames and large field rotation under similar observing conditions and therefore the highest SNR. We use the standard deviation of the three measurements as a proxy for the uncertainty.  
 \label{fig:hr4796spf}}
\end{figure*} 

After verifying that reconstruction of the scattering phase function from PSF subtracted images should be possible, we reconstructed the scattering phase function of the HR 4796A disk from our GPI total intensity images. First, we corrected the images for inverse-square illumination effects. We used our ADI-KLIP measured stellar offset to calculate the deprojected distance from the central star in pixels, $r$, and multiplied by the square of the deprojected distance from the central star ($r^2$). Next, we computed the polar projection for each PSF-subtracted image by averaging the $r^{2}$-scaled surface brightness of the disk within 3 $\degr$ wedge-shaped elliptical apertures at single pixel intervals. Then, we extracted the SPF at each scattering angle from the polar projected images by summing the disk intensity over projected radii extending from $\sim$0$\farcs$18 to $\sim$0$\farcs$31. Finally, we corrected the scattering phase function of the HR 4796 disk at the 1-dimensional level using our isotropic disk model injected into an empty GPI observing sequence and PSF subtracted with Mask-and- Interpolate, RDI-NMF, RDI-KLIP, and/or ADI-KLIP.  More specifically, we divided the HR 4796 disk scattering phase function by that from a similarly extracted isotropic disk for each PSF subtraction technique.

We plot our corrected, empirical H and K1 scattering phase functions in Figure~\ref{fig:hr4796spf}. Unfortunately, the library of reference PSFs is substantially smaller in the J, K1, and K2 bands than in the H band; therefore, reference PSF subtraction (RDI) for J, K1, and K2 is currently not feasible. We find that the K1-Spec Mask-and-Interpolate and ADI-KLIP scattering phase functions are consistent with one another to within our uncertainties at scattering angles 30$\arcdeg$-150$\arcdeg$. In addition, we find that our GPI scattering phase function is consistent with the SPHERE H2 and H3 phase functions previously reported by \citetalias{milli17}. Indeed the GPI K1-Spec phase function exhibits some of the same extremely forward-scattering behavior at small phase angles $<$45$\arcdeg$ and modest backward-scattering at large phase angles $>$90$\arcdeg$. However, the precision at which the phase function is measured at small ($<$40$\arcdeg$) and large ($>$140 $\arcdeg$) phase angles is poorer in the GPI observations despite good observing conditions and a comparable range of parallactic angles. There are many factors that affect high contrast imaging observations including the weather conditions and the instrument performance. 

Our GPI H-Spec Mask-and-Interpolate, RDI-NMF, RDI-KLIP, and ADI-KLIP SPFs are not consistent with one another or with the SPHERE SPF. Instead, all of our reconstructions appear to include an over-correction for the SPF near forward and backward scattering. Since the shorter wavelength data are more susceptible to speckle noise, one possibility is that our empty observing sequences are not sufficiently similar to our science observations to provide an accurate correction. Unfortunately, we can not reconstruct J or K2 SPFs using the PSF subtracted images. The residual speckle noise in the J-band Mask-and-Interpolate image is too high. There is so little GPI data taken using the K2 filter that we could not find a suitable empty observing sequence in which to inject our idealized disk model. 

\subsection{Modeling Analysis} 
\begin{figure}
\includegraphics[width=0.49\textwidth]{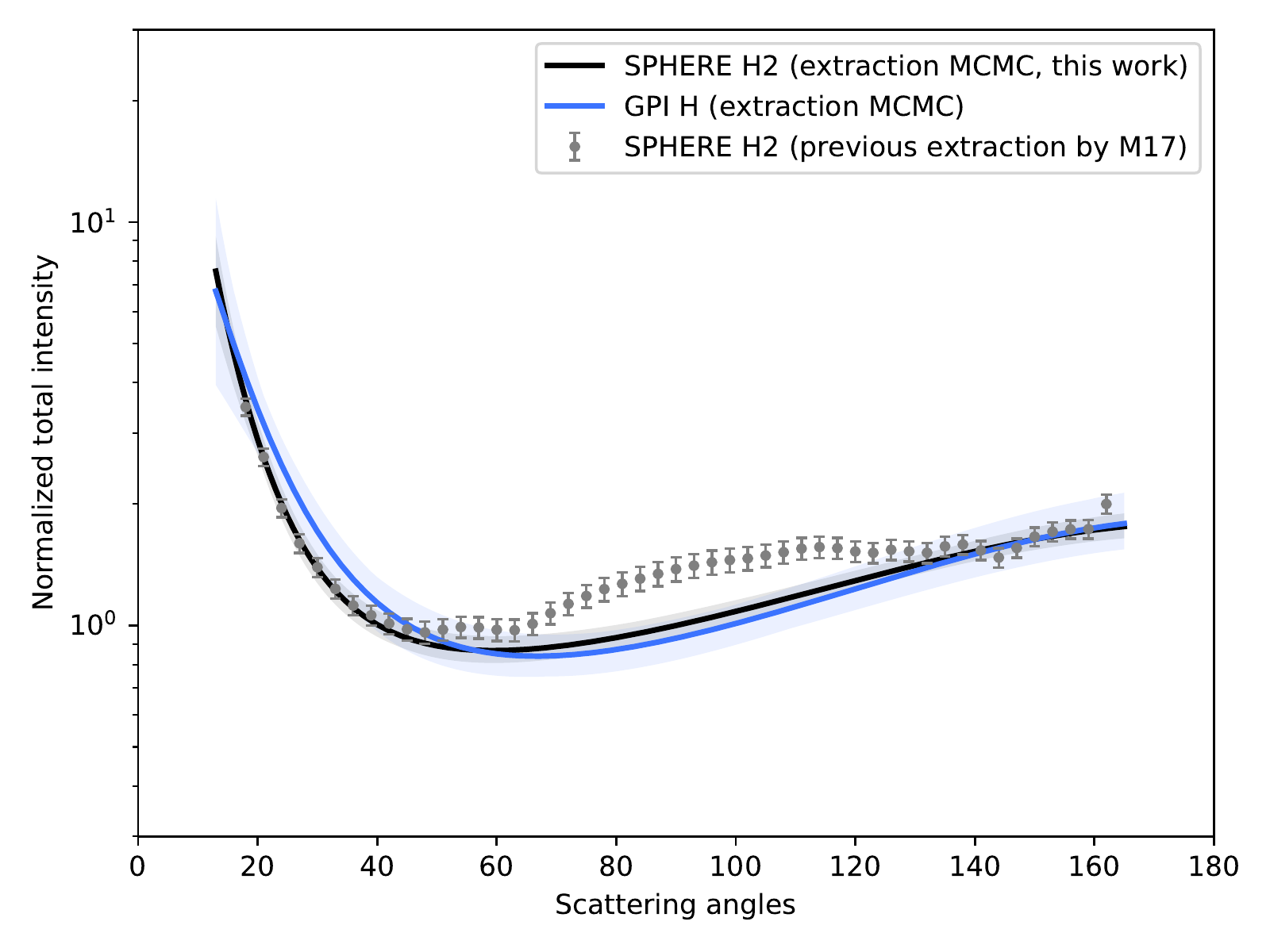}
\figcaption{
 \label{fig:compare_spf} SPHERE H2 Scattering Phase Function for the HR 4796A debris disk extracted by \citetalias{milli17} from a corrected KLIP-ADI PSF-subtracted image (gray error bars) and by our MCMC disk fitting routine from SPHERE H2 (black line) and GPI H-Spec (blue line) observations. The shaded regions in the plots show the uncertainties and are estimated from 1000 randomly chosen SPFs generated by the MCMC sampler.}
\end{figure}

For the GPI H-Spec image, we show the Posterior Distribution Functions for our best-fit model, inferred from our MCMC analysis, assuming the 2 component HG phase function model (Figure~\ref{fig:gpihposteriors}). We list the best-fit parameter values in Table~\ref{tab:mcmc_spf_param}. We estimate asymmetric scattering parameters, $g_1$ = 0.78$\pm$0.04 and $g_2$ = -0.21$\pm$0.01. We plot our best-fit model scattering phase function in blue in Figure~\ref{fig:compare_spf}. This SPF is generally consistent with those inferred from previous ground- and space-based observations \citep[][\citetalias{milli17}]{debes08}. The HR 4796A dust scattering phase function shows very forward scattering behavior a scattering angles $<$40$\arcdeg$, a minimum at scattering angles 30$\arcdeg$ to 70$\arcdeg$, and back scattering at scattering angles $>$70$\arcdeg$. This shape is distinctive compared with SPFs for Solar System dust and other debris disks that typically have a minimum around 90$\arcdeg$ \citep{hughes18}. Our MCMC-exctracted SPF (blue solid line in Figure~\ref{fig:compare_spf}) shows small but significant differences in the shape of the SPF compared with that published by \citetalias{milli17} (grey points), specifically at the position of the ansae ($60\arcdeg < \theta < 120\arcdeg$). To understand this difference, we reanalyze the SPHERE H2 data with our disk fitting method.

\subsection{Reanalysis of the SPHERE H2 data} 
\label{sec:mcmc_sphere}

\begin{figure*}
\centering \includegraphics[width=0.8\textwidth, trim=0 80 0 0, clip]{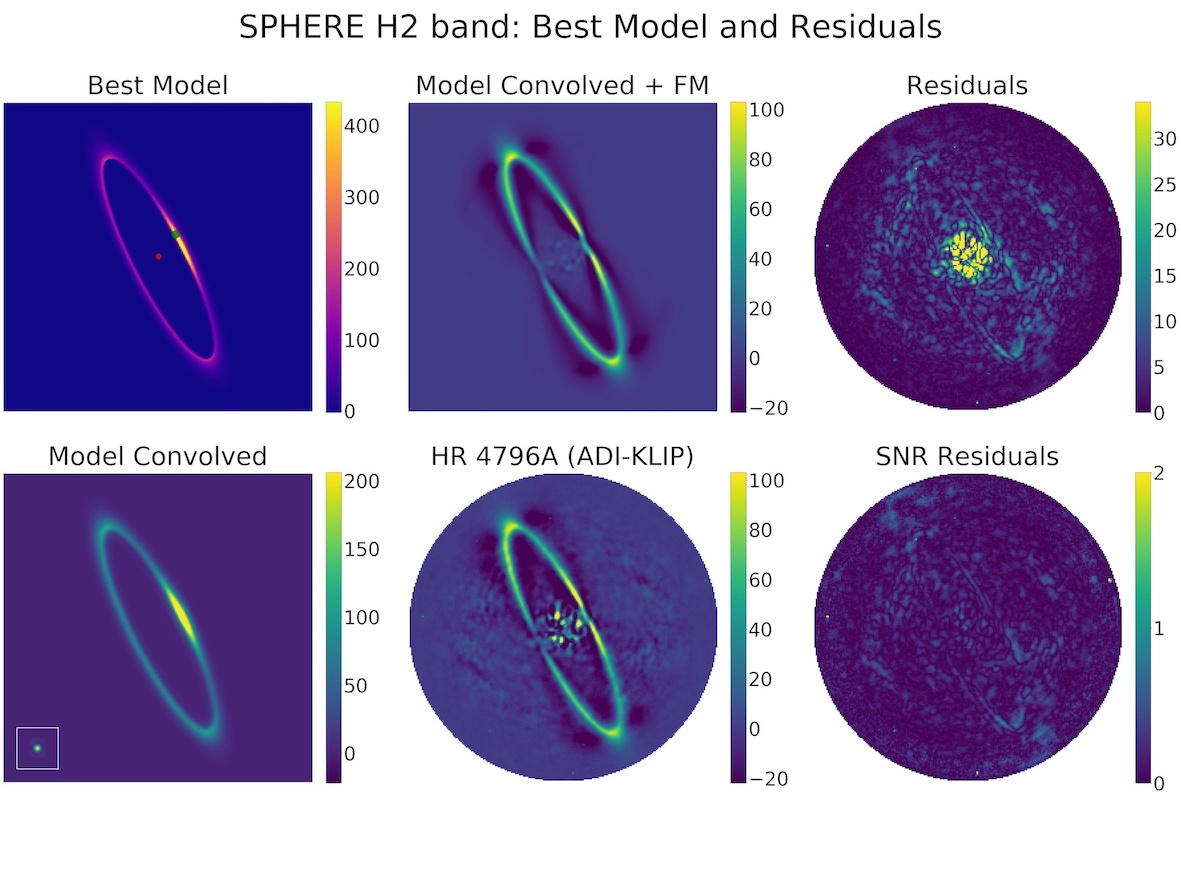}
\figcaption{
 \label{fig:spheremodel} Best-fit model resulting of the MCMC: Same as Figure~\ref{fig:gpihmodel} for SPHERE H2-band.}
\end{figure*} 

We applied our fitting code (already described in Section \ref{subsec:image_fitting}) to the already published SPHERE IRDIS H2 observations of HR 4796A (\citetalias{milli17}). These data were obtained in February 2015\footnote{Based on observations made at the Paranal Observatory under ESO programme 095.C-0298(H)}. The raw frames were sky-subtracted, flat-fielded and bad-pixel corrected using the SPHERE data reduction and handling \citep[DRH,][]{pavlov18} pipeline and star centered using the waffle spots. The uncertainty in the measurement of True North is 0.08$\arcdeg$ for SPHERE/IRDIS \citep{maire16}. The uncertainty in the position of the star is 0.25 pixels (or 3 mas) for SPHERE/IRDIS (\citetalias{milli17}). We re-reduced the data using the {\tt pyKLIP} ADI-KLIP algorithm and the same parameters as for the GPI data (3 KL modes and an exclusion critera of 6$\arcdeg$), resulting in the image shown in Figure \ref{fig:spheremodel} (Bottom-Middle). 

One difference between the GPI-Spec and SPHERE/IRSDIS observations is the availability of a PSF reference. The SPHERE observations did not include satellite spots during the deep coronagraphic sequence; however, the host star was observed out of the coronagraphic mask right before and right after the ADI sequence to facilitate absolute flux calibration. We used these unocculted observations of the host star as a proxy for the PSF and convolved them with our model to estimate the observed astronomical scene. We presented the original KLIP-ADI PSF subtracted SPHERE H2 IRDIS image, the best fit model, the best fit model convolved with the PSF, the forward model of the observing sequence, the residuals for the observing sequence, and the SNR residuals for the observing sequence in Figure~\ref{fig:spheremodel}. We showed the Posterior Distribution Functions for the model parameters estimated from our MCMC analysis, using the 2 component HG phase function, in Figure~\ref{fig:sphereposteriors}.

We found that our SPHERE H2 and GPI H-Spec total intensity best fit models were in very good agreement with one another. The geometric parameters (see Table~\ref{tab:mcmc_geometry_param}) were consistent to within 1$\sigma$ with the exception of the eccentricity ($e$). The smaller SPHERE uncertainties were the result of the higher SNR, especially at small angular separations. The uncertainties on the True North and star position, that were not folded into the reported uncertainties, were 2-3 times larger for SPHERE than for GPI. Even though the SNR was higher in the SPHERE image, the residuals in the SPHERE data were generally smaller than those in the GPI H observations.

The debris disk's geometry and SPF were not expected to change in the intervening year between the observations, especially because the two sets of observations were made in approximately the same spectral band. From our measurements, we concluded that our 11 parameter model did not limit our ability to fit the disk because the fitting procedure produced similar results for the two different observations. We therefore concluded that the differing residuals amplitude between the IRDIS H2 and GPI H-Spec observations stemmed from differences in the fidelity of the PSFs used to simulate our observations. GPI and SPHERE observers used different methods to estimate the instrumental PSFs. GPI observers recreated the PSF from observations of the satellite spots that were spectrally collapsed. This method provides a simultaneous measurement of the PSF, but produces low SNR PSFs. SPHERE observers estimated the PSF from observations of the un-occulted host star taken before and after the observational sequence. This method provides higher SNR PSF measurements but not simultaneous. The smaller residuals in the SPHERE image indicates that, at least in the context of analysis of bright disks such as HR 4796, the second PSF generation method seems to produce better results.

Recent analysis suggests that the brightness asymmetry observed near pericenter can not be entirely attributed to the fact that dust grains at pericenter receive, and therefore scatter, more light compared with those at apocenter. Indeed, \cite{olofsson19} required a density enhancement at pericenter to explain the brighness asymmetry observed in SPHERE observations. In our analysis, we assumed that the dust density was azimuthally symmetric. In this case, (1) any North-South brightness asymmetries in the model are the result of a stellar offset and (2) any additional brightness asymmetry beyond that predicted by our model must be the result of a local density enhancement. We did not find any large asymmetry in our residuals, suggesting that bulk of the disk brightness asymmetry could be explained by star ilumnination differences. However, we did detect a marginal North-South (2$\sigma$) brightness asymmetry in the residuals (e.g. GPI H-Spec and K1-Spec), particularly at the pericenter of the disk (near forward-scattering), that could be consistent with a dust density enhancement, similar to the one measured by \cite{olofsson19}.

 \begin{figure*}
\centering \includegraphics[width=0.8\textwidth]{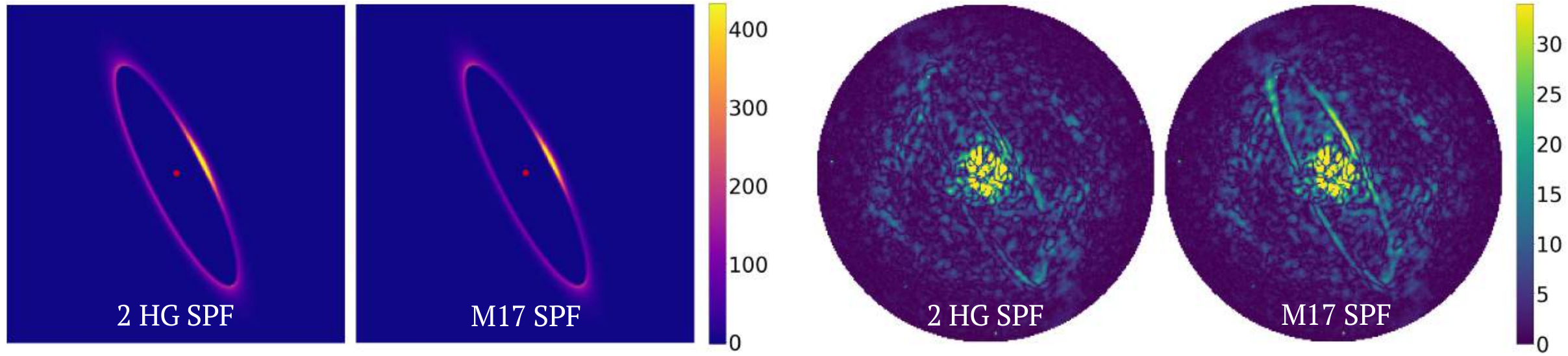}
\figcaption{
 \label{fig:compar_residuals_sphereH2} Comparison of the models (Left) and residuals (Right) for the SPHERE H2 dataset obtained with our MCMC fit method or with the SPF exctracted by Milli et al. 2017.}
\end{figure*} 

The fits to the SPHERE and GPI observation indicated that the SPFs were consistent to within 1 $\sigma$ (Figure~\ref{fig:compare_spf}) with the SPHERE SPF appearing slightly more forward scattering and slightly less backward scattering than the GPI SPF. This difference was seen in the best-fit values of $g_1$ and $g_2$ (see Table~\ref{tab:mcmc_spf_param}). For the SPHERE image, we measured asymmetric scattering parameters $g_1$ = 0.90$\pm$0.03 and $g_2$ = -0.17$\pm0.01$. Although the SPHERE and GPI measured values of $g_1$ and $g_2$ are inconsistent by $>$2$\sigma$, the plotted values of the SPFs are consistent to within the uncertainties. As a result, we concluded that a SPF may be degenerate with values of $g_1$, $g_2$, and $\alpha$. Indeed, our corner plot (see Figure~\ref{fig:sphereposteriors}) showed a strong correlation between $g_1$ and $\alpha$. At present, no alternate prescription has been proposed that removes this degeneracy. As a result, we prefer to compare the SPFs directly and not the SPF fitting parameters $g_1$, $g_2$, and $\alpha$. 

When we compared the SPF that we measured from the SPHERE IRSDIS H2 observations (black solid line in Figure~\ref{fig:compare_spf}) with that published by \citetalias{milli17}, we found significant differences in the shape of the two SPFs. This was surprising because the measurements were made from the exact same observations. When comparing an image of our best model for the SPHERE observations with that of a HR 4796A-like disk with the \citetalias{milli17} scattering phase function (left panel of Figure~\ref{fig:compar_residuals_sphereH2}), we found only minor differences. Visually, the disk with the \citetalias{milli17} SPF appeared only slightly more forward scattering. However, when we compared the residuals of these models after they were convolved with the PSF, forward-modeled, and PSF subtracted right panel of Figure~\ref{fig:compar_residuals_sphereH2}), we found significant differences. Our SPF generally had much smaller residuals than the \citetalias{milli17} SPF. For this reason, we preferred our modeling approach to measuring the SPF compared with extraction and correction from the ADI-reduced images, as described in \citetalias{milli17} and in Section \ref{sec:spf_extracted_from_image}.

We attributed differences in the estimated scattering phase functions to errors in the correction of ADI over subtraction used by \citetalias{milli17} and our team in Section \ref{sec:spf_extracted_from_image}. Previously, \cite{milli12} showed that ADI oversubtraction in disks was dependent on the disk geometry and surface brightness distribution. We concluded that correcting ADI oversubtraction on a complex SPF was challenging. Specifically, using an isotropic SPF model disk that is not representative of the actual disk SPF biased the estimate of the SPF by over correcting the flux at the ansae. Indeed, we concluded that having a good estimate of the disk geometry and SPF is necessary in order to properly account for the complex effects of oversubtraction on disk shape and azimuthal brighness. Furthermore, we discovered that our MCMC sampler approach, using a random exploration of the parameter space with only a few assumptions for the priors, was well suited for this kind of problem. 

In their analysis, \citetalias{milli17} argued that their HR 4796A SPF was too complex to be accurately represented by a 2 component HG phase function. However, in \ref{appendix-3g}, we attempted to reproduce the \citetalias{milli17} SPF using a more complex model (3 component HG phase function). We demonstrated that a more complex model was not necessary and that the SPF was well described using a 2 component HG phase function.
 
\subsection{GPI Color}

\begin{figure}
\centering \includegraphics[width=0.49\textwidth]{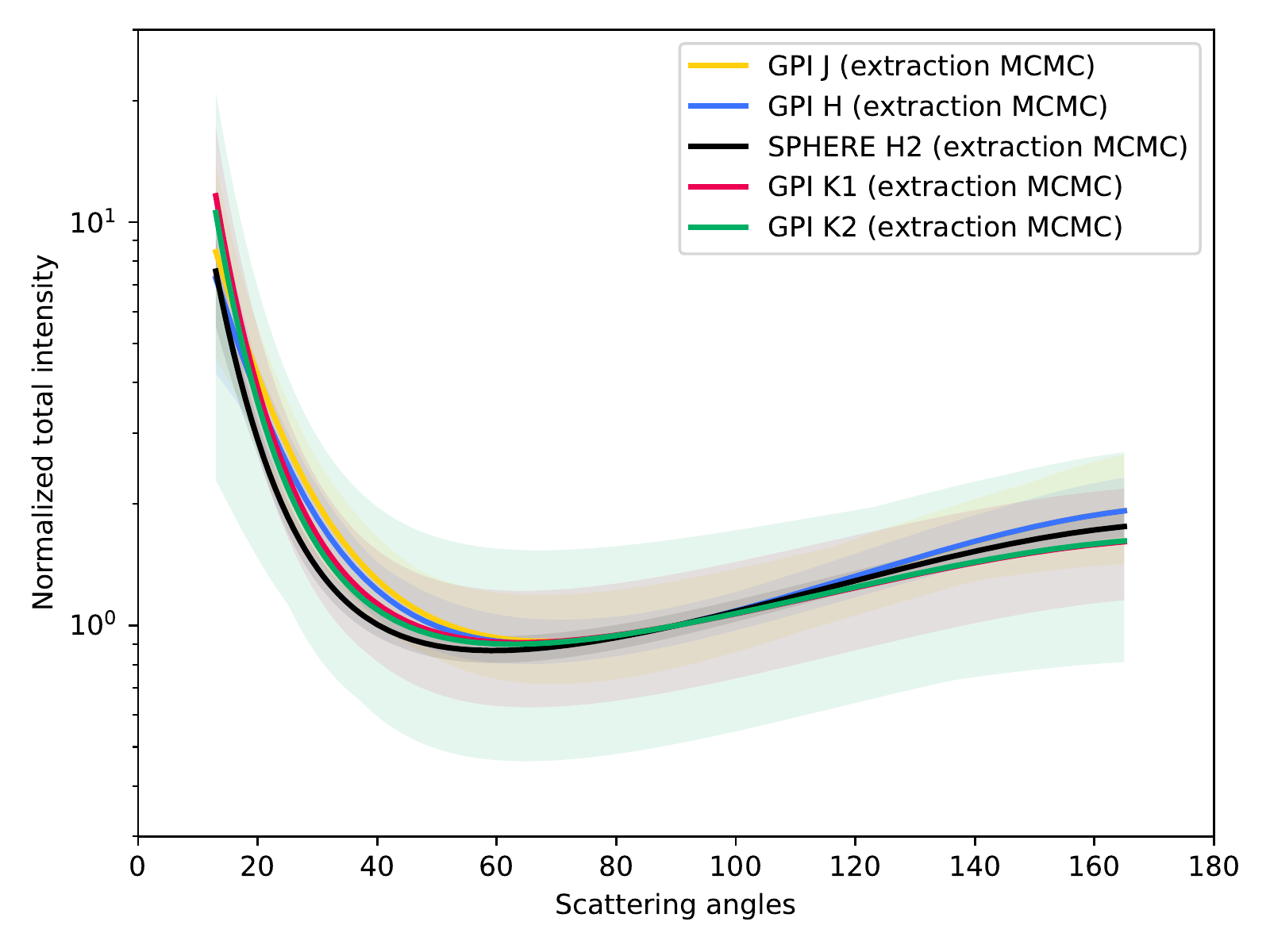}
\figcaption{
 \label{fig:allspf} HR 4796A scattering phase functions estimated by fitting GPI J (brown), H (blue), K1(magenta), and K2 (purple) images. Our analysis does not detect any differences between the shapes of the HR 4796A SPF as a function of near-infrared wavelength within the GPI data. The best values for all of the posteriors are given in Table~\ref{tab:mcmc_spf_param}.} 
\end{figure}

We did not detect any statistically significant differences between the SPFs at J, H, K1 and K2 (see Figure~\ref{fig:allspf}). The formal best fit values for $g_1$, $g_2$, and $\alpha$ were statistically different; however, the curves were consistent with one another to within the uncertainties plotted. For example, the best fit values of $g_1$ and $g_2$ indicate that the grains were more forward scattering at K1 and K2 ($g_1$ = 0.94$\pm$0.05) than at J and H ($g_1$ = 0.80$\pm$0.04) at the 3-4$\sigma$ level. However, the SPFs overlapped to within the uncertainties over the entire range for which the SPFs were measured ($\sim$15$\arcdeg$-155$\arcdeg$, see Figure~\ref{fig:allspf}). Similar to our SPHERE H2 and GPI H-Spec analysis, we concluded that the fitting parameters did not accurately capture the consistencies and inconsistencies between the measurements and that multiple values of $g_1$, $g_2$, and $\alpha$ could be used to represent a single phase function.

The SPF uncertainties were the largest for J and K2 bands and the smallest for SHERE-H2, GPI-H and K1 bands. We attributed the large uncertainties in the J band measurements to the large speckle noise at the shortest wavelengths where the PSF contains the most substructure. We attributed the next largest uncertainties in K2 band to the small throughput of the bandpass and large thermal background that limited the SNR at which the disk was detected. The best performance occured in H band where the PSF was still relatively fine but where the PSF was less complicated.

\input{table4.tex}

\subsection{GPI/STIS Color}
Visual scattered light images of the HR 4796A disk and its interstellar environment have been obtained using the \textit{HST} STIS coronagraph \citep{schneider09,schneider18}. We compared the STIS data with the SPHERE near-infrared scattered light images to estimate the near-infrared-to-visual color. For HR 4796A, the range of scattering angles over which the SPF could be measured was dependent on the instrument Inner Working Angle (IWA); the sampling of the SPF was dependent on the plate scale. HR 4796A was imaged using STIS twice. The first time using WEDGE-A at a location where the wedge is 0.$\arcsec$63 wide \citep{schneider09}. A second time using BAR5 that has a width 0.3$\arcsec$ in conjunction with WEDGE-A at a location where the wedge is 1.0$\arcsec$ wide to enable imaging near and far from the star at high dynamic range \citep{schneider18}. The STIS occultors were wider than the SPHERE Lyot N\_ALC\_YJH\_S coronagraph mask (185 mas) used by \citetalias{milli17}. Unfortunately, STIS's occulter did not provide access to the disk's minor axis. In addition, the STIS plate scale (50.77 mas) was larger than the SPHERE IRDIS plate scale (12.25 mas). The SPHERE PSF was super-sampled for a 8-m telescope operating at near-infrared wavelengths while the STIS PSF was sub-sampled for a 2.5-m telescope operating at visual wavelengths. 

We measured the HR 4796A visual SPF from the more recent STIS BAR5 and WEDGE A-1.0 observations that detect the dust ring at very high SNR. \cite{schneider18} observed the disk at three roll angles during two epochs. During each epoch, they interleaved PSF observations between successive HR 4796A observations to provide the most representative PSF. They performed reference PSF subtraction for each exposure, masking the diffraction spikes, and combining the images to provide a final PSF subtracted image that is minimally impacted by diffraction artifacts. Their approach to PSF subtraction has been highly effective for STIS coronagraphic observations because \textit{HST} and its instruments are extremely stable. We were able to extract the SPF directly from this combined, PSF subtracted image without extensive modeling because \cite{schneider18} did not use Angular Differential Imaging to construct their final STIS PSF subtracted image of HR 4796A. 

\begin{figure}
\plotone{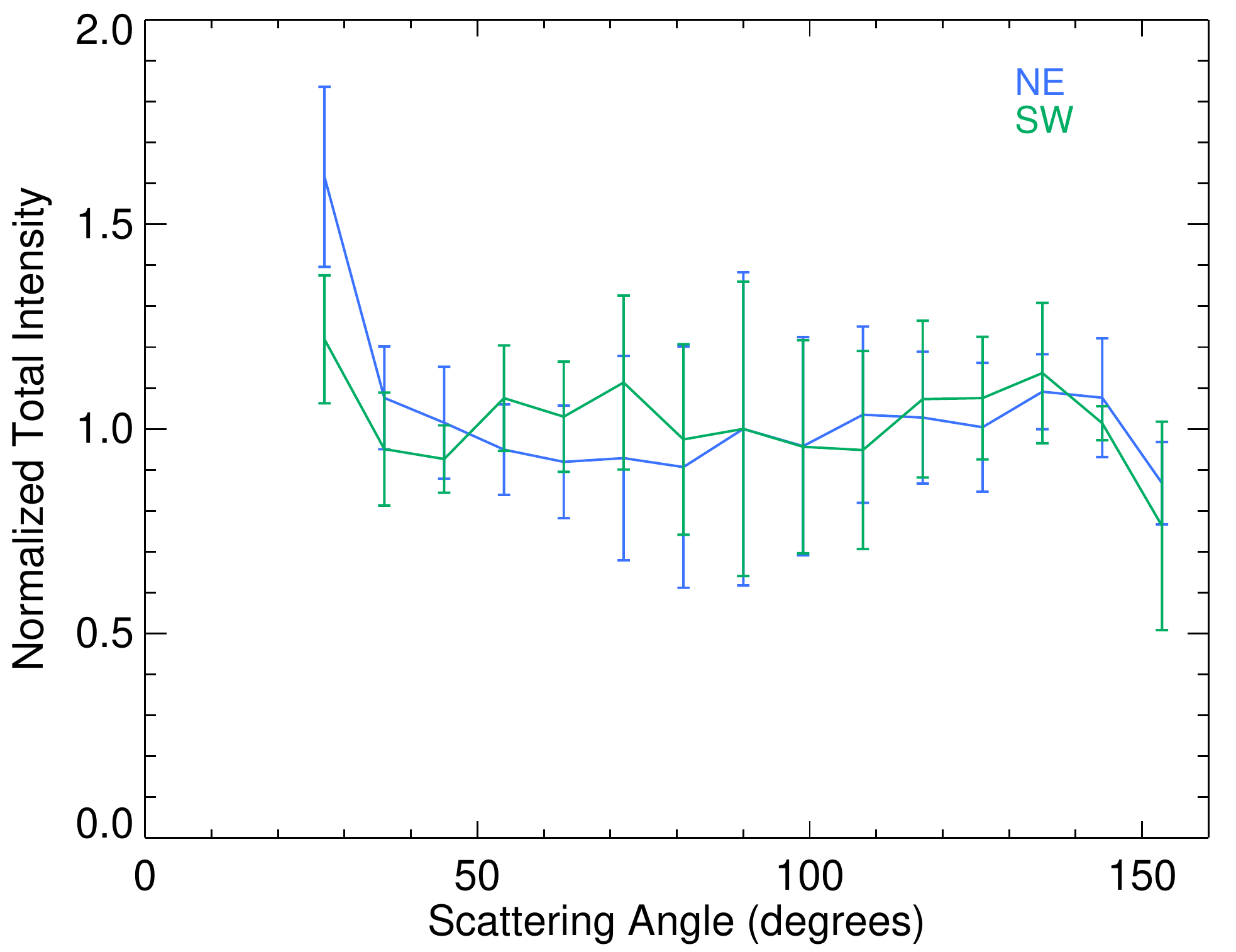}
\figcaption{HR 4796A STIS, visual, total intensity scattering phase function, extracted from high SNR Wedge A-1.0 and BAR5 observations \citep{schneider18}. The 0.15$\arcsec$ IWA does not permit measurement of the disk surface brightness along the minor axis.
\label{fig:reddening}}
\end{figure}

We extracted the STIS SPF for scattering angles 27$\arcdeg$ - 153$\arcdeg$, only a slightly smaller range that the 13.6$\arcdeg$ - 166.6$\arcdeg$ range accessible from our vantage point (Figure~\ref{fig:reddening}). We detected a hint of forward scattering at the smallest scattering angle and a predominantly flat SPF at larger scattering angles.
 
\begin{figure}
\plotone{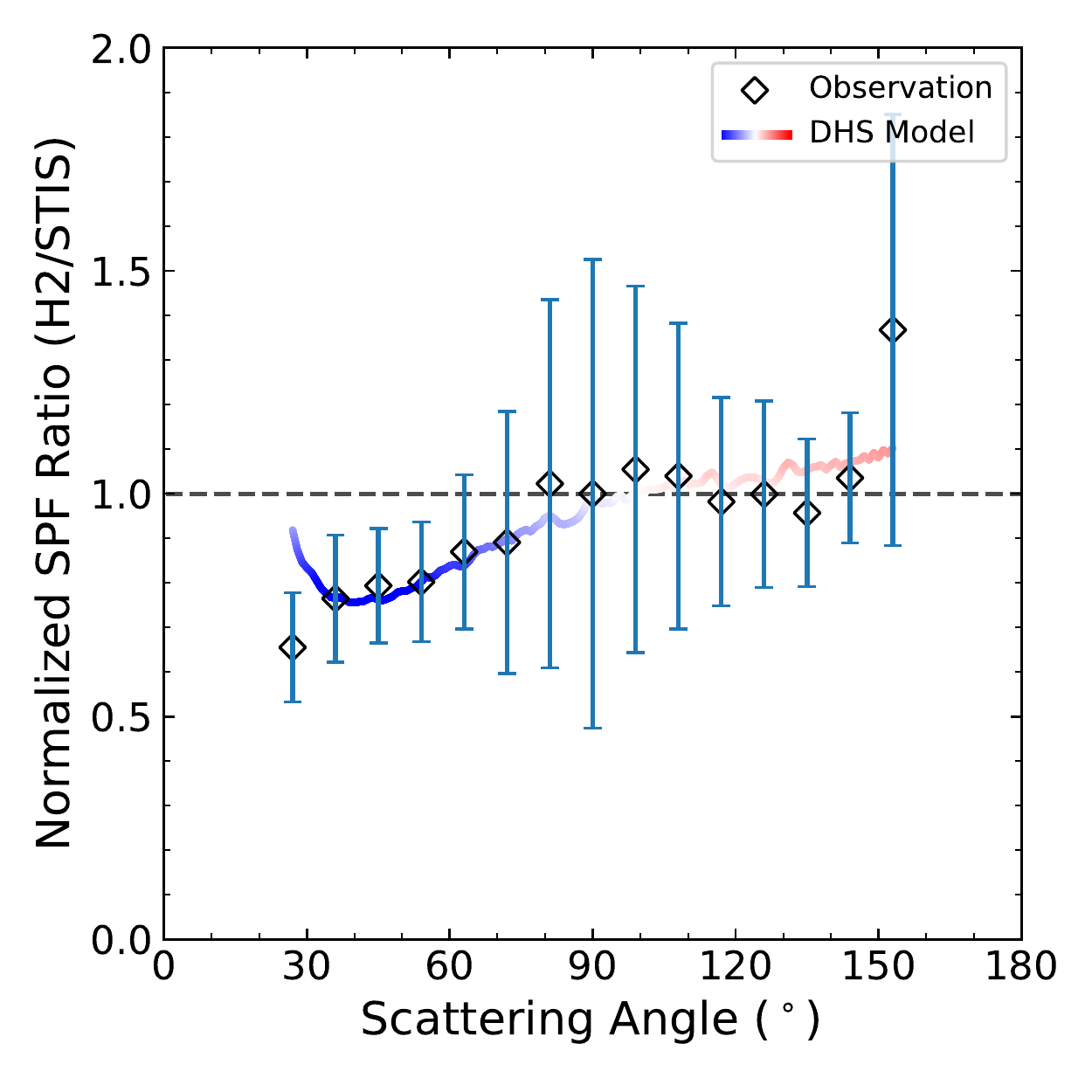}
\figcaption{Ratio of the \textit{VLT}/SPHERE H2 and the  \textit{HST}/STIS SPFs, revealing a slightly blue color for scattering at relatively small phase angles ($<$60$\arcdeg$) and a slightly red color at relatively large phase angles ($>$120$\arcdeg$). Overplotted in red and blue is the expected near-infrared-to-visual color predicted by our DHS model. \label{fig:hr4796spfcolor}}
\end{figure}

We calculated the ratio of the near-infrared to visual SPFs to search more carefully for trends in color as a function of scattering angle. Even though our measurement of the H-band SPF provided an improvement compared with previous measurements, we performed our analysis on the previously published SPHERE H2 SPF derived by \citetalias{milli17} because their analysis did not assume any functional form for the SPF. Further, to facilitate analysis of the two data sets, we interpolated the SPHERE H2 measurements to the same scattering angles as the STIS measurements because the SPHERE measurements are well sampled and smooth. Laboratory measurements of large particles ($>$200 $\mu$m) suggested that the ratio of the reflectances measured at long wavelengths compared to short wavelengths should rise as a function of scattering angle, consistent with the prediction from geometric optics \citep{schroeder14}. We plotted the ratio of the SPHERE H2 and \textit{HST} STIS SPFs in Figure~\ref{fig:hr4796spfcolor}. We found that the ratio of the SPHERE H2 scattering phase function with that of the \textit{HST} STIS scattering phase function has a slightly blue color at relatively small phase angles ($<$60$\arcdeg$) and a slightly red color at relatively large phase angles ($>$135$\arcdeg$), consistent with expectations based on measurements made in the laboratory \citep{schroeder14}.

\begin{figure}
\plotone{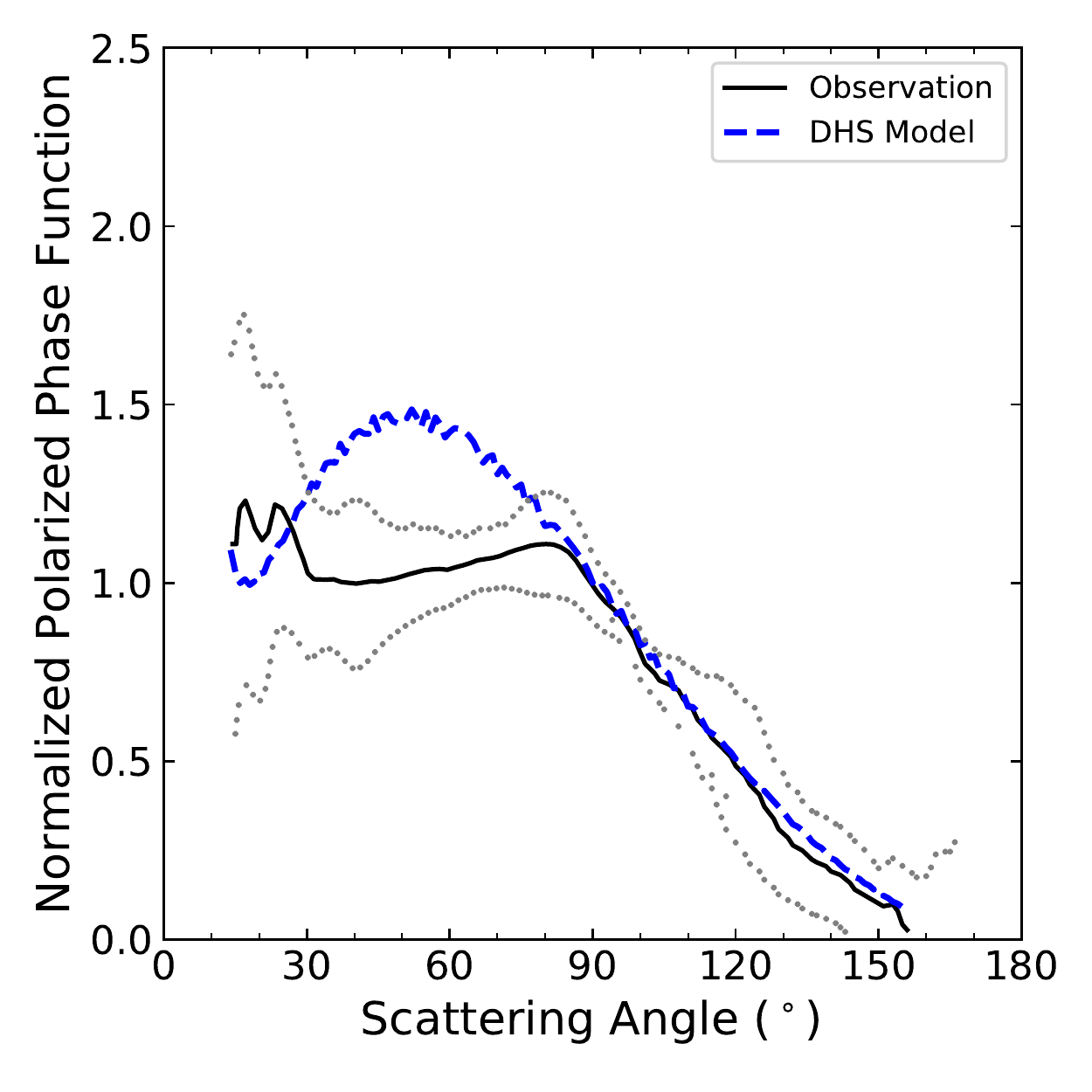}
\figcaption{Observed \textit{VLT}/SPHERE VBB polarized SPF in \citet{milli19} and the prediction by our DHS model. The DHS model is not able to reproduce the observed plateau at ${\sim}30^\circ$ to ${\sim}90^\circ$. \label{fig:hr4796polpf}}
\end{figure} 

\section{Dust Grain Properties}
The SPF can be used to place detailed constraints on the properties of the dust. \citetalias{milli17} modeled their SPHERE HR 4796A SPF using a grid of 7800 dust compositions and sizes assuming Mie Theory and/or the Distribution of Hollow Spheres (DHS). DHS is a theoretical construct that is used to approximate the behavior of aggregates \citep{min05, min16}. \citetalias{milli17} found that the local brightness enhancement on the backside of the disk (close to the ansa) could only be reproduced by DHS models with large silicate grains. Unfortunately, they were unable to reproduce the SPF over the full range of measured phase angles (13.6$\arcdeg$-166.6$\arcdeg$) using either Mie or DHS. Their smallest reduced $\chi^{2}$ = 131. They constructed a best-fit model to the forward scattering part of the SPF ($\leq$45$\arcdeg$) using Mie Theory assuming porous water ice grains (the fraction of void occupied by the ice, $p_{H_{2}O}$ = 90\% and the porosity without ice $P$ = 0.10\%) with a very steep power-law size-distribution, $dn/ds$ $\propto$ $s^{-5.5}$, and a minimum grain size $s_{min}$=17.8 $\mu$m. This model did not provide any back scattering. However, they noted that large grains modeled using Hapke Theory reproduced not only the observed back scattering behavior but also the forward scattering behavior at scattering angles $>$30$\arcdeg$, if they have a size $\sim$30 $\mu$m.

\subsection{Modeling the SPHERE H2 Phase Function}
We used {\tt MCFOST} to model the SPHERE SPF, predicting the SPFs expected from various dust populations, assuming that the grains were well represented by the DHS model. We chose to model the SPHERE SPF so that the results from our modeling effort could be more directly compared with previous modeling efforts.

First, we experimented with fitting by eye. Building on the experience from \citetalias{milli17}, we assumed a large minimum grain size ($\sim$30 $\mu$m) to reproduce the observed back scattering from the more distant side of the disk. This requirement was consistent with the large minimum grain size expected from radiation pressure blow-out ($\sim$10 $\mu$m). We found that we could reproduce the overall shape of the SPF by balancing the grain-size distribution power law with the minimum grain size. However, we could not reproduce the observed dip in the scattering phase function at 30-40$\arcdeg$ if the grains were composed of silicates or water ice only, consistent with the findings of \citetalias{milli17}. We discovered that we could reproduce the dip in the scattering phase function by adding amorphous carbon and metallic iron, common materials observed in the ISM and protoplanetary disks. As a result, we used amorphous silicate, amorphous carbon, and metallic iron in our grain models. This dust composition is consistent with those used in other studies that reproduced the color of the scattered light and/or thermal emission from the disk \citep{rodigas15,debes08}. 

\begin{figure}
\plotone{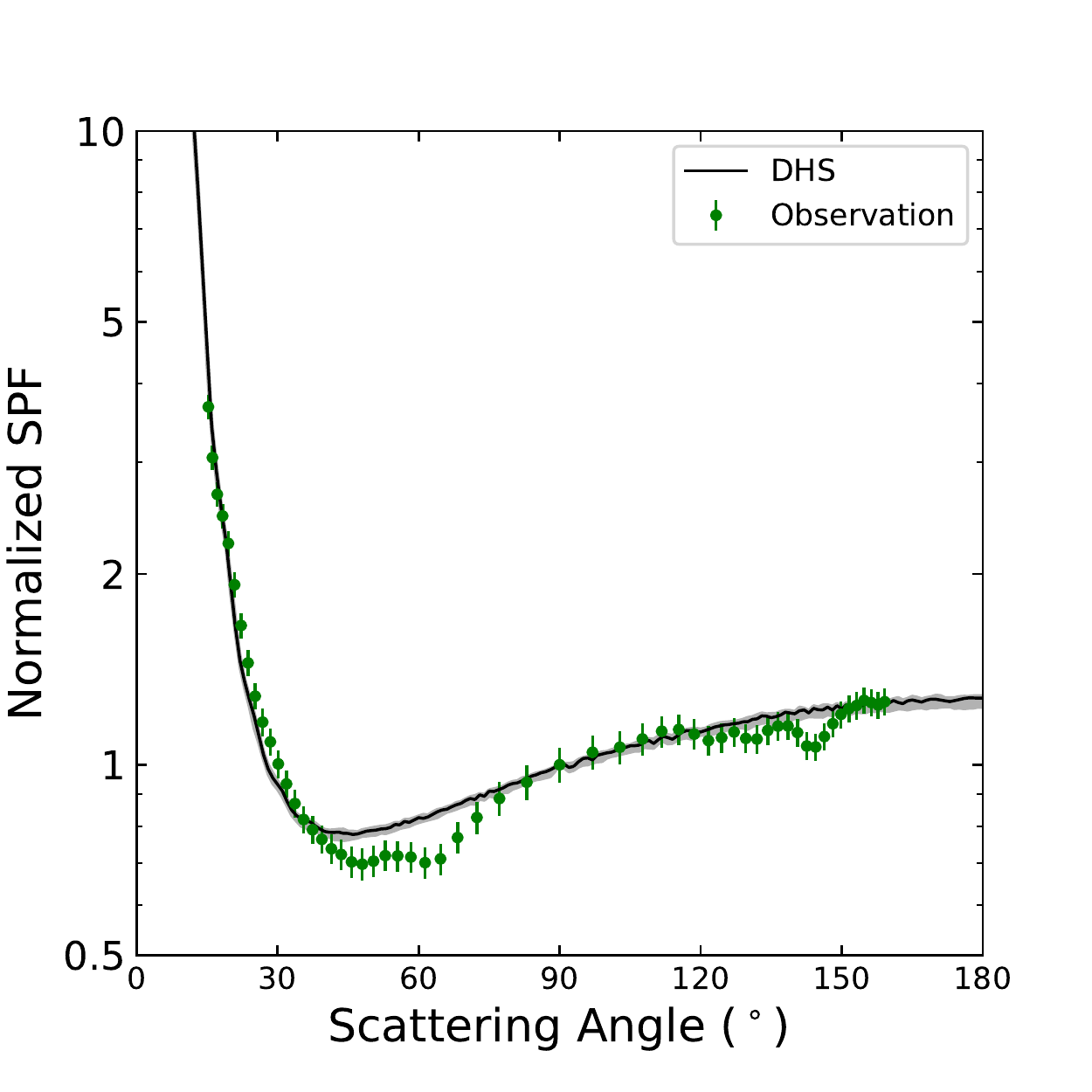}
\figcaption{Comparison of the SPF for our best fitting DHS model (black line) and that extracted from the SPHERE H2 observations by \citetalias{milli17}.
\label{fig:mcfostspf}}
\end{figure}

After we found a reasonable fit by eye, we used a Bayesian analysis to locate the best fit and measure the posterior distribution functions. Specifically, we used the {\tt DebrisDiskFM} \citep{ren19} and {\tt emcee} \citep{foreman-mackey13} packages to distribute the MCMC posterior calculation on different computer nodes. We allowed the maximum void fraction ($V_{max}$), porosity ($P$), amorphous silicate ($f_{Si}$), amorphous carbon ($f_C$), and metallic-iron ($F_{Fe}$) volumes to vary between 0\% and 100\% with the constraint that $f_{Si} + f_{C} + f_{Fe} = 1$. Thus, only $f_{Si}$ and $f_{C}$ are explicitly sampled. We allowed the minimum grain size ($s_{min}$) to vary between 0.5 and 100 $\mu$m but fixed the maximum grain size ($s_{max}$ = 1000 $\mu$m). We allowed the exponent for the power law size distribution to vary between -6 and -3. The best-fitting parameters are listed in Table~\ref{tab:MCFOST} and the resulting SPF is compared to the observed SPHERE scattering phase function in Figure~\ref{fig:mcfostspf}.

\begin{deluxetable}{llcc}[htb!]

\tablecaption{{\tt MCFOST} Dust Model\label{tab:MCFOST}}
\tablehead{
	& Symbol	& \multicolumn{2}{c}{Value}\\
Parameter & 		& Prior\tablenotemark{a}	& Posterior\tablenotemark{b} }
\tablewidth{0pt}
\tablecolumns{3}
\startdata
Scattering Theory &  &  \multicolumn{2}{c}{DHS}  \\
Maximum Void Fraction & $V_{\rm max}$ & $(0\%, 100\%)$ & $76_{-4}^{+6}$\%\\
Porosity & $P$ & $(0\%, 100\%)$ & $15_{-11}^{+12}$\% \\ \hline
Minimum grain size & $s_{\rm min}$ & $(0.5, 100)\,\mu$m & $25_{-4}^{+4}$ $\mu$m \\
Maximum grain size & $s_{\rm max}$ & \multicolumn{2}{c}{$1000\mu$m} \\
Power-law Size Distribution & $\nu$ & $(-6, -3)$ & $-3.74^{+0.12}_{-0.05}$ \\ \hline
Amorphous Silicate Volume\tablenotemark{c} & $f_{\rm Si}$ & $(0\%, 100\%)$ & 42$^{+11}_{-13}$\% \\
Amorphous Carbon Volume\tablenotemark{c} & $f_{\rm C}$ & $(0\%, 100\%)$ & 17$^{+16}_{-12}$\% \\
Metallic Iron Volume\tablenotemark{c} & $f_{\rm Fe}$ & $(0\%, 100\%)$ & 37$^{+15}_{-9}$\% \\  \hline
Reduced Chi-squared Value & $\chi^2_{\nu}$ &  & $2.3$ \\
\enddata
\tablecomments{\tablenotetext{a}{The parameters are limited to 3 decimal digits, with uniform sampling in the prior range (except  $s_{\rm min}$, which is log-uniformly sampled).}
\tablenotetext{b}{$16$th, $50$th, and $84$th percentiles.}}
\end{deluxetable}

Our best fit model is composed of 42$\pm$4\% amorphous silicate, 17$^{+11}_{-13}$\% amorphous carbon, and 37$^{+15}_{-9}$\% metallic iron by volume, suggesting that a substantial fraction of metallic iron was needed to reproduce the shape of the SPHERE H2 SPF. This composition has somewhat less silicates and more iron than estimated for protoplanetary disks (75\% amorphous silicate, 10-15\% amorphous carbon, and 10-15\% iron sulfide, \citealp{min16}). However, we note that the amorphous carbon volume fraction posterior distribution function was not Gaussian, indicating that the porosity and the amount of amorphous carbon was not well constrained. In addition, our best fit model requires a large minimum grain size, $s_{min}$ = 25$\pm$4 $\mu$m, consistent with previous analyses of the SPHERE H2 SPF (\citetalias{milli17}) and somewhat larger than expected from radiation pressure blow-out if the grains are spheres. Our corner plots (see Figure~\ref{fig:mcfostspf}) show that the minimum grain size ($s_{min}$), maximum void fraction ($v_{max}$), and grain size distribution power-law ($\nu$) are correlated with smaller minimum grains requiring grains with larger void fractions and shallower grain size distributions. The large minimum grain size estimates are consistent with new estimates for the minimum grain size using the Discrete Dipole Approximation to more accurately model irregular grains around luminous stars. \cite{arnold19} found that the minimum grains size around A-type stars are expected to be larger than previously predicted using Mie Theory. We expect that the difference between our measured GPI SPF and our model SPF is small relative to the systematic uncertainties in the dust models.

Once we located the best fit for the SPHERE H2 SPF, we predicted the total and polarized intensity SPFs expected for the same grain distribution in the GPI J, H, K1 and K2 and the STIS visual filter. We compared our predictions to the observed (1) SPHERE/IRSDIS H2 to STIS color and (2) SPHERE/ZIMPOL visual polarized intensity phase function. We overlaid our predictions for the H2 to STIS color on the measurements in Figure~\ref{fig:hr4796spfcolor}. We found that our model predictions are consistent with the observations to within 1$\sigma$. SPHERE/ZIMPOL polarized intensity images of HR 4796A in the Very Broad Band indicate that the polarized intensity phase function was relatively uniform over the scattering angles observed \citep{milli19}. Our model indicated that the HR 4796A dust was forward scattering at scattering angles smaller than 13.6$\arcdeg$, the minimum scattering angle accessible due to the disk's 76$\arcdeg$ inclination. 

\section{Discussion}
In our analysis of the GPI HR 4796A observations, we (1) estimated the geometric parameters for the disk, (2) estimated the scattering phase function (SPF) for the grains, and (3) constrain the dust properties (e.g. size, porosity, and composition) assuming that the grains could be well described using a DHS. We tried two approaches for estimating the geometric parameters and the scattering phase function: reconstruction from PSF subtracted images (where we tried multiple PSF subtraction techniques) and recovery by forward modeling the images, assuming a simple geometric model and a two component Henyey-Greenstein SPF. 

Our measurements of the geometric parameters estimated from PSF subtracted images and from forward modeling are consistent with one another, despite very different approaches. Our results gives us confidence that many measurement techniques are robust. Indeed, our results are also broadly consistent with previously published studies, albeit with significantly lower eccentricity. Our results may also suggest that measuring the geometric parameters for low-eccentricity disks may be disproportionately impacted by uncertainties in the stellar position and/or PSF subtraction effects, such as self-subtraction. We confirm the non-zero eccentricity and therefore confirm the likelihood of an underlying planetary mass companion. We find tentative evidence of excess brightness at periastron at the 2$\sigma$ level, consistent with a density enhancement there.

Consistent with \citetalias{milli17}, we measure strong foward scattering, mild back scattering, and a minimum in the scattering phase function at 5 - 60$\arcdeg$ scattering angle. However, the details of the SPF depend sensitively on the technique used. For example, the SPF bump at 90$\arcdeg$ report by \citetalias{milli17} is recovered in our Mask-and-Interpolate SPF reconstruction (in K1-Spec) but not in our forward modeling analysis (of J-, H-, K1-Spec). The discrepancy between the two techniques is unexpected because our model tests indicate that both techniques can be used to robustly retrieve an input SPF. However, our model test were carried out assuming extremely stable observing conditions in which the PSF for the target and the reference were the same. We did not detect any differences between the SPFs measured at J, H, and K1; however, we did detect differences between the near-infrared SPF and the visual SPF measured from \textit{HST} STIS observations. The near-infrared-to-visual color of the scattering phase function is consistent with measurements of very large particles.

We modeled the HR 4796A scattering phase function using {\tt MCFOST}. Previous studies struggled to fit the SPF using Mie Theory and DHS. We chose to focus our modeling efforts on DHS grains because they better approximate aggregate grains that have been lofted off of Solar System minor bodies. Our best-fit {\tt MCFOST} DHS model to the SPHERE H2 SPF broadly predicts (1) the visual SPF measured from \textit{HST} STIS observations (without using the \textit{HST} data in the model) and (2) the visual polarized intensity phase function measured from SPHERE ZIMPOL observations, although it does not predict the exact behavior in the ZIMPOL data near 50$\arcdeg$ scattering angle correctly. Therefore, our analysis is consistent with the presence of large aggregates composed of smaller monomers. Unfortunately, modeling with true aggregate scattering properties is beyond current computational capabilities. 

\begin{figure*}
\plotone{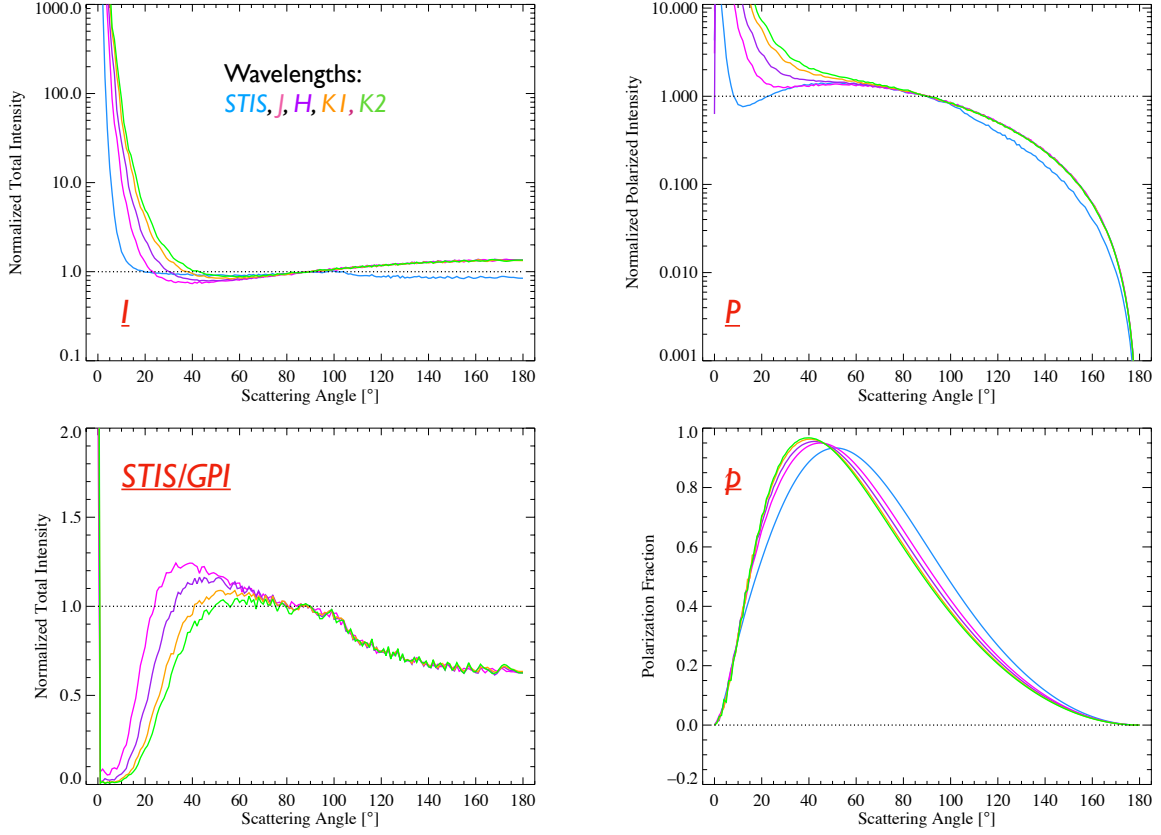}
\figcaption{(Top-Left) DHS model total intensity scattering phase functions shown for \textit{HST} STIS and GPI J, H, K1, and K2. (Top-Right) DHS model polarized intensity scattering phase functions. (c) (Bottom-Left) Ratio of the \textit{HST} STIS and GPI J, H, and K1 scattering phase functions. (Bottom-Right) The polarization fraction ($p$ = $P$/$I$) scattering phase functions.   
\label{fig:dhsjhk}}
\end{figure*} 

In addition, our model predicts changes in the total intensity and polarized intensity SPFs when moving from short to long wavelengths within the near-infrared bands. For example, both the total intensity and polarized intensity SPFs are expected to have increasingly broader forward-scattering cones at longer and longer wavelengths and smaller and smaller shoulders of approximately constant polarized intensity at shorter and shorter wavelengths (see Figure~\ref{fig:dhsjhk}). The lack of detection of these differences within the J, H, and K1 SPFs is probably the result of the low SNR with which we extract the SPFs. A new analysis of the HR 4796A K1-Spec data using NMF with data imputation, where the disk region is first ignored and then empirically recovered using PSF-only signals to minimize reduction bias, indicates that the observations did capture changes in the SPF as a function of wavelength within the IFS data cube \citep{ren20}.

Although not discussed here, GPI's Polarimetric mode can measure the polarized intensity at J, H, and K1 bands to place additional constraints on the grain properties. This mode uses a Wollaston prism to simultaneously divide the light into the Ordinary and Extraordinary beams and differencing the Ordinary and Extraordinary images to obtain the PSF subtracted polarized light image. Thus, the polarized intensity image better preserves the disk geometry and surface brightness distribution and achieves a contrast performance $\sim$10 times better than ADI \citep{perrin15}. This improved performance may allow the detection of the trends in polarized intensity predicted by our model.

\section{Conclusions}
We have obtained Gemini Planet Imager J, H, K1, and K2-Spec observations of the iconic HR 4796A debris disk. We find the following:

1. The H-band Scattering Phase Function (SPF) is forward scattering at small scattering angles with a minimum near 60$\arcdeg$ and backward scattering at larger scattering angles, reaching a plateau at $\sim$150$\arcdeg$, indicating that it is slightly less forward scattering than previously reported (\citetalias{milli17}).

2. The J, H, K1, and K2 Scattering Phase Functions are consistent with one another to within the uncertainties of the observations, suggesting there is no evidence for changes in the SPF as a function of wavelength at near-infrared wavelengths. 

3. The scattering phase function is well modeled assuming a smooth, offsetted, axis symmetric ring, indicating that there are no large asymmetries within the dust distribution. However, a marginal excess in the residuals at pericenter could be consistent with a pericenter dust enhancement recently reported by \cite{olofsson19}.

4. The SPHERE H2 total intensity scattering phase function is well modeled using large, irregular grains composed of amorphous silicate, amorphous carbon, and metallic iron. The large minimum grain size is consistent with expectations for irregular grains based on Discrete Dipole Approximation calculations.

5. This dust grain model also reproduces the ratio of the SPHERE H2 to STIS SPFs and the shape of the visual polarized intensity phase function, reinforcing the conclusion that the disk is composed of large, irregular grains.

\acknowledgements
Based on observations obtained at the Gemini Observatory, which is operated by the Association of Universities for Research in Astronomy, Inc., under a cooperative agreement with the NSF on behalf of the Gemini partnership: the National Science Foundation (United States), the National Research Council (Canada), CONICYT (Chile), Ministerio de Ciencia, Tecnolog\'{i}a e Innovaci\'{o}n Productiva (Argentina), and Minist\'{e}rio da Ci\^{e}ncia, Tecnologia e Inova\c{c}\~{a}o (Brazil). This material is based upon work supported by the National Science Foundation under Astronomy and Astrophysics Grant No.~1616097 (JM, BR) and No.~1518332 (TME, RJDR, JRG, PK, GD) and NASA grants NNX15AC89G and NNX15AD95G/NExSS (TME, RJDR, GD, JJW, PK). JM acknowledges support for part of this work was provided by NASA through the NASA Hubble Fellowship grant \#\textit{HST}-HF2-51414 awarded by the Space Telescope Science Institute, which is operated by the Association of Universities for Research in Astronomy, Inc., for NASA, under contract NAS5-26555. This work benefitted from NASA's Nexus for Exoplanet System Science (NExSS) research coordination network sponsored by NASA's Science Mission Directorate. This research has made use of the VizieR catalogue access tool, CDS,  Strasbourg, France (DOI: 10.26093/cds/vizier)). The original description of the VizieR service was published in A\&AS 143, 23 \citep{ochsenbein00}.  This research project (or part of this research project) was conducted using computational resources (and/or scientific computing services) at the Maryland Advanced Research Computing Center (MARCC). This paper is dedicated to UCLA  Professor Michael Jura who not only discovered the very bright infrared excess associated with HR 4796A but also taught a generation of young astronomers how to think about debris disks. 
 
\software{Gemini Planet Imager Data Pipeline (\citealt{perrin14, perrin16}, \url{http://ascl.net/1411.018}), pyKLIP (\citealt{wang15}, \url{http://ascl.net/1506.001}), numpy, scipy, Astropy \citep{astropy2018}, matplotlib \citep{matplotlib2007, matplotlib_v2.0.2}, emcee (\citealt{foreman-mackey13}, \url{http://ascl.net/1303.002}), corner (\citealt{corner18}, \url{http://ascl.net/1702.002}), DebrisDiskFM \citep{ren19}}. 

\appendix

\section{Two component Henyey-Greenstein vs Three component Henyey-Greenstein Scattering Phase Function}\label{appendix-3g}
Our analysis clearly recovered a different SPF than the one published in \citetalias{milli17} (see Fig. \ref{fig:compare_spf}). We were concerned that the main reason for this difference was that our model was too simple. The SPF extracted by \citetalias{milli17} directly from the image may have been too complex to be reproduced by a two component Henyey-Greenstein Phase Function. In this appendix we show that increasing the level of complexity of the scattering phase function does not improve the fit and therefore that a 2 component Henyey-Greenstein Phase Function is sufficient to describe the SPF in these data. Specifically, we introduced a third HG component to our SPF and compare to previous result. Equation \ref{eq:2g_spf} now becomes:
\begin{equation}
\label{eq:3g_spf}
\begin{split}
p_3(g_1,\alpha_1,g_2,\alpha_2, g_3,\theta) =& \alpha_1 HG(g_1, \theta) + \alpha_2 HG(g_2, \theta) \\
&+(1- \alpha_1 - \alpha_2) HG(g_3, \theta).
\end{split}
\end{equation}

\begin{figure}
\centering \includegraphics[width=0.49\textwidth]{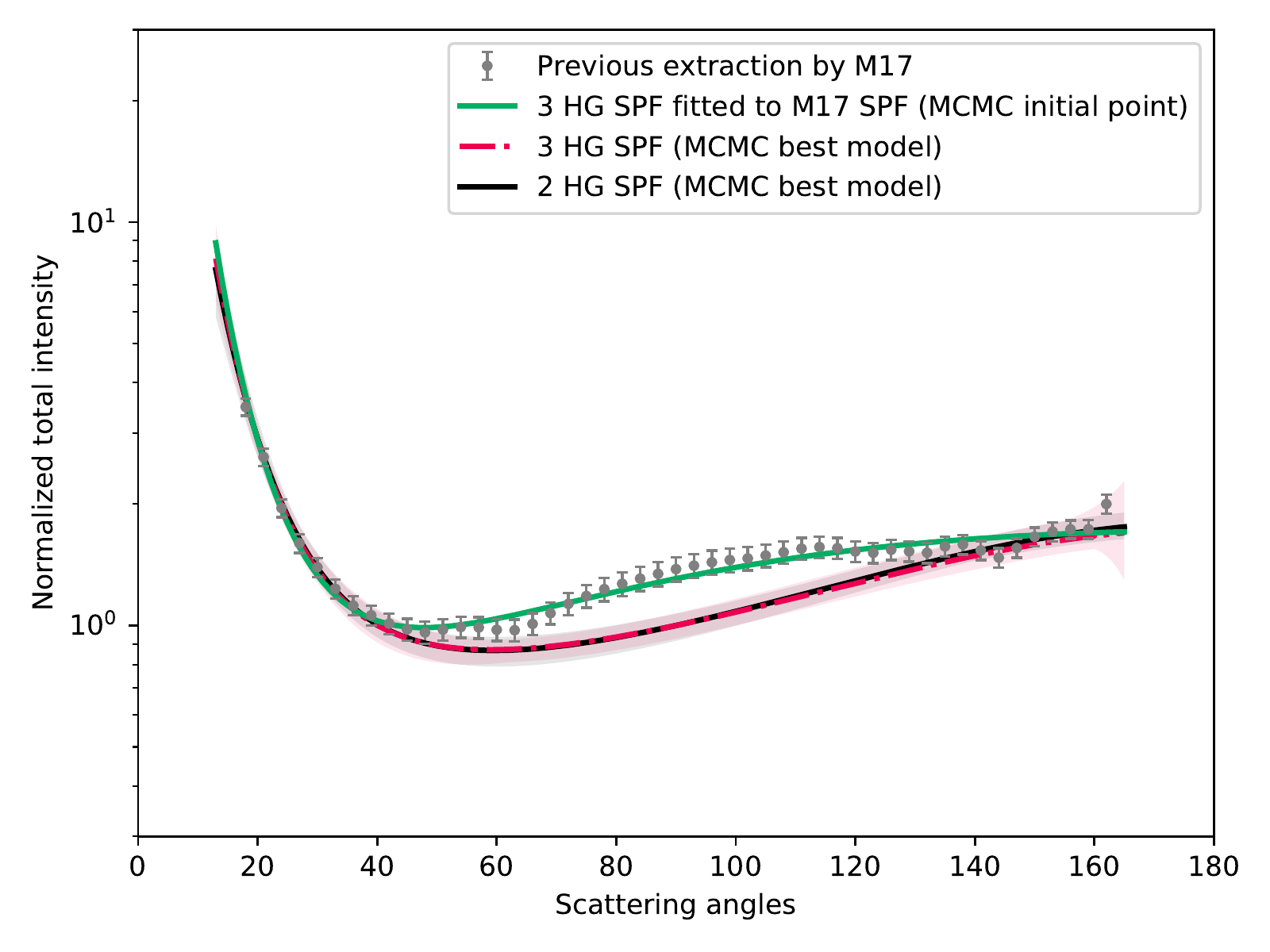}
\figcaption{
 \label{fig:compare_3g_SPF} SPHERE H2 SPF: comparison of exctraction methods. The gray points and error bars are the SPF previously extracted by \citetalias{milli17} from a cADI PSF-subtracted image and then corrected by dividing by a forward model. We fit a 3 component HG function to this SPF (green solid line). We over-plot SPFs obtained by the MCMC fitting extraction method on the SPHERE H2 data with a 3 component HG (red dashed line) and a 2 component HG (black solid line). The shaded regions are plotted by measuring the SPF of 1000 randomly chosen accepted SPF from the MCMC sampler.}
\end{figure}

Using the SPF exctracted by \citetalias{milli17}, we fitted a 3 component Henyey-Greenstein function (green solid line in \ref{fig:compare_3g_SPF}). This shows that except for a few points, the general shape exctracted by \citetalias{milli17} can be reproduced by the 5 parameters ($g_1$,$g_2$,$g_3$,$\alpha_1$,$\alpha_2$) of a 3 components Henyey-Greenstein function. We then ran the fitting extraction method on the SPHERE H2 data using the same model as previously but with a 3 component Henyey-Greenstein function (13 parameter model instead of 11 parameter model). We chose as initial points the SPF parameters which best fitted the SPF exctracted by \citetalias{milli17} (green solid line) to be sure the MCMC would consider this part of the parameter space.

The resulting best fit 3 component Henyey-Greenstein SPF is shown in \ref{fig:compare_3g_SPF} (red dashed line). The black solid line is the 2 component HG function extracted using this same method (see Section \ref{sec:mcmc_sphere}). First, this shows that the best fits from the 2 component HG function MCMC and the 3 component HG function MCMC almost totally overlap.  The only difference is at very small scattering angle ($<18\arcdeg$) where the 3 HG component SPF is very slightly more forward scattering than 2 HG component SPF. However, considering the error bars in this region of the image, we considered that for this disk and for the range of scattering angle probed with this observation ($13.5\arcdeg$ to $166.5\arcdeg$), a 2 HG component (3 parameters) can produce a function complex enough to reproduce the actual scattering phase function. For this reason, we only used a 2 HG component in this study for all the data. Second, the 3 component HG function fitted to the the SPF extracted by \citetalias{milli17} (green solid lime) is within the parameter range accessible (i.e. within the prior) of the MCMC sampler. However, once again, it is significantly different than the one favored by the sampler a posteriori, which is why we favored ours for the theoretical analysis.

\section{Scattering phase function: extraction of known parameters}
\label{appendix-injection}


Accurately measuring the SPF is very challenging. In this paper, we used two different methods to estimate the SPF: a direct extraction from the image and an image-based forward modeling fitting in an MCMC framework. To ensure that these techniques provide correct results on the SPF, we performed a series of tests by injecting a model disk in an empty GPI dataset. This Appendix shows the results of those tests. 

\subsection{Validating the Reconstruction Analysis}
\label{appendix-injection-with-exctraction}

\begin{figure*}
\plottwo{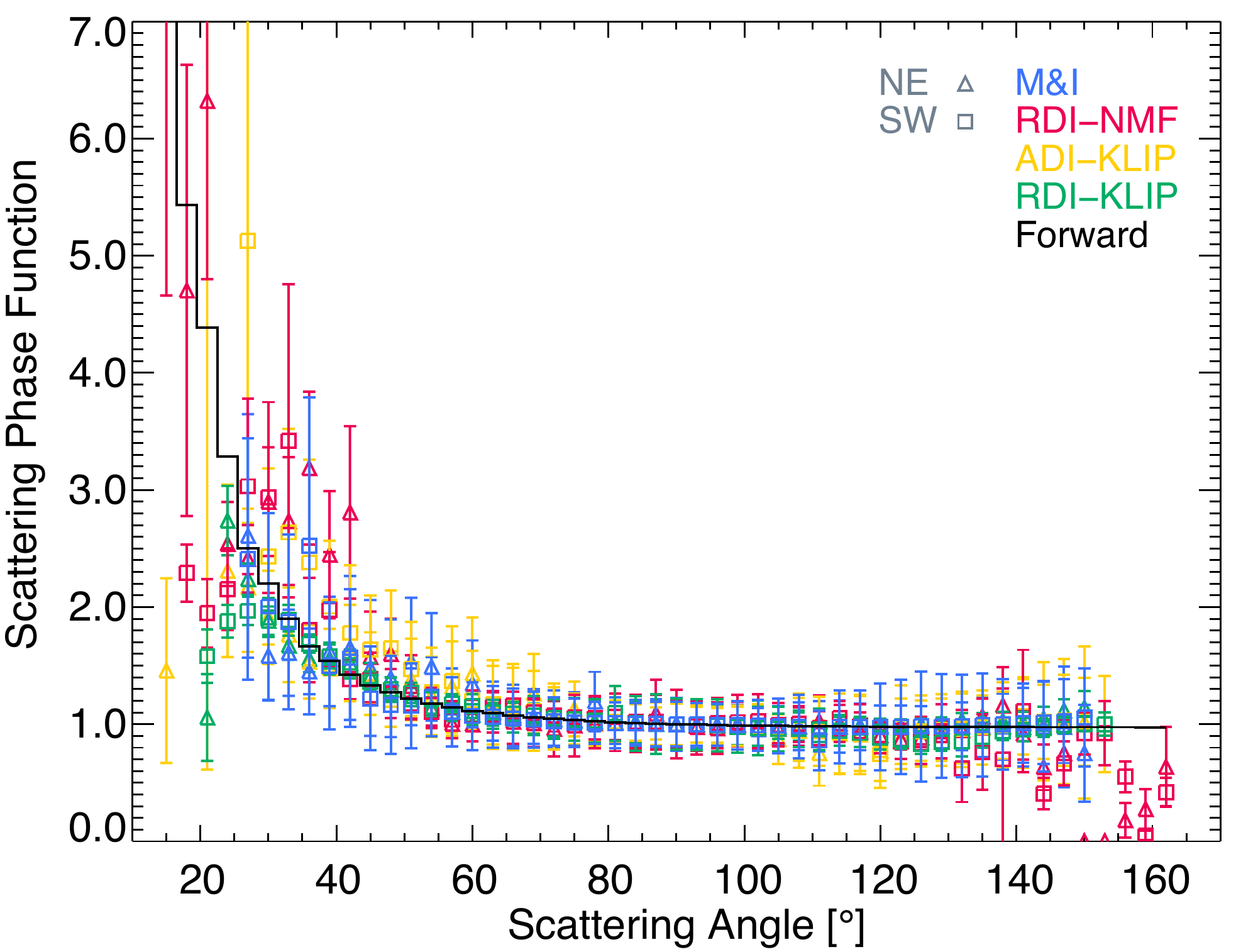}{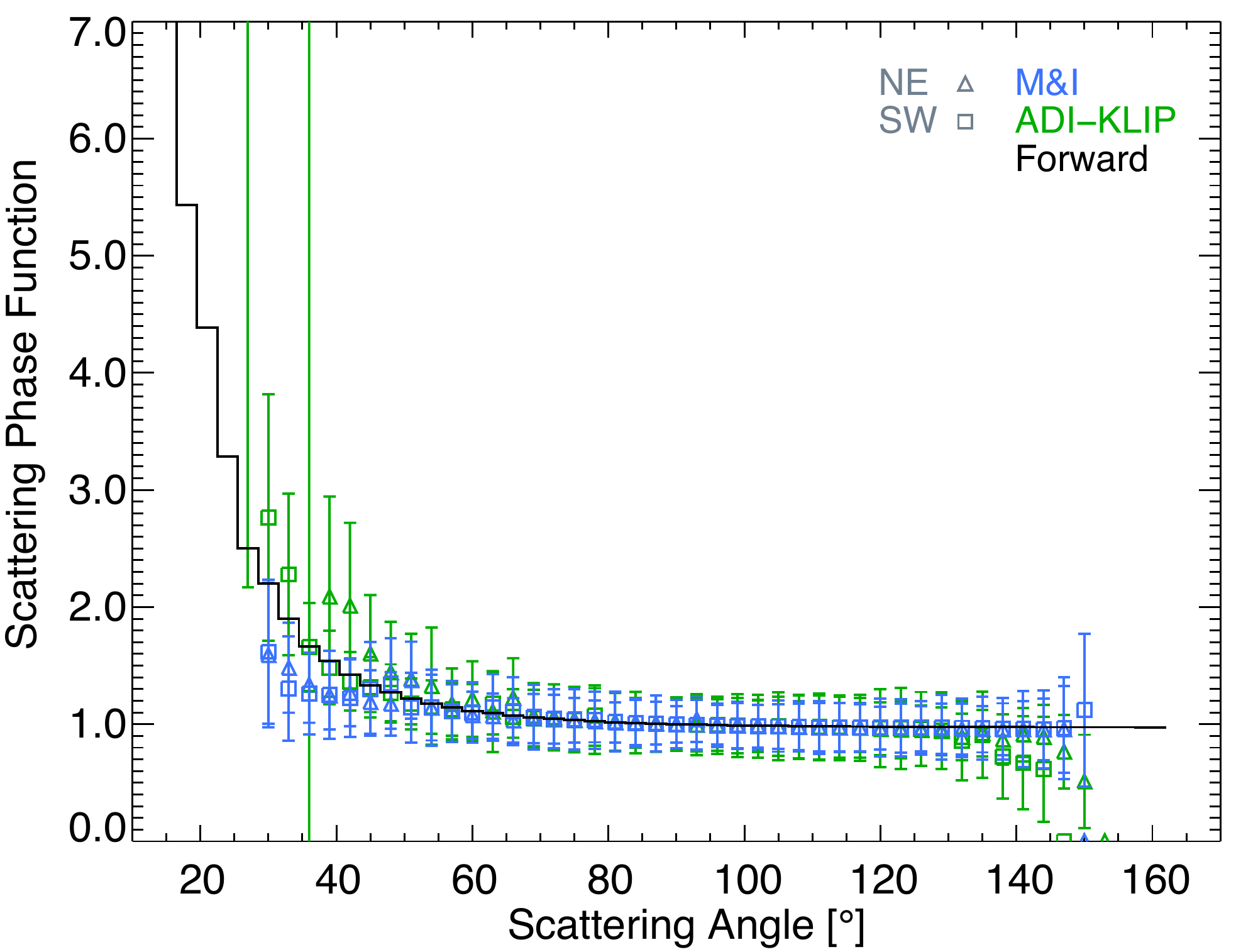}
\figcaption{Scattering phase functions reconstructed from a forward-scattering, HR 4796A-like disk at (left) H-band and (right) K1-band. The phase functions extracted using Mask-and-Interpolate (blue), RDI-NMF (pink), ADI-KLIP (yellow), and RDI-KLIP (green) are over plotted with the assumed scattering phase function (black) for comparison. Unfortunately, the library of reference PSFs is substantially smaller in J and K1 than in H; therefore, reference PSF subtraction for J- and K1 bands is currently not feasible. 
\label{fig:miespf}}
\end{figure*}

We demonstrate that reconstruction of the HR 4796A SPF should be feasible by recovering the SPF from an idealized HR 4796A-like disk with strong forward-scattering from a disk-injected, empty, observing sequence. We use {\tt MCFOST} to simulate a disk with spherical grains composed of 70\% silicates and 30\% metallic iron by volume, a grain size distribution, $dn/da$ $\propto$ $a^{-4}$, a minimum grain size $a_{min}$ = 5 $\mu$m, and a maximum grain size, $a_{max}$ = 1 mm. We inject the HR 4796-like disk into an empty GPI observing sequence using the same process described for the isotropic disk. As with the isotropic disk, we convolve the {\tt MCFOST} disk image with an unblocked, Spec observation of HD 118335 and inject our convolved model into an empty GPI sequence. We PSF subtract the injected, forward-scattering disk sequence, and reconstruct the SPF from the PSF subtracted image. Finally, we divide the SPF extracted from the injected, forward-scattering, disk sequence by that extracted from the injected, isotropic, disk sequence. We find that we can reconstruct the Mie disk SPFs in H and K1 bands (see Figure~\ref{fig:miespf}) at scattering angles, 40$\arcdeg$ $<$ $\theta$ $<$ $140$ $\arcdeg$. Our process fails to accurately recover the SPF at smaller and larger scattering angle where the disk has a smaller angular offset from the coronagraphic spot and the image suffers from larger speckle noise. Thus, we anticipate that reconstruction of the HR 4796A SPF at scattering angles, 40$\arcdeg$ $<$ $\theta$ $<$ $140$ $\arcdeg$ should be feasible.

\subsection{Validating the Modeling Analysis}
\label{appendix-injection-with-mcmc}

\begin{figure*}
\centering \includegraphics[width=0.8\textwidth, trim=0 80 0 0, clip]{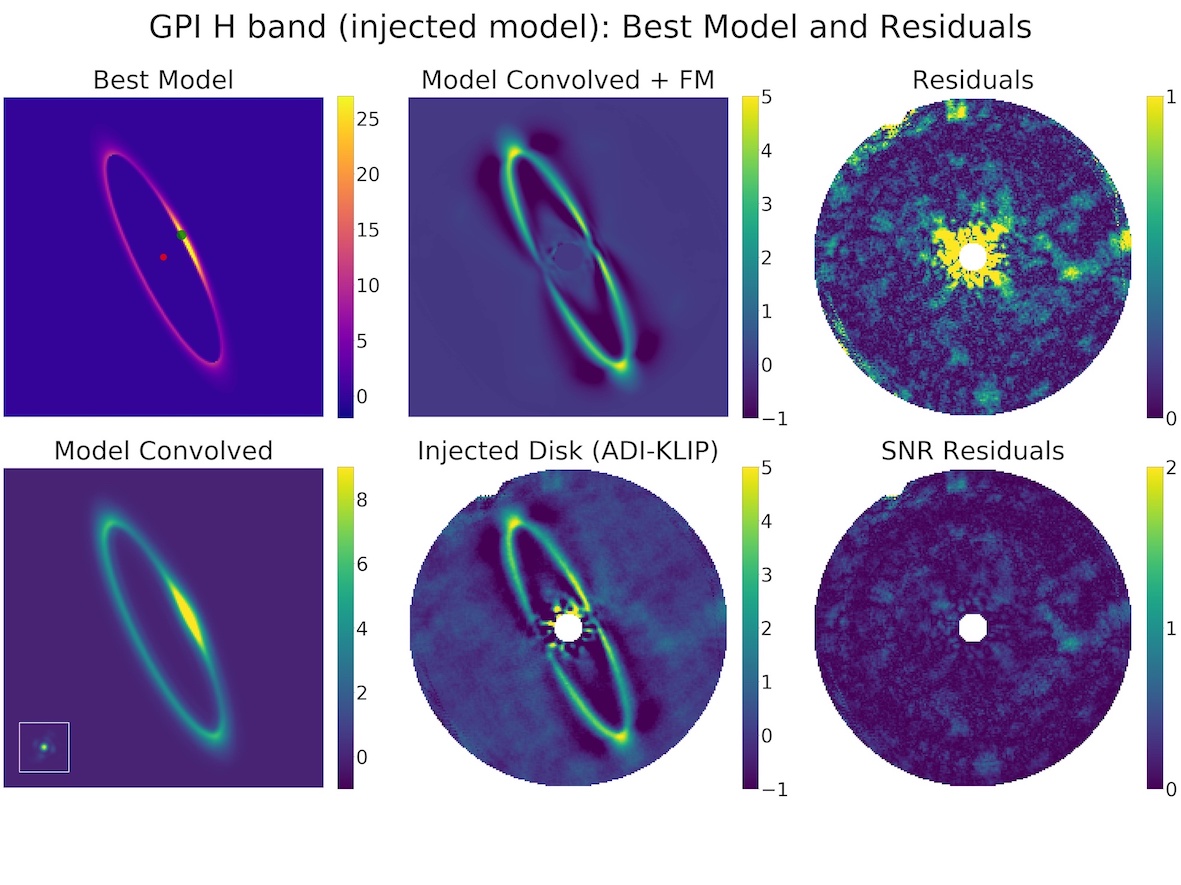}
\figcaption{
 \label{fig:testmodels} Best-fit model resulting of the MCMC: Same as Figure~\ref{fig:gpihmodel} for an HD 48524 H-band dataset with an injected HR 4796A-like disk with known parameters.}
\end{figure*} 

We constructed an image assuming a disk with an inner radius, $R_1$ = 75 AU, an outer radius, $R_2$ = 100 AU, a power-law surface distribution, $\beta$ = 12.4, asymmetric scattering parameters, $g_1$ = 0.825 and $g_2$ = 0.201, a relative weight, $\alpha$ = 0.298, an inclination, $i$ = 76.8$\arcdeg$, a position angle, $PA$ = 26.64$\arcdeg$, stellar positional offsets, $dx$ = -2.0 AU and $dy$ = 0.94 AU, and a flux normalization, $N$ = 80 (hereafter, ``injected parameters''). Next, we injected the model into an empty GPI H-Spec observational sequence, with the same parallactic angles as in our HR 4796A dataset. This sequence was obtained with GPI the star HD 48525, on January 28th, 2018. Finally, we reduced this sequence using the same ADI-KLIP method as for the GPI HR 4796A observations, producing the image shown in Fig. \ref{fig:testmodels} (Bottom-Middle) and applied our MCMC method to extract parameters from this reduction (hereafter ``recovered parameters''). 

Our goal for this test was to show that disk modeling recovers the injected parameters to within $1 \sigma$ of the values assumed. However, during initial testing, we discovered that the MCMC error bars were too small (the injected parameters were originally within 2 or 3$\sigma$ of the recovered parameters), suggesting that we underestimated the noise in our uncertainties maps (probably because we made an axisymetric noise assumption). Therefore, we multiplied our noise map by a scalar factor of 4 to obtain PDFs with a Gaussian shape and parameters well within 1$\sigma$ of the injected parameters. We estimated the scalar factor with which to multiple the noise by requiring that the amplitude of the noise in our residual maps has SNR$<$2. For our data, we typically multiplied our uncertainties maps by factors of 3-5. Scalar multiplication of the noise map does not change the maximum value of the likelihood, but does produce larger error bars.

\begin{figure*}
\plotone{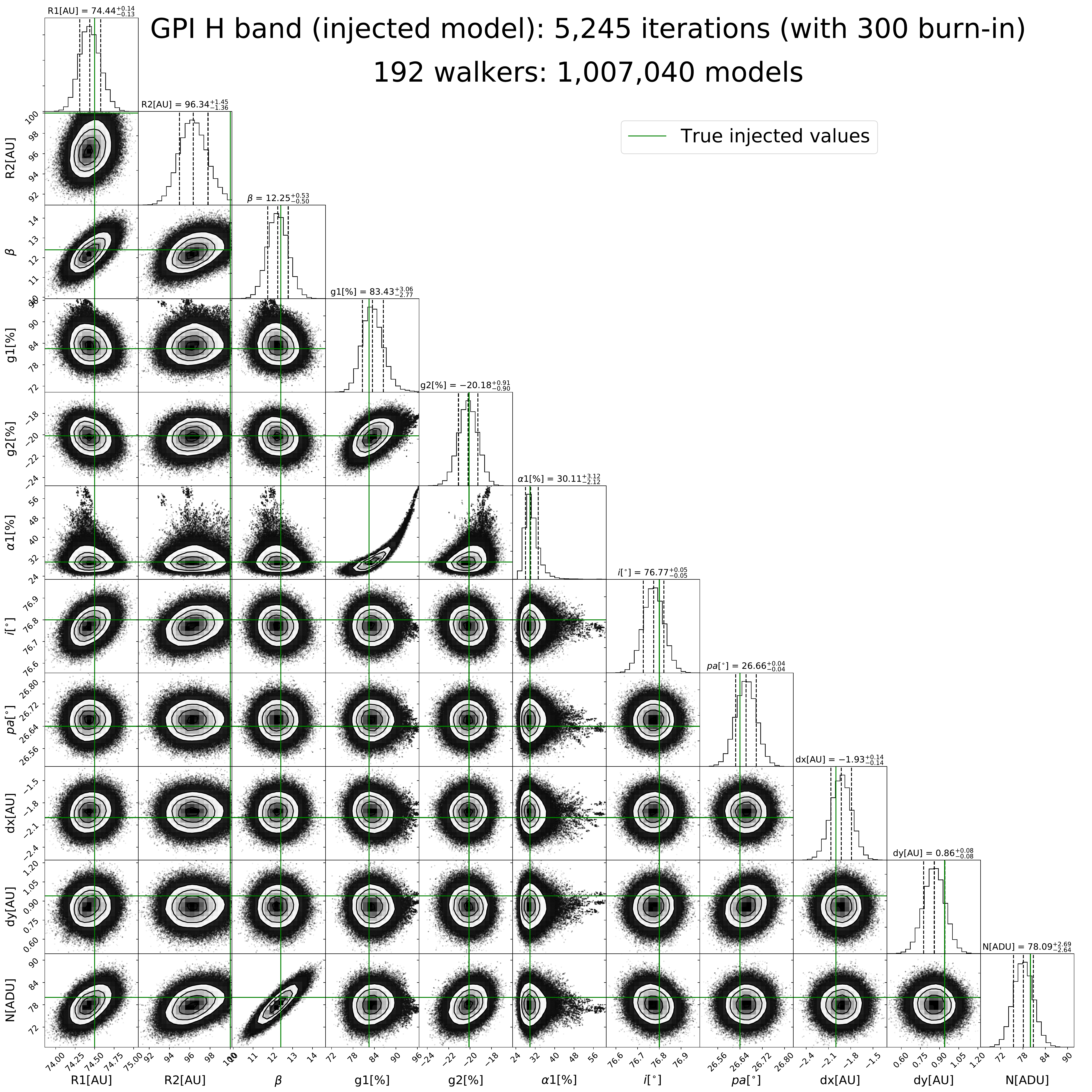}
\figcaption{
 \label{fig:testposteriors} MCMC posterior distributions recovered for a HR 4796A-like disk injected in a HD 48525 dataset with a 2 component HG phase function. The diagonal histograms show the posterior distributions of each parameter marginalized over all other parameters. In each plot, the dashed lines show the 16\%, 50\%, and 84\% percentiles. The off-diagonal plots display the joint probability distributions with contour levels at the same percentiles. For each posterior, the parameter ``true'' value of the injected disk is overplotted in green.}
\end{figure*} 

\begin{figure}
\centering \includegraphics[width=0.49\textwidth]{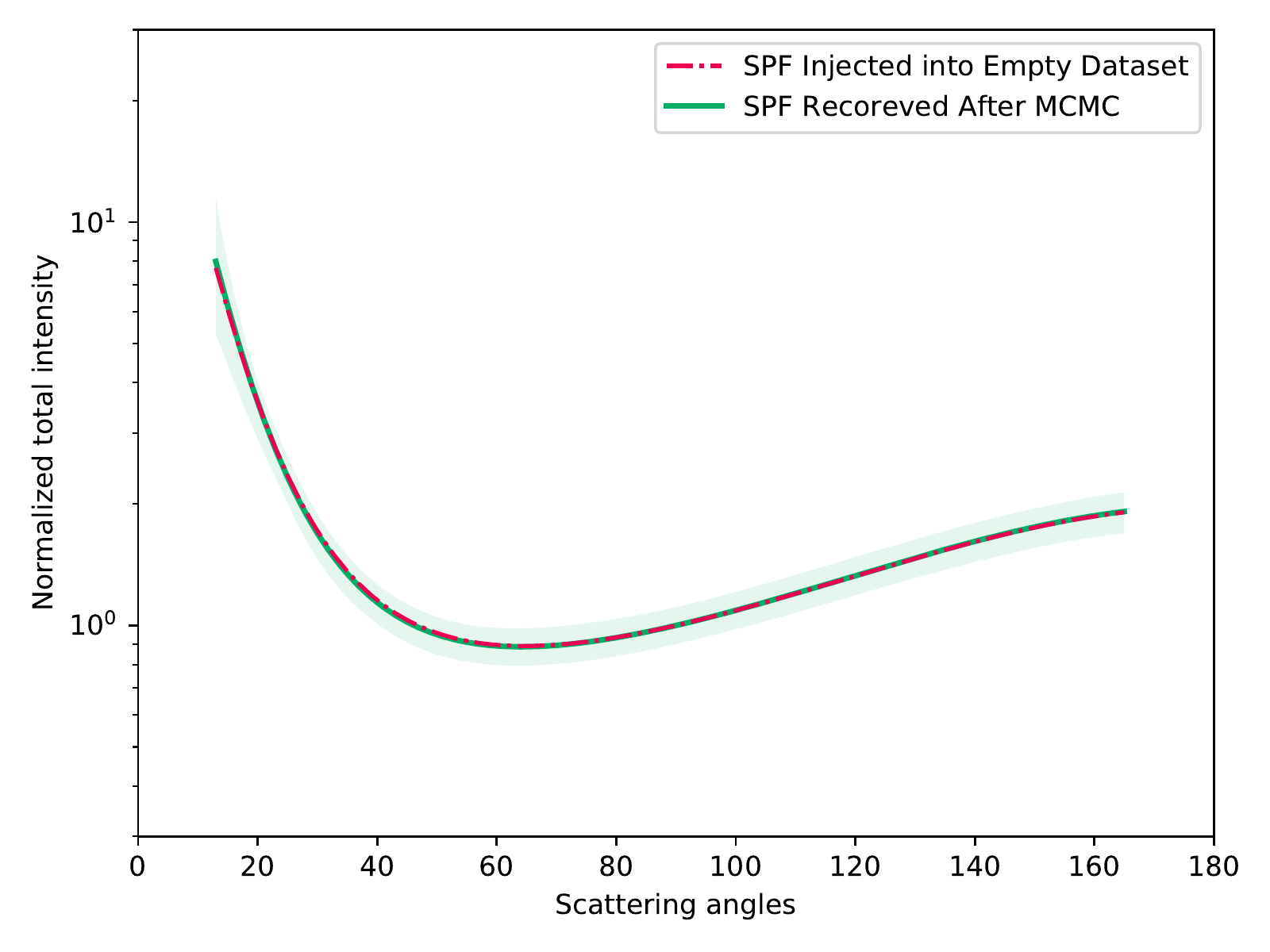}
\figcaption{
 \label{fig:recovered_spf} Scattering Phase Function injected into disk image (red dashed line) compared with the estimated phase function recovered from disk fitting (green solid line). The shaded regions in the plots show the uncertainties and are estimated from 1000 randomly chosen SPFs generated by the MCMC sampler.}
\end{figure}

In this initial test, we recovered the values for all of the injected parameters (except $R_2$) to within 1$\sigma$ (see Figure~\ref{fig:testposteriors}, injected parameter values are plotted as green lines), with a noise map multiplied by a scalar factor of 4. Only the value recovered for the outer radius ($R_2$ = 96$\pm$1 AU) was $>$4$\sigma$ different from that assumed in the model, which was expected (see previous section). We perfectly recovered the injected SPF. In the left panel of Figure~\ref{fig:recovered_spf}, we show the injected (red dashed line) and recovered SPFs (green solid line shows the SPF and the shaded green area its associated uncertainty). For this PSF plot and all SPF plots here after, the shaded regions represent the region covered by 1000 randomly chosen SPFs from the MCMC sampler, after convergence of the MCMC.

In thissimple test, the model science images were generated us-ing same code that was used to simulate the disk in ourretrievals  and  the  disk  was  completely  removed  in  theresiduals images. However, our simple 11 parameter diskmodel did not reproduce the complex structure observedat high SNR in our HR 4796A images. We attributed thechallenges in our disk modeling to two main limitations.First, we assumed that the dust density was azimuthallysymmetric within the disk.  However, the residual mapsshowed significant local differences in the surface bright-ness along the major axis that could not be explained bythe SPF because the SPF is symmetric with respect to the minor-axis.  Second, our satellite spot PSF had relativelylow SNR. To build SNR, we averaged all of the satellitespots observations within an observational sequence.  Asa result, we did taken into account changes in observingconditions on fast timescales.

\section{MCMC detailed results}
\label{appendix-figsmcmc}

This Appendix shows the products of the MCMC for all dataset used in this paper. We shows the corner plots for GPI J-, H-, K1-, K2-band and SPHERE H2-band and the best model for GPI J-, K1- and K2-band.

\begin{figure*}
\plotone{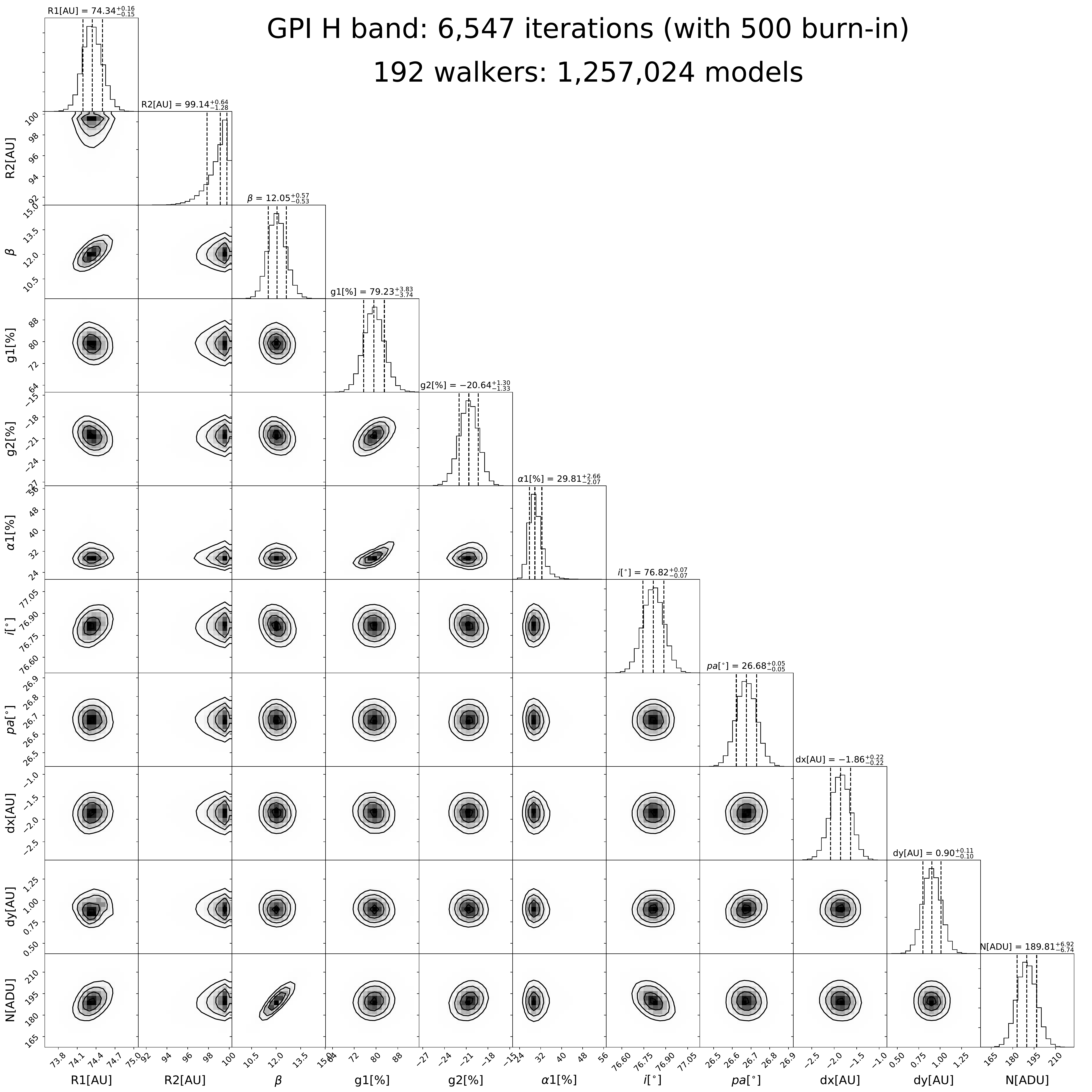}
\figcaption{
 \label{fig:gpihposteriors} Same as Figure~\ref{fig:testposteriors} for ADI-KLIP GPI H-band of the HR 4796A debris disk.}
\end{figure*} 

\begin{figure*}
\plotone{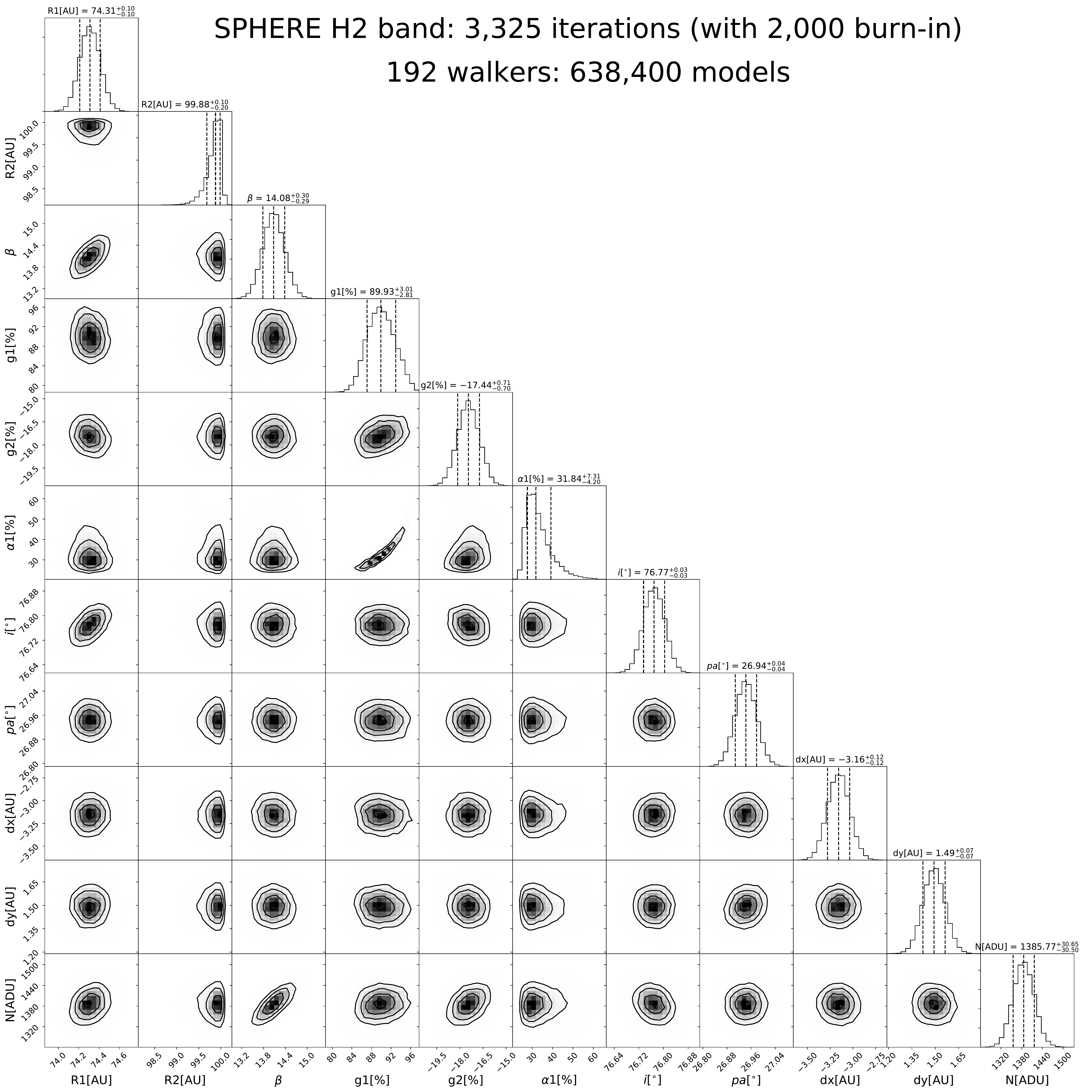}
\figcaption{
 \label{fig:sphereposteriors} Same as Figure~\ref{fig:testposteriors} for ADI-KLIP SPHERE H2-band of the HR 4796A debris disk.}
\end{figure*}

\begin{figure*}
\centering \includegraphics[width=0.6\textwidth, trim=0 80 0 0, clip]{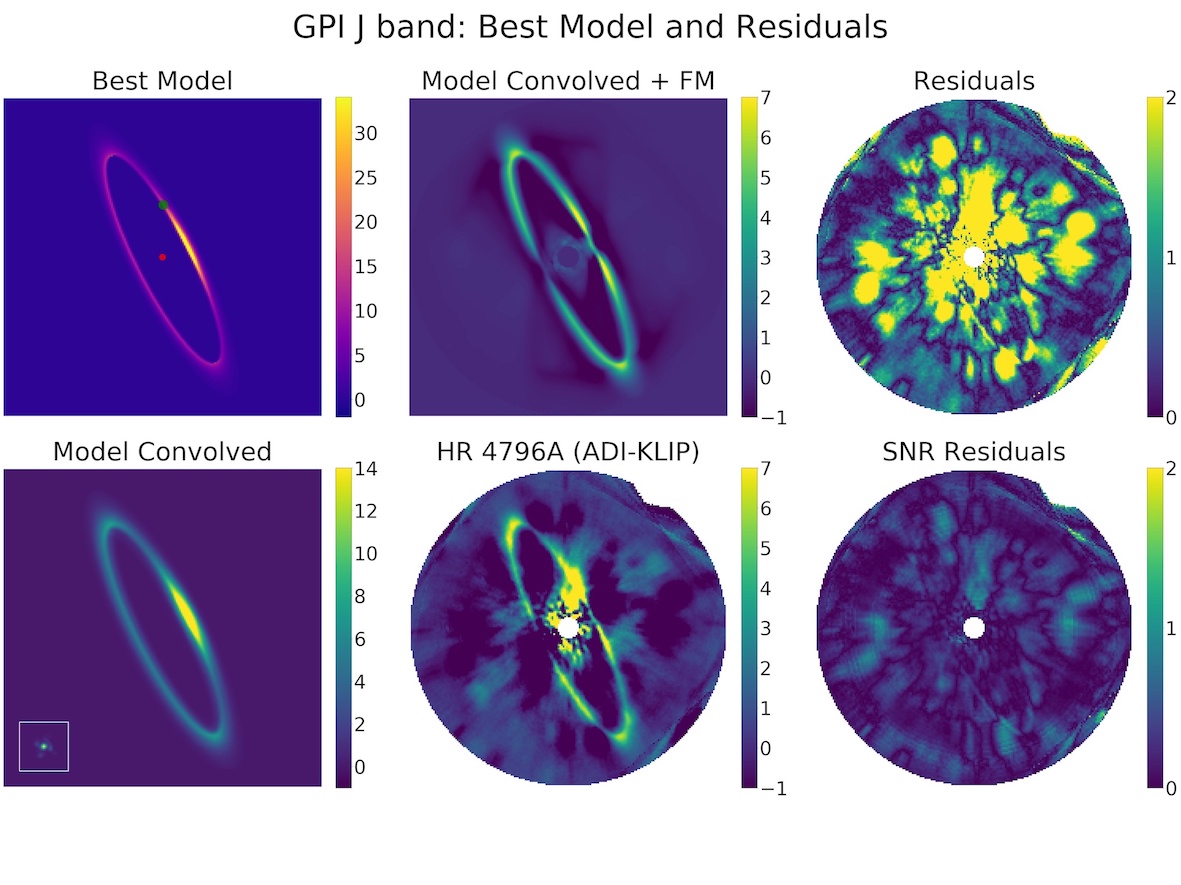}
\figcaption{
 \label{fig:gpijmodel} Best-fit model resulting of the MCMC: Same as Figure~\ref{fig:gpihmodel} for GPI J-band.}
\end{figure*} 

\begin{figure*}
\centering \includegraphics[width=0.75\textwidth]{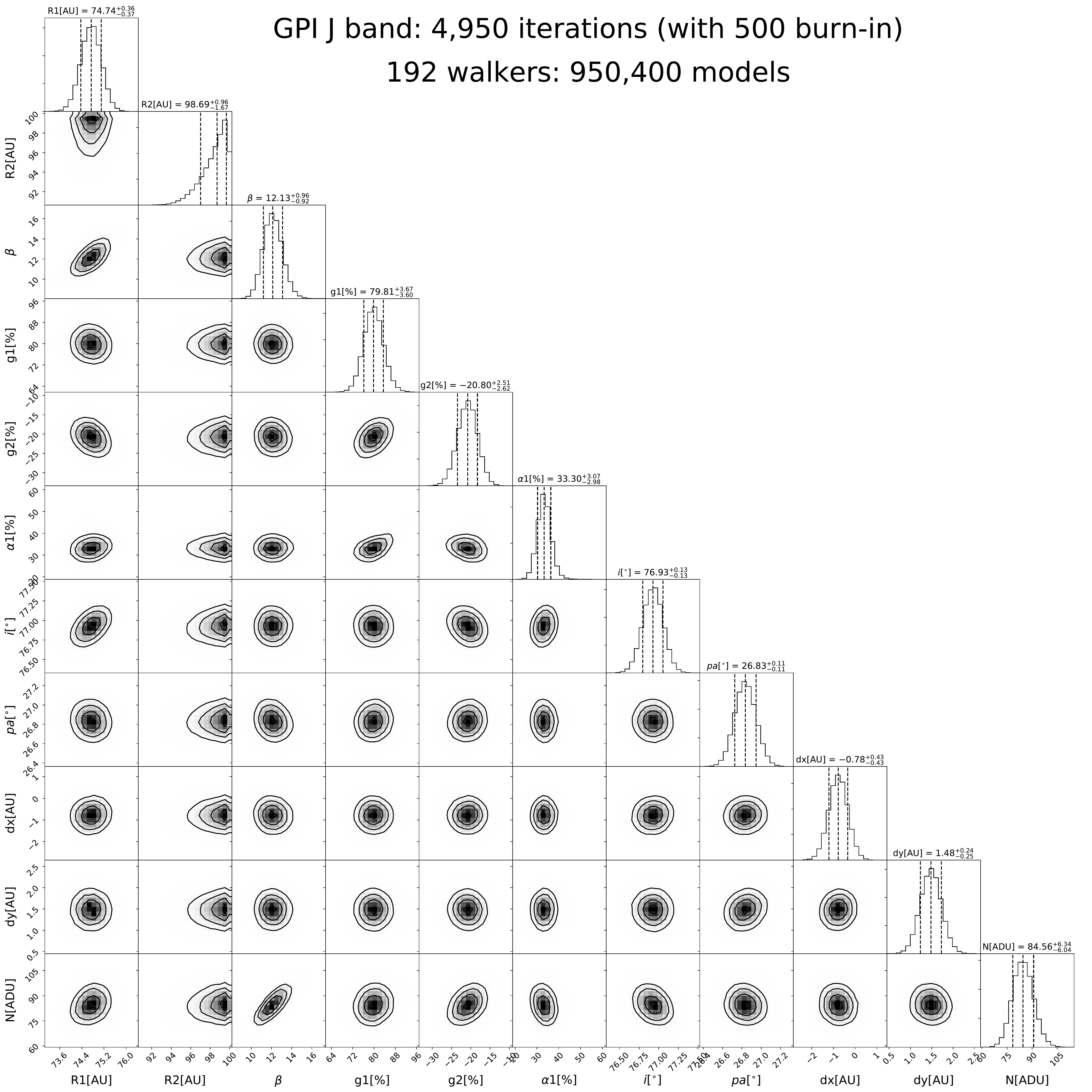}
\figcaption{
 \label{fig:gpijposteriors} Same as Figure~\ref{fig:testposteriors} for ADI-KLIP GPI J-band of the HR 4796A debris disk.}
\end{figure*}

\begin{figure*}
\centering \includegraphics[width=0.6\textwidth, trim=0 80 0 0, clip]{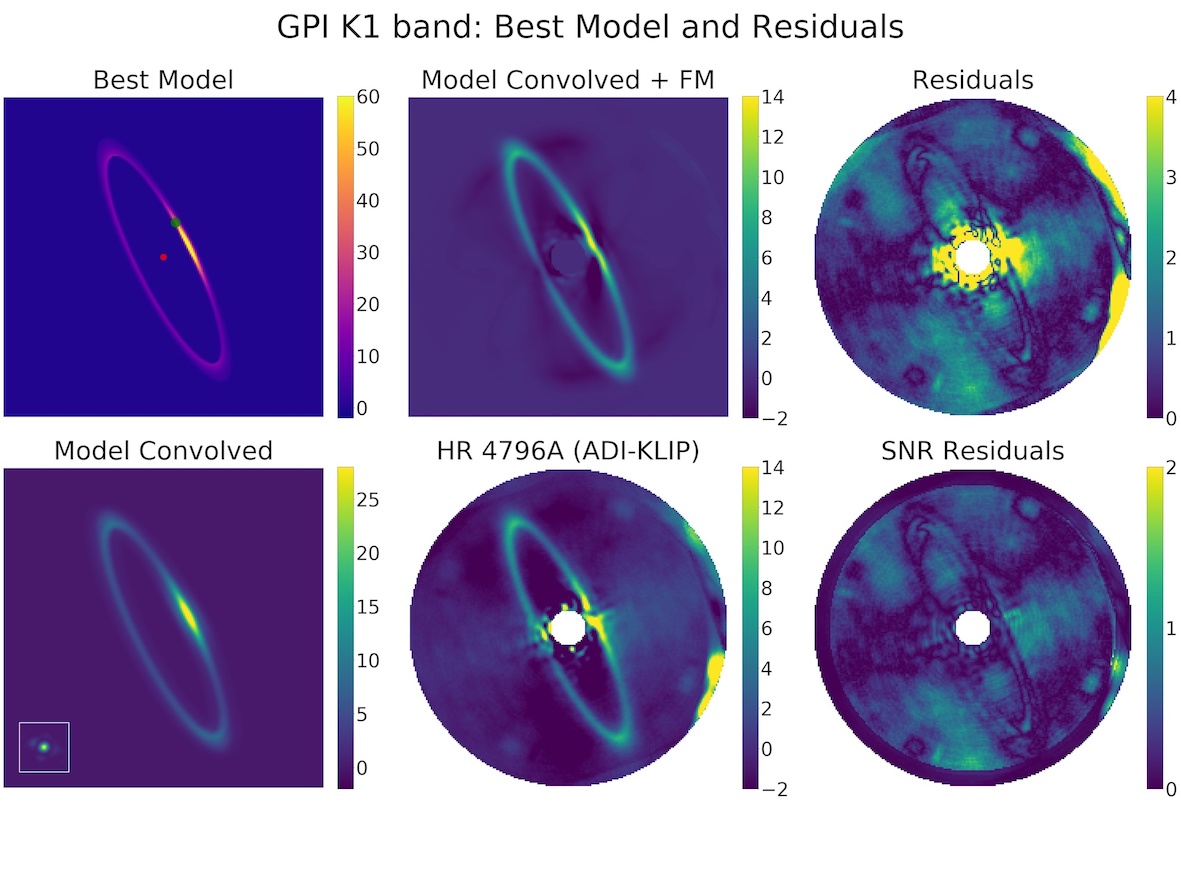}
\figcaption{
 \label{fig:gpik1model} Best-fit model resulting of the MCMC: Same as Figure~\ref{fig:gpihmodel} for GPI K1-band.}
\end{figure*} 

\begin{figure*}
\centering \includegraphics[width=0.75\textwidth]{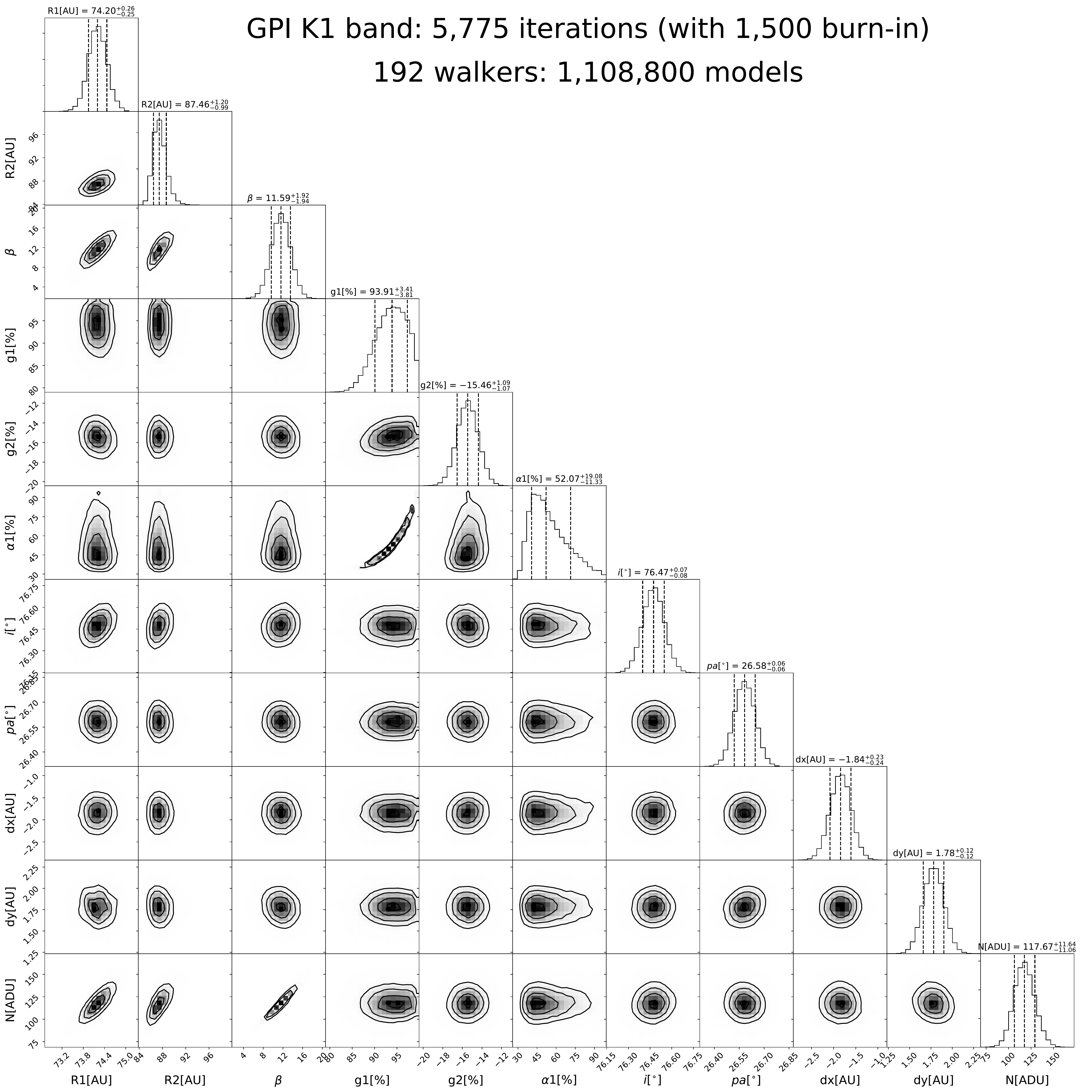}
\figcaption{
 \label{fig:gpik1posteriors} Same as Figure~\ref{fig:testposteriors} for ADI-KLIP GPI K1-band of the HR 4796A debris disk.}
\end{figure*}

\begin{figure*}
\centering \includegraphics[width=0.6\textwidth, trim=0 80 0 0, clip]{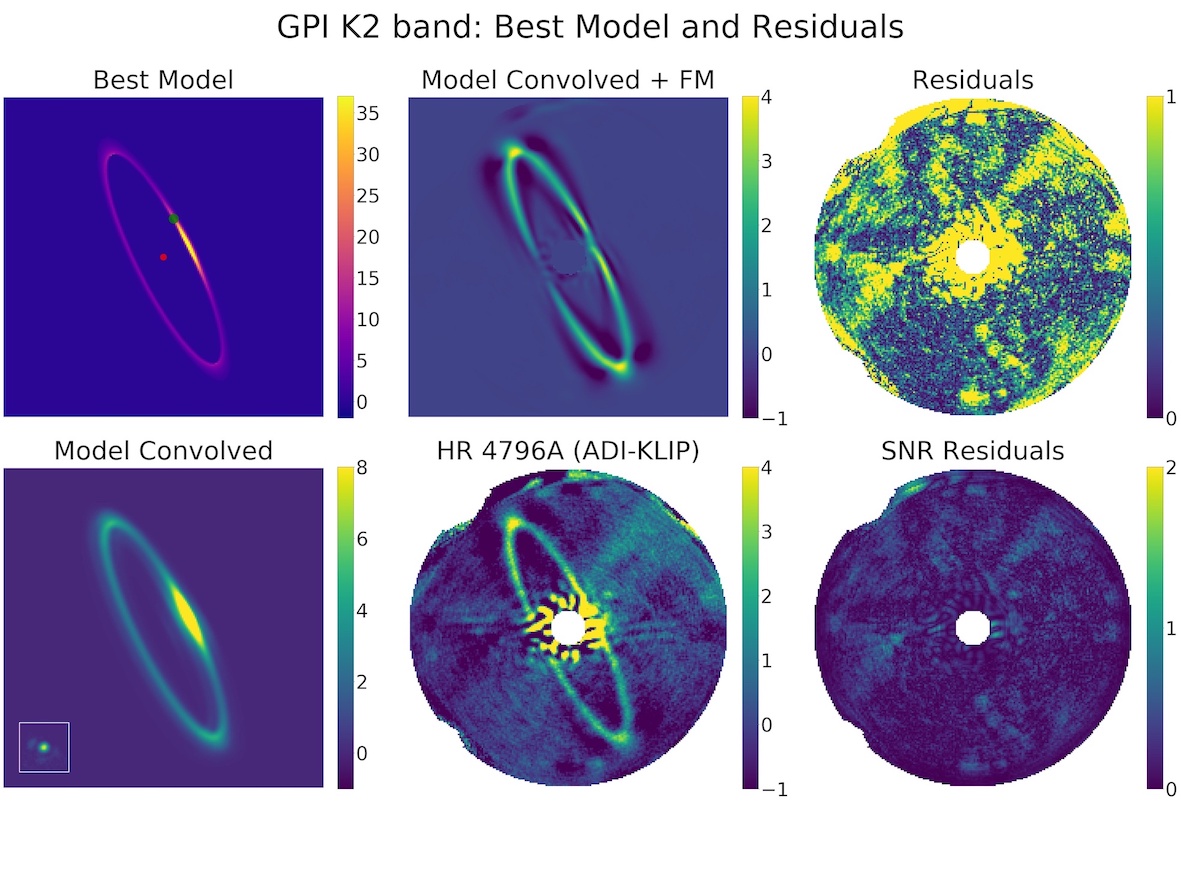}
\figcaption{
 \label{fig:gpik2model} Best-fit model resulting of the MCMC: Same as Figure~\ref{fig:gpihmodel} for GPI K2-band.}
\end{figure*} 

\begin{figure*}
\centering \includegraphics[width=0.75\textwidth]{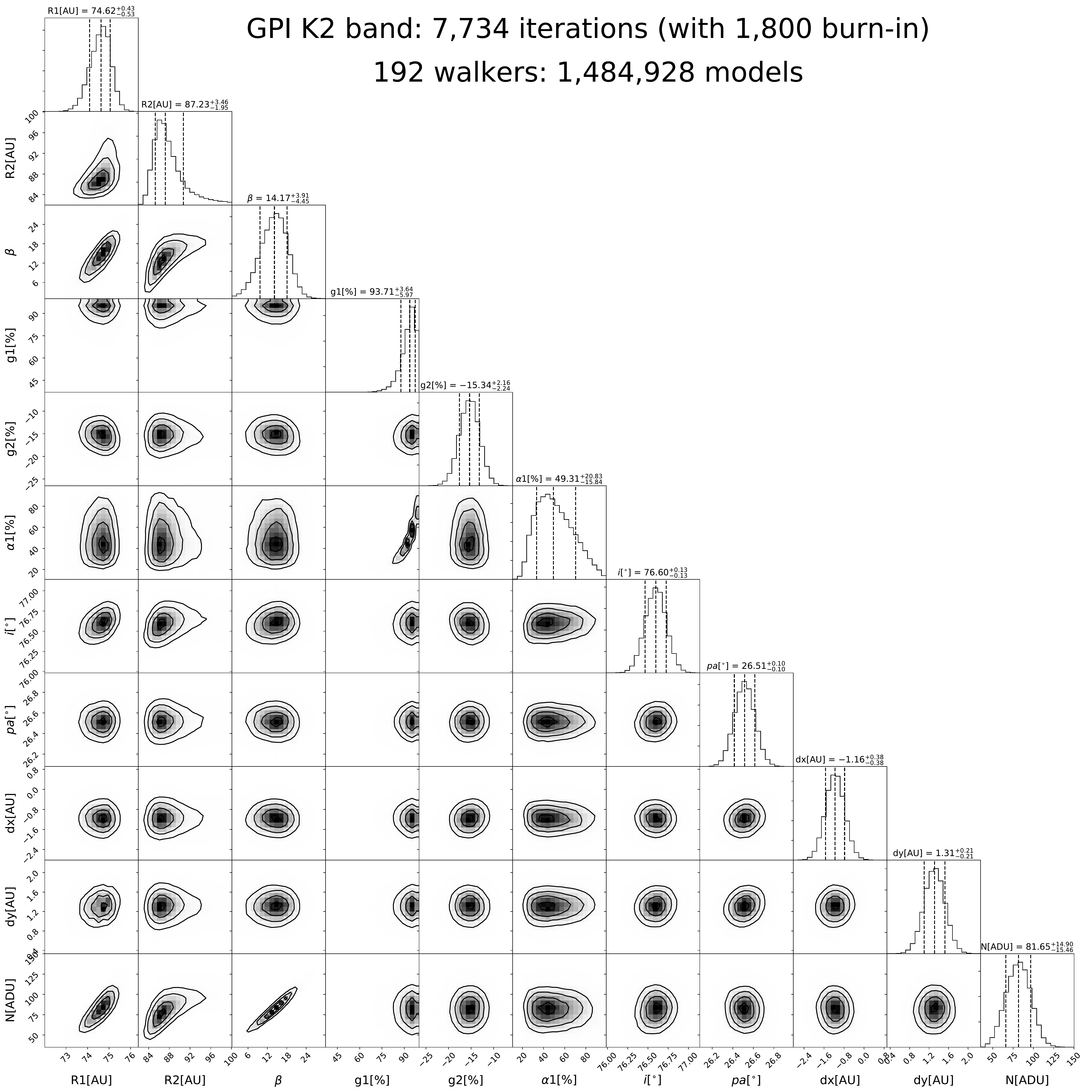}
\figcaption{
 \label{fig:gpik2posteriors} Same as Figure~\ref{fig:testposteriors} for ADI-KLIP GPI K2-band of the HR 4796A debris disk.}
\end{figure*}

\bibliography{chen2020}

\end{document}